\documentclass[aoas,preprint]{imsart}

\RequirePackage[OT1]{fontenc}
\RequirePackage{amsthm,amsmath}
\RequirePackage{natbib}
\RequirePackage[colorlinks,citecolor=blue,urlcolor=blue]{hyperref}

\usepackage{amssymb,amsfonts}
\usepackage{graphicx,psfrag,epsf}
\usepackage{enumerate}
\usepackage{booktabs}
\usepackage{longtable,array}
\usepackage{tabularx}
\usepackage{colortbl}
\usepackage{dblfloatfix}
\usepackage{xcolor}
\newcolumntype{A}{>{\hsize=0.9\hsize}X}
\newcolumntype{B}{>{\hsize=0.4\hsize}X}
\newcolumntype{C}{>{\hsize=1.2\hsize}X}

\usepackage{cleveref}

\usepackage[ruled,norelsize,vlined]{algorithm2e}
\SetAlgoSkip{}
\newcommand{\beginsupplement}{%
        \setcounter{table}{0}
        \renewcommand{\thetable}{S\arabic{table}}%
        \setcounter{figure}{0}
        \renewcommand{\thefigure}{S\arabic{figure}}%
     }

\usepackage{comment}
\usepackage{xurl}
\arxiv{}

\startlocaldefs
\numberwithin{equation}{section}
\theoremstyle{plain}

\newtheorem{lemma}{Lemma}[section]
\newtheorem{prop}{Proposition}[section]
\endlocaldefs

\begin{document}

\begin{frontmatter}
\title{Multivariate mixed membership modeling: Inferring domain-specific risk profiles}
\runtitle{Multivariate mixed membership modeling}

\begin{aug}
  \author{\fnms{Massimiliano} \snm{Russo}\thanksref{t1}\ead[label=e1]{m\_russo@hms.harvard.edu}},
\author{\fnms{Burton H.} \snm{Singer}\thanksref{t2}\ead[label=e2]{second@somewhere.com}}\\
\and
\author{\fnms{David B.} \snm{Dunson}\thanksref{t3}
\ead[label=e3]{third@somewhere.com}
\ead[label=u1,url]{http://www.foo.com}}

\affiliation{Harvard Medical School, and Dana-Farber Cancer Institute\thanksmark{t1}, University of Florida\thanksmark{t2} and Duke University\thanksmark{t3}}
\address{
Massimiliano Russo \\
Harvard-MIT Center for Regulatory Science,\\ Harvard Medical School,\\
and Department of Data Science,\\ Dana-Farber Cancer Institute,\\
Boston, MA 02115\\
\printead{e1}\\
}

\address{
Burton H. Singer \\ 
Emerging Pathogens Institute\\  and Department of Mathematics \\ 
University of Florida \\ 
Gainesville, FL 32610
}

\address{David B. Dunson \\ 
  Department of Statistical Science \\ 
  Duke University  \\
214 Old Chemistry Bldg, Box 90251 \\
Durham, NC 27708-0251}
\end{aug}

\begin{abstract}
 Characterizing the shared memberships of individuals in a classification scheme poses severe interpretability issues, even when using a moderate number of classes (say 4). Mixed membership models quantify this phenomenon, but they typically focus on goodness-of-fit  more than on interpretable inference. To achieve a good numerical fit, these  models may in fact require many extreme profiles, making the results difficult to interpret. We introduce a new class of multivariate mixed membership models that, when variables can be partitioned into subject-matter based domains, can provide a good fit to the data using fewer profiles than standard formulations. The proposed model explicitly accounts for the blocks of variables corresponding to the distinct domains along with a cross-domain correlation structure, which provides new information about shared membership of individuals in a complex classification scheme. We specify a multivariate logistic normal distribution for the membership vectors, which allows easy introduction of auxiliary information leveraging a latent multivariate logistic regression. A Bayesian approach to inference, relying on P\'olya gamma data augmentation, facilitates efficient posterior computation via Markov Chain Monte Carlo. We apply this methodology to a spatially explicit study of malaria risk over time on the Brazilian Amazon frontier.
\end{abstract}

\begin{keyword}
\kwd{Admixture model}
\kwd{Contingency table}
\kwd{Latent Dirichlet allocation}
\kwd{Multivariate categorical data}
\kwd{Multivariate logistic normal distribution}
\newline
\vspace{0.3cm}
\thanksmark{t1} 
\href{mailto:m_rurro@hms.harvard.edu}{m\_russo@hms.harvard.edu}
\end{keyword}

\end{frontmatter}

\section{Introduction} \label{sec1}
{\em Mixed membership (MM)} modeling began in response to difficulties in achieving crisp classification of individuals on the basis of assessments of many characteristics about them~\citep{woodbury1978}. MM also proved useful for identifying the driving forces of a specific outcome when they are expressed by multiple potentially influencing features, no combination of which occurred with high frequency in the overall population~\citep{berkman1989}. More recently MM has been used in a variety of contexts including text analysis~\citep{blei2003}, medicine~\citep{erosheva2007}, and several studies of social interactions~\citep[e.g.,][]{airoldi2005,airoldi2008,kao2018}, among many others. An extensive review on this class of models can be found~\citet{handbook2014}.

Algorithms for MM analyses usually begin by fitting a set of $H$ pure, or ideal, types summarizing high dimensional discrete-valued data, and assigning probabilities for levels of each variable to be members of each pure type. With a set of pure types at hand, it is useful to think of them as vertices of a unit simplex. Then each individual's response vector is associated with a point inside or on the boundary of the simplex. Each point is given a set of degree of similarity scores, the score vector, $\boldsymbol{\lambda_i} =(\lambda_{i1},\ldots,\lambda_{iH})^T$, such that $0 < \lambda_{ih} <1 $ and $\sum_{h=1}^H \lambda_{ih} =1$, that represent location in the simplex. If an individual has, for example, 5 non-zero elements in the score vector, each representing relative proximity to a different pure type, then the individual shares characteristics with 5 pure-types. If all individuals in a population have response vectors that are assigned a score of 1, relative to some pure type, then crisp classification has occurred, with the pure types associated with one or more individuals being the categories in a classification scheme. When individuals have more than one component of their score vector positive, they share conditions represented by each of the pure types to which they have some similarity, which is particularly appealing when an exact grouping is difficult if not impossible to obtain, as for example in identification of disease risks~\citep[e.g.,][]{chuit2001,decastro2006} or political ideology~\citep[e.g.,][]{gross2012}.

If many individuals have score vectors with $4$ or more non-zero components, then it becomes difficult, in almost any application, to write a coherent sentence describing what this complex set of shared memberships actually means. This is a reflection of the intrinsic limitations on human capacity for understanding many distinct ideas simultaneously~\citep{miller1956}, particularly when these are not easily summarized in a plot or a table. When most individuals only have two non-zero components in their score vectors---i.e. they are located on an edge in the unit simplex with pure types defined as the vertices---then they share conditions with a particular pair of pure types, and interpretable description tends to be straightforward. To-date many published MM analyses have a number of pure types ranging from $10$~\citep[e.g.,][]{erosheva2005} to several hundreds~\citep[e.g.,][]{griffiths:2004}. Curiously, considerations on interpretability  have mostly been avoided by there being almost no discussion of the sets of shared memberships. Most of the emphasis has gone to descriptions of the pure types; a notable exception is~\cite{erosheva:2004}. From our perspective, this is avoiding one of the most informative, and even motivating, features of MM representations. Hence, it is desirable to employ a small number of profiles, e.g. $H < 4$. In epidemiology applications, we frequently use $H = 2$, with the two profiles corresponding to high and low risk. The weight vector $\boldsymbol{\lambda}_i$  then corresponds to values in $(0, 1)$ summarizing the degree of risk to which individual $i$ is exposed.

If we are to accurately represent the dependence structure in most epidemiological data, usually more than two profiles are needed, since goodness-of-fit and interpretability are conflicting factors. A possible way to improve interpretability is to block variables into distinct domains---e.g. human behavioral, physical environmental, and climatic in infectious disease epidemiological studies, and then carry out standard MM analyses on each domain separately for the same set of individuals, with the number of pure types $H$ forced to be $2$ or $3$. The use of the same individuals across models induces a correlation structures in the score vectors, yielding new information about the phenomena under investigation that is not at all transparent from conventional MM specifications~\citep[e.g.,][]{chuit2001,singerinter}. To-date no formalization of this kind of correlation structure exists.

The main aims of this paper are to: (1) specify a new class of {\em Multivariate
  Mixed Membership (MMM)} models that explicitly include the classification of blocks of variables corresponding to distinct subject matter domains and the cross-domain correlation structure; (2) apply the MMM framework to the problem of characterizing malaria risk on the Brazilian Amazon frontier. This problem has been studied previously~\citep{decastro2006,castro2007}, but with less sophisticated tools.

We address (1) by linking group-specific MM models through dependence in the membership scores. We show that this model require fewer profiles to characterize the joint probability mass function underlying the data, relaxing the constraints of the standard mixed membership model formulation. Additionally, we propose a novel joint distribution defined on a product space composed of simplices, leading to an easy-to-implement Gibbs sampler for posterior computation, based on P\'olya gamma data augmentation~\citep{polson2013}. The proposed framework allows simple inclusion of subject and group-specific covariates leveraging multiple latent logistic regression.

The paper is structured as follows. In Section~\ref{sec:mix_mem} we present a brief review of mixed membership models and their connection with tensor decompositions. In Section~\ref{sec:mult_mix_mem} we introduce our MMM generalization of such models and describe some of their key properties. Section~\ref{sec:mult_log_norm} introduces a multivariate distribution defined on a product space of simplices. In Section~\ref{sec:post_comp} we provide technical details on posterior computation. In Section~\ref{sec:simul} we study the performance of our model under different simulation scenarios, and in Section~\ref{sec:app} we apply the model to the problem of characterizing malaria risk over time at a colonization project on the Brazilian Amazon frontier.

\section{Mixed membership models and tensor decompositions}
\label{sec:mix_mem}
Given a collection of categorical random variables $(X_{i1},\ldots,X_{ip})^T$ for $i=1,\ldots,n$ and $j=1,\ldots,p$ such that $X_{ij} \in \{1,\ldots,d_j\}$, a mixed membership model can be defined as follows:
\begin{eqnarray}
    X_{ij} \mid Z_{ij} = h, \boldsymbol{\theta}_h^{(j)} &\sim& \mbox{Cat}(\theta^{(j)}_{h1},\ldots,\theta^{(j)}_{hd_j}), \nonumber\\
    Z_{ij} \mid \boldsymbol{\lambda}_i &\sim& \mbox{Cat}(\lambda_{i1},\ldots,\lambda_{iH}),
    \label{eq:hsfm} \\
    \boldsymbol{\lambda}_i &\sim&   P, \nonumber
\end{eqnarray}
where  $\lambda_{ih} = \mbox{pr}(Z_i =h)$, $\theta^{(j)}_{hk} = \mbox{pr}(X_{ij} = k \mid Z_i = h)$ for $h= 1,\ldots,H,$ $k =1,\ldots,d_j$ and $j=1,\ldots,p$, while $P$ is the distribution of the membership score vector associated with each observation $i$. Popular choices for the distribution $P$ include Dirichlet~\citep{blei2003} and logistic normal~\citep{lafferty2006}. From model~\eqref{eq:hsfm} we can notice that there is a population level assumption, i.e. the population is composed of $H$ subpopulations, and an individual level assumption, for which each subject has a degree of similarity with the type $h$ expressed by $\lambda_{ih}$.

The kernel probabilities $\theta^{(j)}_{hk}$ express the probability of observing the $k$-th category for the $h$-th profile, while the vector $\boldsymbol{\lambda}_i$ represents subject $i$ and quantitatively describes the individual’s degree of similarity to each of the $H$ subpopulations. Geometrically, it locates individual $i$ in a unit simplex whose vertices are identified with the $H$ subpopulations. Leveraging the local independence assumption in model~\eqref{eq:hsfm} the probability distribution for the generic subject $i$ can be expressed, integrating out the latent variable $\boldsymbol{Z}_{i}= (Z_{i1},\ldots,Z_{1p})^T$, as
\begin{eqnarray}
    \mbox{pr}(X_{i1} = x_1,\dots,X_{ip} = x_p \mid \boldsymbol{\lambda}_i, \boldsymbol{\theta}) &=&
    \prod_{j=1}^p \sum_{h=1}^H \lambda_{ih} \theta^{(j)}_{hx_j} \nonumber \\
                                                                                                &=&\sum_{h_1 = 1}^H\cdots \sum_{h_p
         =1}^H  \prod_{j=1}^p \lambda_{ih_j}
     \theta^{(j)}_{h_jx_j},
\end{eqnarray}
which is a product of conditionally independent mixture models. The population model can be retrieved integrating out $\boldsymbol{\lambda_i}$ with respect to its distribution $P$
\begin{eqnarray}
    \mbox{pr}(X_1 = x_1,\dots,X_p = x_p \mid \boldsymbol{\theta}) = \sum_{h_1=1}^H\cdots \sum_{h_p=1}^H a_{h_1\ldots h_p}
    \prod_{j=1}^p \theta^{(j)}_{h_jx_j},
    \label{eq:sfm}
\end{eqnarray}
where $a_{h_1 \ldots h_p} = \mathbb E_P [\lambda_{ih_1} \cdots \lambda_{ih_p}]$ is the expectation of the product of the score vector elements over $P$. Depending on the choice of $P$, the expectation $a_{h_1 \ldots h_p}$ may or may not have a closed form expression.

Equation~\eqref{eq:sfm} is an instance of a Tucker tensor decomposition~\citep[e.g.,][]{kolda2009tensor}, and is a flexible representation for the probability mass function of unordered categorical random variables, since there always exists an $H$ such that any probability mass function can be characterized as in~\eqref{eq:sfm}. Additionally, representation~\eqref{eq:sfm} typically requires a smaller $H$ than a standard discrete mixture model representation~\citep[see for example][]{bhattacharya2012}.

Moreover, equation~\eqref{eq:sfm} can be interpreted as a constrained discrete mixture model with $H^p$ latent components. In fact, the core tensor $ \mathcal A =\{a_{h_1,\ldots,h_p} ; h_j=1,\dots,H; j=1,\ldots p \}$ is specified to be a cubic symmetric tensor. A cubic tensor is a tensor having all modes with the same dimension, while a symmetric tensor, sometime referred to as super symmetric, is the direct generalization of a symmetric matrix in tensor algebra. Formally, given a vector of indices $\mathbf{h} = (h_1,\ldots,h_p)^T$ and defining $\mathfrak{S}_{\mathbf{h}}$ to be the space of all permutation of $\mathbf{h}$, we have that $a_{\mathbf{h}} = a_{\sigma(\mathbf{h})}$ for all $\sigma \in \mathfrak{S}_{\mathbf{h}}$.  This definition implies that just $H^{\bar{p}}/p!$ elements out of the $H^p$ are distinct, where $H^{\bar{p}} = H(H+1)\cdots (H - p -1)$ is the rising factorial. It is easy to see that in $2$-dimensional space the previous definition reduces to the usual symmetric matrix (i.e. equal to its transpose) and that $H^{\bar{2}}/2! = H(H+1)/2$.

Such constraints derive from the exchangeability assumption for the profile probabilities in~\eqref{eq:hsfm}~\citep[e.g.,][]{erosheva2007}. When compared to an unconstrained discrete mixture model, the effect of such constraints is to increase the value of $H$ needed to fully characterize the probability distribution underlying the data. Independent of applications and issues of subject-matter interpretability, which are not mathematical concerns, increasing $H$ as needed poses no particular problem. However, if $H$ is constrained a priori, our representation can lead to an unsatisfactory approximation of the probability mass function. To deal with this issue, we propose a generalization of the above approach relaxing the constraints imposed on the latent part of the model. 

\section{A multivariate mixed membership model}
\label{sec:mult_mix_mem}
We assume, a priori, that variables can be divided into distinct groups which, in applications, are identified with different subject-matter domains. Let $\boldsymbol{g} = (g_1, \ldots , g_p)^T$ be an indicator vector for groups of variables, where $g_j \in \{1,\dots,G\}$  for $j=1,\dots,p$. Each subject is endowed with $G$ membership score vectors $({\boldsymbol{\lambda}_i^{(1)}}^T,\ldots,{\boldsymbol{\lambda}_i^{(G)}}^T)^T$ such that $\sum_{h=1}^H \lambda^{(g)}_{ih} = 1$ for $g=1,\dots,G$. Note that the sum of the membership scores for the different domains is not equal to $1$, i.e. $\sum_{g =1}^G \lambda_{ih}^{(g)} \neq 1$, for $h=1,\ldots,H$. 

The proposed model can be expressed in the following hierarchical form:
\begin{eqnarray}
    X_{ij} \mid Z_{ij} = h, \boldsymbol{\theta}^{(j)}_h &\sim& \mbox{Cat}(\theta^{(j)}_{h1},\ldots,\theta^{(j)}_{hd_j}), \nonumber\\
    Z_{ij} \mid \boldsymbol{\lambda}^{(g_j)}_i &\sim& \mbox{Cat}(\lambda^{(g_j)}_{i1},\ldots,\lambda^{(g_j)}_{iH}),
    \label{eq:hmmm} \\
    ({\boldsymbol{\lambda}_i^{(1)}}^T,\ldots,{\boldsymbol{\lambda}_i^{(G)}}^T)^T &\sim&   P. \nonumber
\end{eqnarray}
As in model~\eqref{eq:hsfm}, representation~\eqref{eq:hmmm} relies on conditional independence of the observed variables given the profile labels; in fact, the latent variables $Z_{ij}$ are conditionally independent given the mixed membership scores $ \bar{\boldsymbol{\lambda}}_i = (\boldsymbol{\lambda}_i^{(1)},\ldots,\boldsymbol{\lambda}_i^{(G)})^T$:
\begin{eqnarray}
\mbox{pr}(X_{i1} = x_1,\dots,X_{ip} = x_p \mid \bar{\boldsymbol{\lambda}}_i, \boldsymbol{\theta})  &=& 
    \sum_{h_1 =1}^H \cdots \sum_{h_p =1}^H  \prod_{j=1}^p \lambda^{(g_j)}_{ih_j} \theta^{(j)}_{h_j x_j} .
    \label{eq:condmmm}
\end{eqnarray}
Integrating out the scores $\bar{\boldsymbol{\lambda}}_i$ from equation~\eqref{eq:condmmm}, we obtain the population level model:
\begin{eqnarray}
    \mbox{pr}(X_1 = x_1,\dots,X_p = x_p \mid \boldsymbol{\theta}) = \sum_{h_1=1}^H\cdots \sum_{h_p=1}^H \bar{a}_{h_1\ldots h_p} \prod_{j=1}^p \theta^{(j)}_{h_jx_j}.
    \label{eq:spem}
\end{eqnarray}
Although equation~\eqref{eq:spem} seems identical to equation~\eqref{eq:sfm}, the elements of the core tensors are different, as are the imposed constraints. The core tensor $\bar{\mathcal A} = \{ \bar{a}_{h_1,\dots,h_p}, h_j=1,\ldots H; j =1,\ldots,p\}$ is not a symmetric tensor, but it has some equality constraints on the elements.  To describe these constraints, we can define a group preserving permutation space; specifically, given the vector of indices $\boldsymbol{h} = (h_1,\ldots,h_p)^T$, and a group indicator vector $\boldsymbol{g}=(g_1,\ldots,g_p)^T$, the group preserving permutation space $\mathfrak{S}^{\boldsymbol{g}}_{\boldsymbol{h}}$ is such that the effect of $\bar{\sigma} \in \mathfrak{S}^{\boldsymbol{g}}_{\boldsymbol{h}}$ is to permute the elements of a vector within the groups, leaving the group structure unchanged. It immediately follows that $\mathfrak{S}^{\boldsymbol{g}}_{\boldsymbol{h}}$ is a well defined group since it is  closed under composition, while also respecting  associativity, identity and invertibility properties~\citep[e.g.,][]{artin1991}.

The core tensor $\bar{A}$ can be defined as a group symmetric tensor, meaning that given a multivariate index $\boldsymbol{h}$ we have $\bar{a}_{\boldsymbol{h}} = \bar{a}_{\bar{\sigma}(\boldsymbol{h})}$, for all  $\bar{\sigma} \in \mathfrak{S}^{\boldsymbol{g}}_{\boldsymbol{h}}$. A symmetric tensor can be viewed as group symmetric with only one group, or can be defined such that it is group symmetric for any possible group configuration $\boldsymbol{g}$. Following the same logic, model~\eqref{eq:hmmm} can be seen as a sub-model of~\eqref{eq:condmmm} having just one group ($G=1$) or the  score vectors $\boldsymbol{\lambda}^{(g)} = \boldsymbol{\lambda}^{(g^\prime)}$ for all $g\neq g^\prime$.

\begin{figure}[h!]
    \centering
    \includegraphics[width =\textwidth]{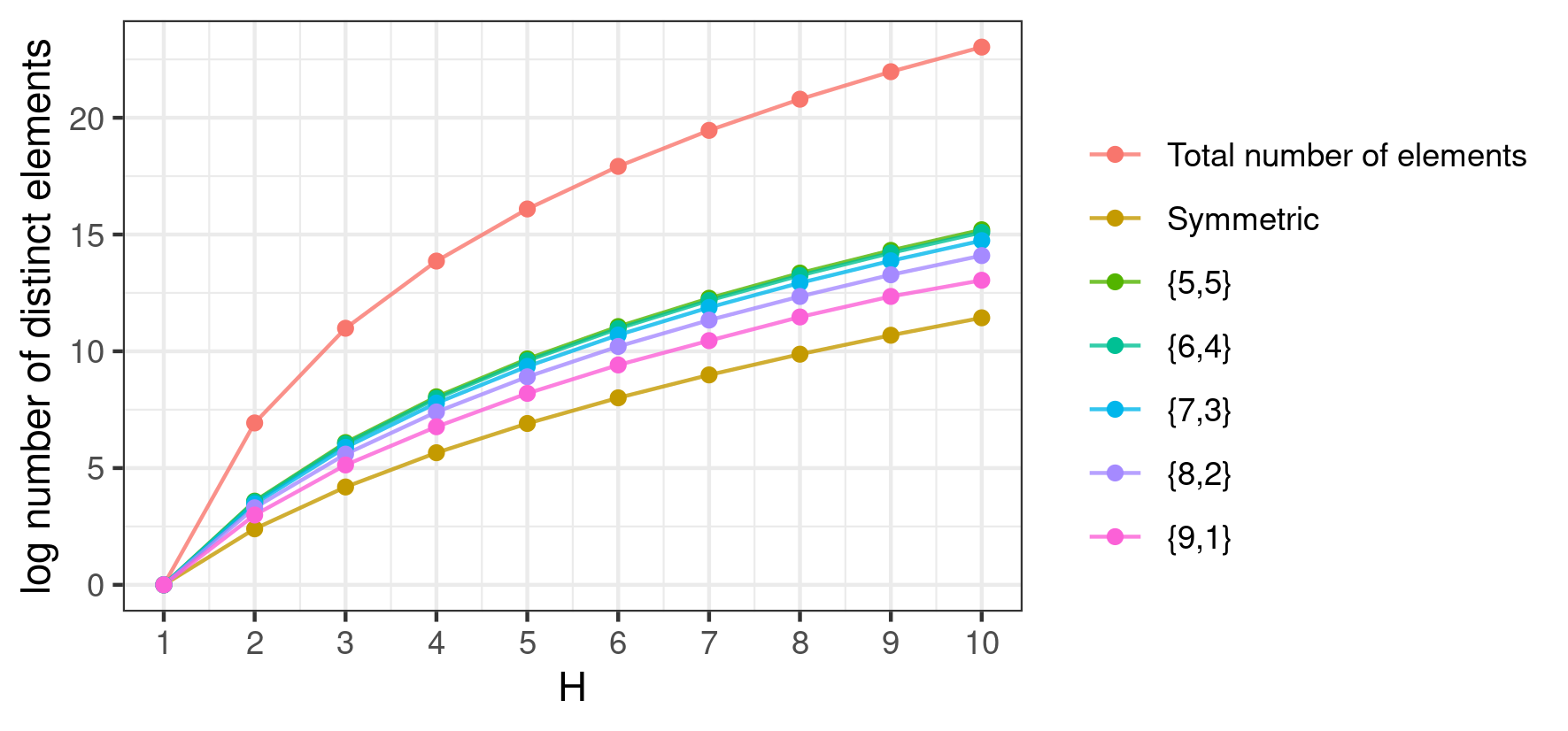}
    \caption{Logarithm of the number of distinct elements in a cubic tensor of dimension $H^{10}$,
        for symmetric and group symmetric tensors for all configurations of two groups.}
    \label{fig:log_num}
\end{figure}

The number of distinct elements in $\bar{\mathcal{A}}$ is given by $\prod_{g=1}^G H^{\bar{p_g}}/p_g!$, which is considerably larger than in the symmetric tensor case, as can be seen from Figure~\ref{fig:log_num}. Moreover, as a consequence of Lemma~\ref{th1}, for any fixed $H$, equation~\eqref{eq:spem} is never worse than equation~\eqref{eq:sfm} in approximating the `true' probability mass function generating the data.

\begin{lemma}
    Let $\pi_0$ be a probability tensor of dimension $d_1 \times \ldots \times d_p$, $\pi^{\text{gsym}}$  and $\pi^{\text{sym}}$ be, respectively, the best group symmetric and symmetric  multi-rank $H$ approximations of $\pi_0$. Then $\|\pi_0 -  \pi^{\text{gsym}}\|_{F} \leq \|\pi_0 - \pi^{\text{sym}} \|_{F}$, where $\| \cdot \|_{F}$ denotes the Frobenius norm.
    \label{th1}
\end{lemma}

The proof immediately follows after noticing that  $\pi^{\text{gsym}}$ by definition minimizes $\|\pi_0 -  \pi^{\text{gsym}}\|_{F}$ under the constraint $\bar{a}_{\boldsymbol{h}} = \bar{a}_{\bar{\sigma}(\boldsymbol{h})}$ for all $\bar{\sigma} \in \mathfrak{S}^{\boldsymbol{g}}_{\boldsymbol{h}}$, and that $\pi^{\text{sym}}$ can be obtained by solving the same problem with additional equality constraints on the element of $\bar{\mathcal{A}}$ such that, $\bar{a}_{\boldsymbol{h}} = \bar{a}_{\boldsymbol{h^\prime}}$ if $\boldsymbol{h} = \sigma(\boldsymbol{h^\prime})$ for a $\sigma \in \mathfrak{S}$. $\qedsymbol$

Lemma~\ref{th1} implies that incorporating group-specific membership scores leads to a population level model~\eqref{eq:spem} with less replicated elements in the core tensor compared to~\eqref{eq:sfm}, for any fixed $H$. Hence, if we fix $H = 2$ or $3$ to ensure interpretability of shared membership score vectors, we will tend to produce a better fit to the data by using group-specific scores than in modeling a single global score vector. Note that Lemma~\ref{th1} does not imply that the MMM specification requires less parameters than MM to characterize the p.m.f., because of the additional group-specific  membership parameters. To complete a specification of the MMM model, it remains for us to choose an appropriate distribution $P$.

\section{The multivariate logistic normal distribution}
\label{sec:mult_log_norm}
Letting $S_H = \{\boldsymbol{x} \in [0,1]^H : \sum_{h=1}^H x_h =1 \}$ denote the $H-1$ probability simplex, we aim to define a joint distribution on the product space $S =  S_{H_1} \otimes \cdots \otimes S_{H_G}$. To achieve this goal, we start from a distribution on  $\mathbb R^{\sum_{g=1}^G (H_g-1)}$, mapping to $S$ via an appropriate transformation. Potentially any continuous multivariate distribution can be used, but we focus on the multivariate Gaussian distribution to retain simplicity and flexibility.
Let $\boldsymbol{Y}  = ({\boldsymbol{Y^{(1)}}}^T,\dots,{\boldsymbol{Y^{(G)}}}^T)^T $ be a multivariate normal distribution of dimension $\sum_{g=1}^G(H_g-1)$ with mean vector $\boldsymbol{\mu} =({\boldsymbol{\mu^{(1)}}}^T,\dots,{\boldsymbol{\mu^{(G)}}}^T)^T$, where $\boldsymbol{\mu}^{(g)} \in \mathbb R^{H_g -1}$, and covariance matrix $\boldsymbol{\Sigma}$. We consider the transformed vector $\boldsymbol{X} = ({\boldsymbol{X^{(1)}}}^T,\dots,{\boldsymbol{X^{(G)}}}^T)^T $, whose elements can be defined as $X^{(g)}_h = \exp\{ Y^{(g)}_h \}[1 + \sum_{k=1}^{Hg-1} \exp \{Y^{(g)}_k \}]^{-1}$ for $h=1,\ldots,H_g-1$ and  $g =1,\ldots,G$, with $X^{(g)}_{H_g} = [1 + \sum_{k=1}^{Hg-1} \exp\{Y^{(g)}_k)\}]^{-1}$.

It is easy to show that $\boldsymbol{X} \in S$ and that the Jacobian matrix of the transformation is block diagonal having  determinant given by $[\prod_{g=1}^{G} \prod_{h=1}^{H_g} X^{(g)}_{h}]^{-1}$. The probability density function of the resulting distribution is 
\begin{eqnarray}
    f_X(\boldsymbol{x};\boldsymbol{\mu},\boldsymbol{\Sigma}) = \frac{ \exp \left\{ -\frac 12 (\boldsymbol{x}^\star - \boldsymbol{\mu})^T \boldsymbol{\Sigma}^{-1}
        (\boldsymbol{x}^\star - \boldsymbol{\mu}) \right\}}{(2\pi)^{\sum_{g=1}^G(H_g -1)/2} |\boldsymbol{\Sigma}|^{1/2}
        \prod_{g=1}^G \prod_{h=1}^{H_g}  x^{(g)}_{h} },
    \label{eq:profmod}
\end{eqnarray}
where $\boldsymbol{x}^\star = \mbox{vec}\left(\left\{ \log( x_h^{(g)}/x^{g}_{H_g}),\mbox{ for } h=1,\dots,(H_g-1);g=1,\dots,G\right\}\right)$. Each of the group marginals $\boldsymbol{X^{(v)}}$ has a logistic normal distribution with parameters $\boldsymbol{\mu^{(v)}}$ and $\boldsymbol{\Sigma^{(v)}}$, where $\boldsymbol{\Sigma^{(v)}}$ is the block of the matrix $\boldsymbol{\Sigma}$ corresponding to the $v$-th group. We refer to~\eqref{eq:profmod} as the {\em Multivariate Logistic Normal Distribution (MLND)}, as it is a multivariate generalization of the logistic normal used in~\citet{lafferty2006}.
  
Distribution~\eqref{eq:profmod} can be alternatively  derived as a compound distribution from a collection of independent logistic normal distributions and a multivariate normal for the mean vectors, as stated in Proposition ~\ref{prop:comp}.

\begin{prop}
Let $\boldsymbol{X} = (\boldsymbol{X^{(1)}}^T,\ldots,\boldsymbol{X^{(G)}}^T)^T \in S$ such that $\boldsymbol{X^{(g)}}\mid \boldsymbol{\mu^{(g)}} \sim \mbox{LogitNormal}(\boldsymbol{\mu^{(g)}},\boldsymbol{\Sigma^{(g)}})$ independently for $g=1,\ldots,G$, and let $\boldsymbol{\mu} = (\boldsymbol{\mu^{(1)}}^T,\ldots,\boldsymbol{\mu^{(G)}}^T) \sim \mathcal N(\boldsymbol{\mu_0},\boldsymbol{\Sigma_0}) $. Then $\boldsymbol{X} \sim \mbox{MLND} (\boldsymbol{\mu_0},\boldsymbol{\tilde{\Sigma}})$, where $\boldsymbol{\tilde{\Sigma}} = \boldsymbol{\Sigma_0} + \mbox{block}(\boldsymbol{\Sigma^{(1)}},\ldots,\boldsymbol{\Sigma^{(G)}})$. 
    \label{prop:comp}
\end{prop}

Following~\citet{atchison1980}, we consider a class of distribution preserving transformations, useful to maintain some invariance properties of the induced distribution. According to our problem, we additionally restrict our attention to the sub-class of group preserving transformations (e.g., group permutation defined in Section~\ref{sec:mult_mix_mem}).

\begin{prop}
    Let $\boldsymbol{X} = ({\boldsymbol{X^{(1)}}}^T,\dots,{\boldsymbol{X^{(G)}}}^T)^T  \sim \text{MLND}(\boldsymbol{\mu},\boldsymbol{\Sigma})$ and $\boldsymbol{B}$ a $Q\times \sum_{g=1}^G (H_g -1)$ block diagonal matrix, having diagonal blocks $\boldsymbol{B}^{(g)}$ of dimension $q_g \times (H_g -1)$ for $g=1\dots,G$, then the  $Q \times G$ dimensional vector $\bold{X}^\prime$ whose elements are defined as 
\begin{eqnarray*}
        {x^\prime}^{(g)}_{q} &=& \prod_{h=1}^{H_g -1} \left(
            \frac{x^{(g)}_{h}}{x^{(g)}_{H_g}}\right)^{b^{(g)}_{qh}} 
        \left[1 + \sum_{k=1}^{q_g} \prod_{h=1}^{H_g-1}  \left(
                \frac{x^{(g)}_{h}}{x^{(g)}_{H_g}}\right)^{b^{(g)}_{kh}} 
 \right]^{-1},\end{eqnarray*}
 for $q=1,\ldots,q_g$ and $g=1,\ldots G$, has distribution $\boldsymbol{X^\prime} \sim \text{MLND}\left(  \boldsymbol{B\mu} ,  \boldsymbol{ B\Sigma B}^T\right)$.
\label{prop:1}
\end{prop}

The diagonal block structure of matrix $\boldsymbol{B}$ in Proposition~\ref{prop:1} ensures that the transformation preserves the same group structure of the original vector; for a general matrix $\boldsymbol B \in \mathbb R^{Q\times \sum_{g=1}^G (H_g -1)}$,  $\boldsymbol{X^\prime}$ is still distributed as an MLDN, but categories in different groups can be merged. Proposition~\ref{prop:1} implies that the MLDN distribution is invariant with respect to permutations of the labels, and allows easy computation of the joint distribution of the vector $\boldsymbol Y$ when some categories are merged.

The proposed MLND distribution has finite moments, but these moments in general do not have an analytic form.  However, we can obtain simple expressions for moments related to log-odds and odds ratios both between and across the groups.
For example, letting
\begin{eqnarray*}
  \mbox{m}_l (h,g;h^\prime,g^\prime) &=& \mathbb{E}\left[ \log\left( \frac{X^{(g)}_h/X^{(g)}_{H_g}}{ X^{(g^\prime)}_{h^\prime}/X^{(g^\prime)}_{H_{g^\prime}}} \right) \right], \quad \mbox{and}
    \\
  \mbox{m}_o (h,g;h^\prime,g^\prime) &=& \mathbb{E}\left[\left( \frac{X^{(g)}_h/X^{(g)}_{H_g}}{ X^{(g^\prime)}_{h^\prime}/X^{(g^\prime)}_{H_{g^\prime}}} \right) \right],
\end{eqnarray*}
we have
\begin{eqnarray}
   \qquad \mbox{m}_l (h,g;h^\prime,g^\prime) &=& \mu^{(g)}_{h} - \mu^{(g^\prime)}_{h^\prime},\nonumber \\
    \qquad \mbox{m}_o (h,g;h^\prime,g^\prime) &=& \exp\left\{ \mu^{(g)}_{h} - \mu^{(g^\prime)}_{h^\prime} +
            \frac{1}{2}
            \left[ \Sigma^{(g)}_{h  h} +  \Sigma^{(g^\prime)}_{h^\prime  h^\prime} - 2
                \Sigma^{(g,g^\prime)}_{h h^\prime} \right]
       \right\},
\label{eq:moments}
\end{eqnarray}
where with an abuse of notation we indicate with $\Sigma^{(g,v)}_{h k}$ the element in position $(h,k)$ of the non diagonal block of $\boldsymbol{\Sigma}$ corresponding to the groups $g$ and $v$. Higher order moments can also be computed relying on normal and log-normal distribution properties.

From equations~\eqref{eq:moments} we can notice that the log-odds of the elements in different groups are linearly related. Moreover, when applied to multivariate mixed membership models with $H_g=2$, log-odds and odds ratios give important insights on which group is more important in characterizing high and low risk conditions. Additionally, the elements of $\boldsymbol \Sigma$, or of the corresponding correlation matrix $\mathbf C$, can be used to assess if membership scores are independent across domains, or if a single MM model is sufficient to describe the latent structure.  In fact if $C^{(g,v)} = 0$, the model reduces to independent MM models for the domains $g$ and $v$.

For $H_g=2$, this hypothesis can be checked by choosing a hyperprior for $\boldsymbol\Sigma,$  and inspecting the credible interval for $\Sigma^{(g,v)}$, or $C^{(g,v)},$ for a specified credible level;
if the credible interval includes $0$, separate MM models can be a viable alternative to a full MMM. 
Similarly, if the posterior correlation concentrates near $-1$ or $1$, a single vector for the membership scores is sufficient to describe the considered data.

\section{Posterior computation}
\label{sec:post_comp}
We propose an algorithm to simulate from the posterior of model~\eqref{eq:hmmm}, with $({\boldsymbol{\lambda}_i^{(1)}}^T,\dots,{\boldsymbol{\lambda}_i^{(G)}}^T)^T  \sim \mbox{MLND}(\boldsymbol{\mu},\boldsymbol{\Sigma})$ defined in~\eqref{eq:profmod}. We focus on the special case where $H_g = 2$ for $g = 1,2,\ldots, G$. Generalization to more pure types can be obtained by iterating the proposed Polya gamma data augmentation on all the conditional log-odds~\citep{polson2011}, or alternatively relying on the stick breaking parameterization of the multinomial likelihood of~\citet{linderman2015}.

We begin by specifying conjugate prior distributions for all the parameters in the model. For the kernel probabilities we set $\boldsymbol{\theta^{(j)}_h} \sim \text{Dir}(\alpha^{(j)}_1,\ldots, \alpha^{(j)}_{d_j})$, for the hyperparameter $\boldsymbol{\mu} \sim \mathcal N (\boldsymbol{\mu_0},\boldsymbol{\Sigma_0})$ and for the covariance matrix $\boldsymbol{\Sigma} \sim \mathcal{IW}(\nu_0,\boldsymbol{\Psi}_0)$. Parameters can be updated by iterating the steps in Algorithm~\ref{alg:sim}.
In step $5$ we update the mean vector of a multivariate Gaussian distribution: at this step we can substitute a multivariate regression to account for covariate effects. 

\begin{algorithm}
\DontPrintSemicolon
\def\spacingset#1{\renewcommand{\baselinestretch}{#1}\small } \spacingset{0.9}
 \caption{Posterior Computation for the MMM model}
  \label{alg:sim}
 \Begin{
  {\bf [1]} Update the kernel probabilities

 \For(){j=1:p \& h=1:2}
 {
 \begin{algomathdisplay}
 \boldsymbol{\theta^{(j)}_h} \mid - \sim \mbox{Dir}\bigg(\alpha^{(j)}_{1} + \sum_{i : z_{ij} =h} I(x_{ij} = 1),\ldots, \alpha^{(j)}_{d_j} + \sum_{i : z_{ij} =h} I(x_{ij} =d_j)\bigg),
\end{algomathdisplay}
 where  $I(\cdot)$ is the indicator function. 
}

 {\bf [2]} Considering the model $\mbox{pr}(Z_{ij}=2 \mid  \lambda^{(g_j)}_{i}) = \lambda^{(g_j)}_i$, we can sample the profile indicator with probability 

 \For(){i=1:n \& j=1:p}{
\[    \mbox{pr}(Z_{ij} = 2\mid - ) = \frac{\lambda^{(g_j)}_{i} \theta^{(j)}_{2x_{ij}} } {
(1- \lambda^{(g_j)}_{i})\theta^{(j)}_{1x_{ij}}+ \lambda^{(g_j)}_{i} \theta^{(j)}_{2x_{ij}}}.\]
}

{\bf [3]} We make use of Polya gamma data augmentation to retrieve conjugacy between binomial and logistic normal distributions. We consider the augmented variables
    
\For(){ i=1:n \& g=1:G}{
    \[ \omega^{(g)}_i \mid - \sim \mbox{PG}(p_g, \mbox{logit}(\lambda^{(g)}_{i})),\]
where $p_g = \sum_{j=1}^p I(g_j = g)$ is the number of variables in $g$-th group for $g=1,\ldots,G$.
}

{\bf [4]} We define $k^{(g)}_i = \sum_{j=1}^{p_g} I(Z_{ij}=2) - p_g/2$, and we have that 
the vector $(\boldsymbol{k_i/\omega_i}) = (k_i^{(1)}/\omega^{(1)}_i,\ldots,k_i^{(G)}/\omega^{(G)}_i)^T \mid \boldsymbol{\lambda_i} \sim \mathcal N(\boldsymbol{\mbox{logit}(\lambda_i)}, \mbox{diag}(1/\omega^{(1)}_i,\ldots,1/\omega^{(G)}_i))$ and we can update the membership scores from 

\For(){i=1:n}{\[
    \boldsymbol{\lambda_i} \mid \boldsymbol{\mu}  \sim \mbox{MLND}(\boldsymbol{\mu^\star}, \boldsymbol{\Sigma^\star}),\]

where $\boldsymbol{  \Sigma^\star}    =    \left( \mbox{diag}(\omega^{(1)}_i,\ldots,\omega^{(G)}_i)+ \boldsymbol{\Sigma^{-1}} \right)^{-1}$ and $
    \boldsymbol{  \mu^\star} = \boldsymbol{\Sigma^\star ( \Sigma^{-1}\mu  + k_i)}$.
  }
    {\bf [5]} We can update the vector $\boldsymbol{\mu}$ integrating out the membership scores vectors $\boldsymbol{\lambda_i}$; we have that the vector $(\boldsymbol{k_i/\omega_i}) \mid \boldsymbol{\mu} \sim \mathcal N(\boldsymbol{\mu},\boldsymbol{\Upsilon^{-1}_i})$, where $\boldsymbol{\Upsilon^{-1}_i} = (\mbox{diag}(1/\omega^{(1)}_i,\dots,1/\omega^{(G)}_i) +\boldsymbol{\Sigma})$,   and hence the full conditional is given by
    \[\boldsymbol{\mu} \mid - \sim \mathcal N(\boldsymbol{\mu^{*}},\boldsymbol{\Sigma^{*}}),\]
    where
    $\boldsymbol{\Sigma^{*}} =  \left(\sum_{i = 1}^n\boldsymbol{\Upsilon_i}  + \boldsymbol{\Sigma^{-1}_0} \right)^{-1}$ and $\boldsymbol{\mu^{*}} =  \boldsymbol{\Sigma^{*}}\left( \sum_{i=1}^n \boldsymbol{\Upsilon_i k_i/\omega_i} + \boldsymbol{\Sigma_0^{-1} \mu_0} \right)$.

    {\bf [6]}We finally update the covariance matrix and its parameters from the full conditional 

    \[\boldsymbol{\Sigma} \mid - \sim \mathcal{IW}\bigg( \nu_0 +n, \boldsymbol{\Psi}_0 + \sum_{i=1}^n(\mbox{logit}(\boldsymbol{\lambda_i}) -\boldsymbol{\mu})(\mbox{logit}(\boldsymbol{\lambda_i}) -\boldsymbol{\mu})^T  \bigg).\]
 }
\end{algorithm}
 
Potentially for our MMM model, as in other MM models and more broadly for mixture models, we may encounter label switching. This occurs when the extreme profiles change their meaning across MCMC iterations. Although including information on variable partitions can reduce identifiability issues~\citep[e.g.,][]{xu2017}, the invariance with respect to group transformations of the MLND distribution (Proposition~\ref{prop:1}) makes labels exchangeable. When label switching occurs, post processing should be used to appropriately align the  MCMC samples~\citep[see for example][]{stephens2002}. However, such post processing was not applied in any of the simulated data we report below, as trace plots showed no evidence of label switching in the MCMC samples (refer to Figure~{S1} in \ref{suppA} for an example).

\section{Simulation study}
\label{sec:simul}
We analyze different simulation scenarios in evaluating the performance of our approach. We consider different probability distribution functions for the membership scores $P$, relying on hierarchical representation~\eqref{eq:hmmm} to generate the data.  The goal in defining these scenarios is to assess whether the proposed model can characterize generative mechanisms having broadly different properties. We compare our results with the standard admixture formulation implemented in the R package \texttt{mixedMem}, using separate models for each group. This package relies on a Variational EM algorithm, approximating the posterior distribution of the latent memberships and selecting hyperparameters through a pseudo MLE procedure~\citep[refer to][for more details]{MMpack}. We quantified uncertainty in the estimates using the bootstrap procedure in \citet{chen2018}. Additional comparisons with an MCMC implementation of the same model are provided in~\ref{suppA}.

We initially assume that $H_g = 2$ is the `true' number of extreme profiles, presenting four different scenarios, while in a second Section we consider the misspecified case $H_g>2$.  The code to reproduce our simulations, together with broader implementation of Algorithm~\ref{alg:sim}, can be found at~\url{https://github.com/rMassimiliano/MMM-tutorial}.

\subsection{Number of profiles correctly specified}
 We consider $G=2$ groups, $n=1000$ subjects,  $p_g=5$ categorical variables, having $d_j = d =4$ levels and $H_g=2$ profiles for $g=1,2$. We simulate data from categorical distributions, whose probabilities are drawn from a Dirichlet distribution with parameters $\varphi^{(g)}_h$ having values 
$\varphi_1^{(1)}  =(10,3,2,1)^T$,
$\varphi_2^{(1)}  =(1,1,1,11)^T$,
$\varphi_1^{(2)}  =(5,5,1,0)^T$ and
$\varphi_2^{(2)}  =(1,1,1,8)^T$. 
In the first simulation scenario, we let the probability density function for the joint distribution of the score vectors $(\lambda_i^{(1)},\lambda_i^{(2)})^T$ be a bivariate normal distribution truncated over the unit square, having parameter $\boldsymbol{\mu} = (0.5,0.5)^T $ and $\mbox{vec}(\boldsymbol{\Sigma}) = ( \Sigma_{11}, \Sigma_{21}, \Sigma_{12}, \Sigma_{22})^T = (0.05,0.02,0.02,0.05)^T$. This formulation induces positive dependence between the two scores with their distribution having ellipsoid contours truncated at the borders. In the second simulation scenario, we consider the distribution proposed in Section~\ref{sec:mult_log_norm} with $\boldsymbol{\mu} = (-1.2,1)^T$ and $\mbox{vec}(\boldsymbol{\Sigma}) = (3.0,-2.4,-2.4,3.5)^T$. In the third scenario, we rely on the generative mechanism~\eqref{eq:hsfm}, having profile distribution shared by all variables; we generate this profile from a uniform distribution. Finally, in the fourth simulation scenario, we consider $P$ to be the product of two independent uniforms, forcing independence in the variables belonging to different groups, which translates into the case in which two separate models for the groups represents the correctly specified model.

We perform posterior inference under the proposed model~\eqref{eq:hmmm} with priors defined in Section~\ref{sec:post_comp}, setting $\alpha^{(j)}_1 =\dots=\alpha^{(j)}_{d_j} = 1/d_j$ for $j=1,\ldots,p$, we consider $\boldsymbol{\mu_0} = (0,0)^T$, $\boldsymbol{\Sigma_0} = \boldsymbol{I}$, $\nu_0 =2$ and $\Psi_0 = \boldsymbol{I}$. We maintained these default hyperparameters in all our simulation cases, collecting $5000$ Gibbs samples from Algorithm~\ref{alg:sim}. Trace plots suggest convergence is reached by a burn-in of $1000$.
\begin{figure}[h!]
    \centering
    \includegraphics[width = \textwidth]{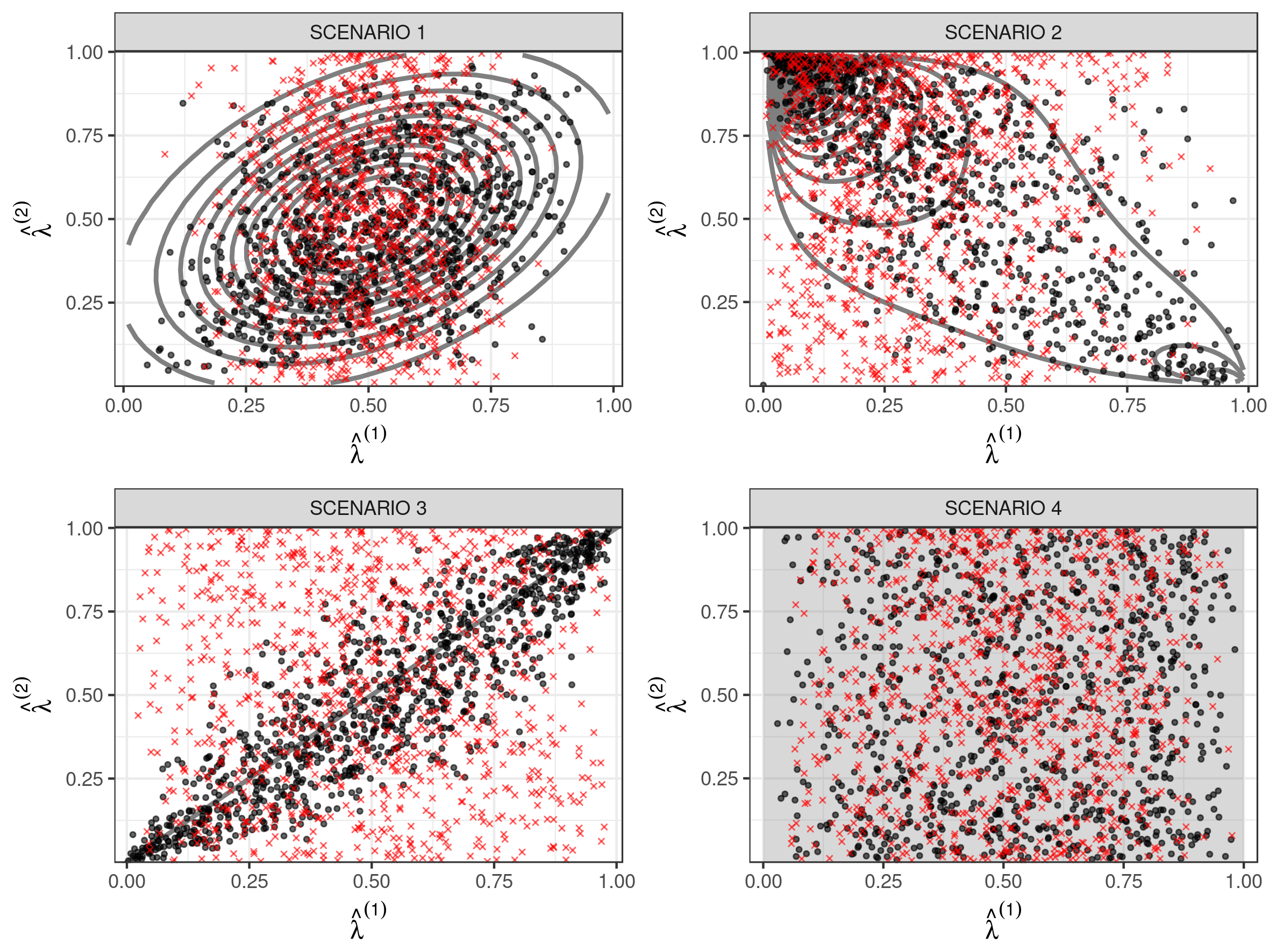}
    \caption{1000 samples from estimated membership scores distribution from model~\eqref{eq:hmmm} (black dots) and
        \texttt{mixedMem} (red crosses). Grey area represents the contour of the true profiles distribution.}
    \label{fig:mmm1}
\end{figure}

 Figure~\ref{fig:mmm1} shows the estimated profile distribution $P$ for all simulated scenarios, comparing results with the use of two separate MM models. Despite the challenging scenarios and the misspecification of the profile distribution, our proposed approach is able to  reconstruct the latent mechanism underlying the profiles in a satisfactory way.

In evaluating subject-specific  estimates of the scores $(\lambda^{(1)}_i,\lambda^{(2)}_i),$ we rely on the mean L1 distance relative to the `true' values. We obtain good results in retrieving the `true' membership vectors in all simulation scenarios (Table~\ref{tab:MSE_sim}) as the proposed approach always produces better or comparative results to the standard mixed membership model implemented in the package \texttt{mixedMem}. 
\begin{table}[h!]
    \centering
\def\spacingset#1{\renewcommand{\baselinestretch}%
{#1}\small\normalsize} \spacingset{1.0}
\caption{Mean (and standard deviation) of the L1 distance of the individual membership scores $(\lambda^{(1)}_i,\lambda^{(2)}_i)$ and their `true' values in all simulation scenarios.}
    \label{tab:MSE_sim}
    \vspace{0.2cm}
 \scalebox{0.7}{
\begin{tabular}{lllll}
\toprule
               & SCENARIO 1 & SCENARIO 2 & SCENARIO 3 & SCENARIO 4\\
\midrule
 MMM g = 1      &0.132(0.096) &0.126(0.097) &0.122(0.090) &0.162(0.106) \\
 MMM g = 2      &0.130(0.094) &0.134(0.103) &0.117(0.095) &0.138(0.105) \\
 mixedMem g = 1 &0.139(0.096) &0.148(0.110) &0.174(0.113) &0.174(0.113) \\
 mixedMem g = 2 &0.140(0.104) &0.162(0.119) &0.147(0.108) &0.141(0.103) \\
\bottomrule
\end{tabular}
}
\end{table}

\begin{figure}[h!]
    \centering
    \includegraphics[width=\textwidth]{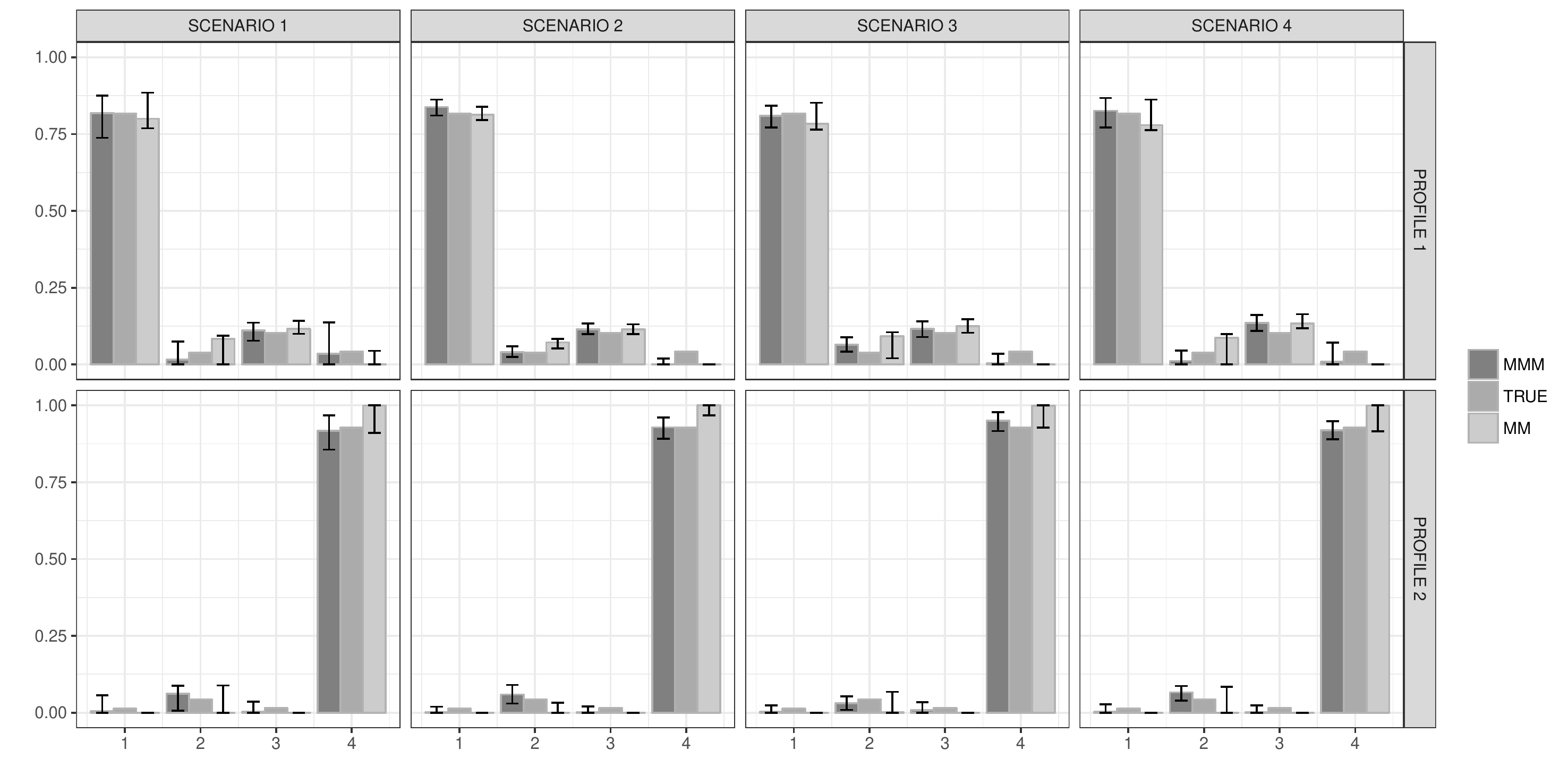}
    \caption{True values of the estimated profiles $\boldsymbol{\theta}^{(j)}_h$ for $h=1,2$ of a representative variable in group $g_j = 1$. Bars represent 0.1 and 0.9  posterior quantiles for our MMM model and bootstrap 0.8 confidence intervals for the MM model estimated with the \texttt{mixedMem} package.}
     \label{fig:sim_kern1}
\end{figure}

\begin{figure}[h!]
    \centering
    \includegraphics[width=\textwidth]{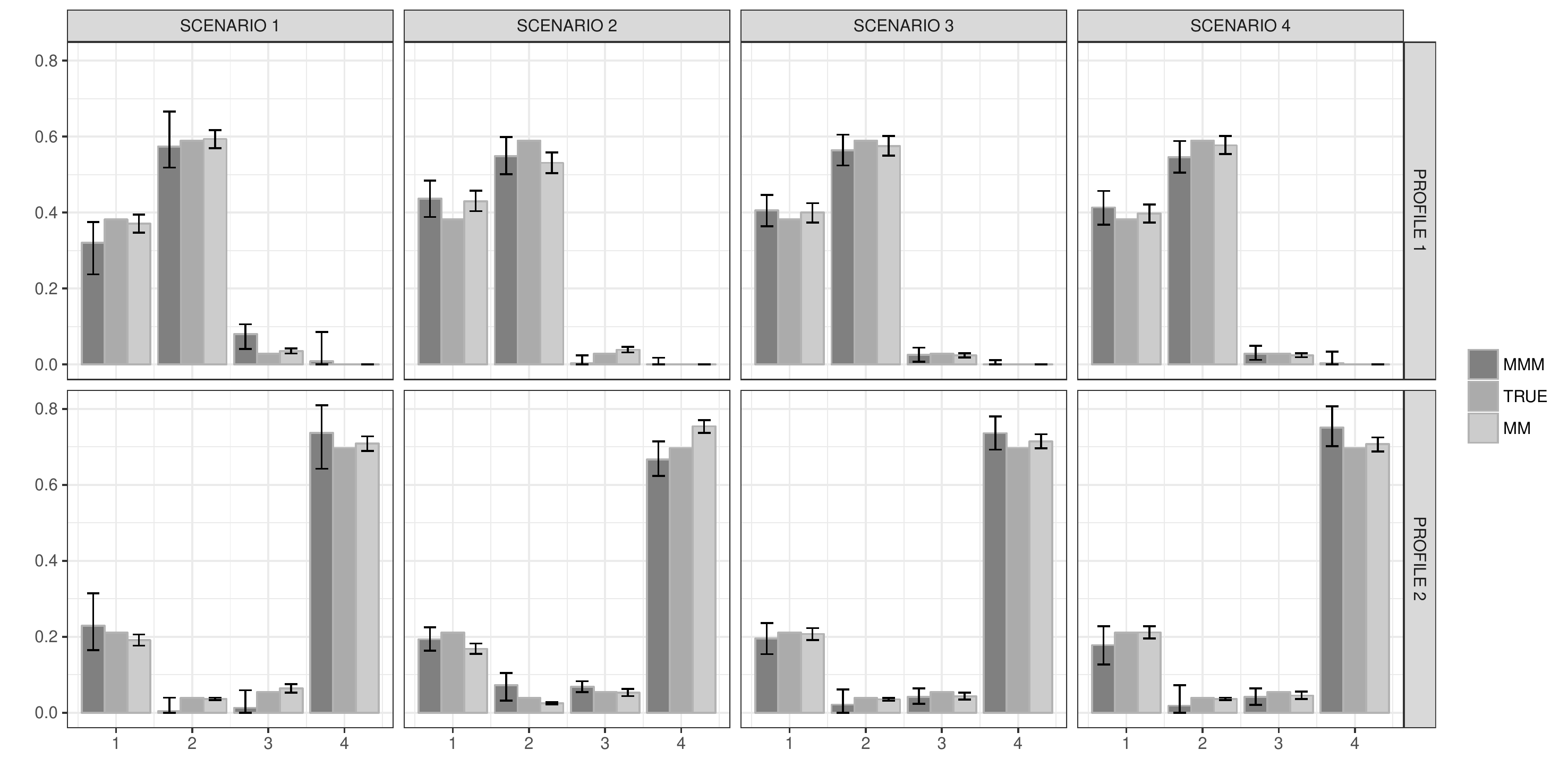}
    \caption{True values of the estimated profiles $\boldsymbol{\theta}^{(j)}_h$ for $h=1,2$ of a representative variable in group $g_j = 2$. Bars represent 0.1 and 0.9  posterior quantiles for our MMM model and bootstrap 0.8 confidence intervals for the MM model estimated with the \texttt{mixedMem} package.}
     \label{fig:sim_kern2}
\end{figure}

Figures~\ref{fig:sim_kern1} and \ref{fig:sim_kern2} show posterior estimates and credible intervals for the kernel parameters $\boldsymbol{\theta}^{(j)}_h$ for selected variables in both scenarios. We notice that our proposed approach robustly estimates kernels in the considered scenarios. Contrarily in some cases, the MM model underestimates variability. This behavior is evident in the lower part of Figures~\ref{fig:sim_kern1} where MM produces confidence intervals for $\theta^{(j)}_{21}$, $\theta^{(j)}_{22}$, $\theta^{(j)}_{23}$ collapsing to $0$ inappropriately. This underestimation of the kernel parameters variability is probably due to the variational approximation of their posterior distribution. In fact, when using MCMC to approximate the posterior of two separate MM models, we do not observe a collapse in the variability. Instead, this variability tends to be overestimated in comparison with our proposed MMM model (see Figures S4 and S5 in \ref{suppA}).

The proposed MMM model also presents a good fit to the data with L1-norm between the estimated and empirical proportions close to 0.2 for all marginal and bivariate distributions (see Figure S1 in~\ref{suppA}).

\subsection{Misspecification: more than two pure types}
In this Section we consider a scenario in which generative model~\eqref{eq:hmmm} has more than $2$ types, while retaining the proposed inference model with $H_g=2$ for $g=1,2$. The key idea is to understand how the model is able to approximate the profiles in a lower dimensional space, and compare this approximation with that for the standard MM model. Specifically, we consider as generative mechanism a $G=2$ group model with $H^0_1=4$. Kernels for the first group are fixed as 
$\varphi^{(1)}_1 = (0.85,0.05,0.05,0.05)^T$, 
$\varphi^{(1)}_2 = (0.05,0.85,0.05,0.05)^T$, 
$\varphi^{(1)}_3 = (0.05,0.05,0.85,0.05)^T$ and  
$\varphi^{(1)}_4 = (0.05,0.05,0.05,0.85)^T$, while membership scores $ (\lambda^{(1)}_{i1},\ldots \lambda^{(1)}_{i4})^T \sim \mbox{Dirichlet}(0.25,0.25,0.25,0.25)$. For the second group we consider instead the same mechanism used in scenario 4 with $H^0_2 = 2$, enforcing no dependence in the scores distribution. This scenario is constructed to favour the use of two separate MM models, having no dependence across the groups and a Dirichlet distribution for the profiles.

\begin{figure}[h!]
    \centering
    \includegraphics[width = \textwidth]{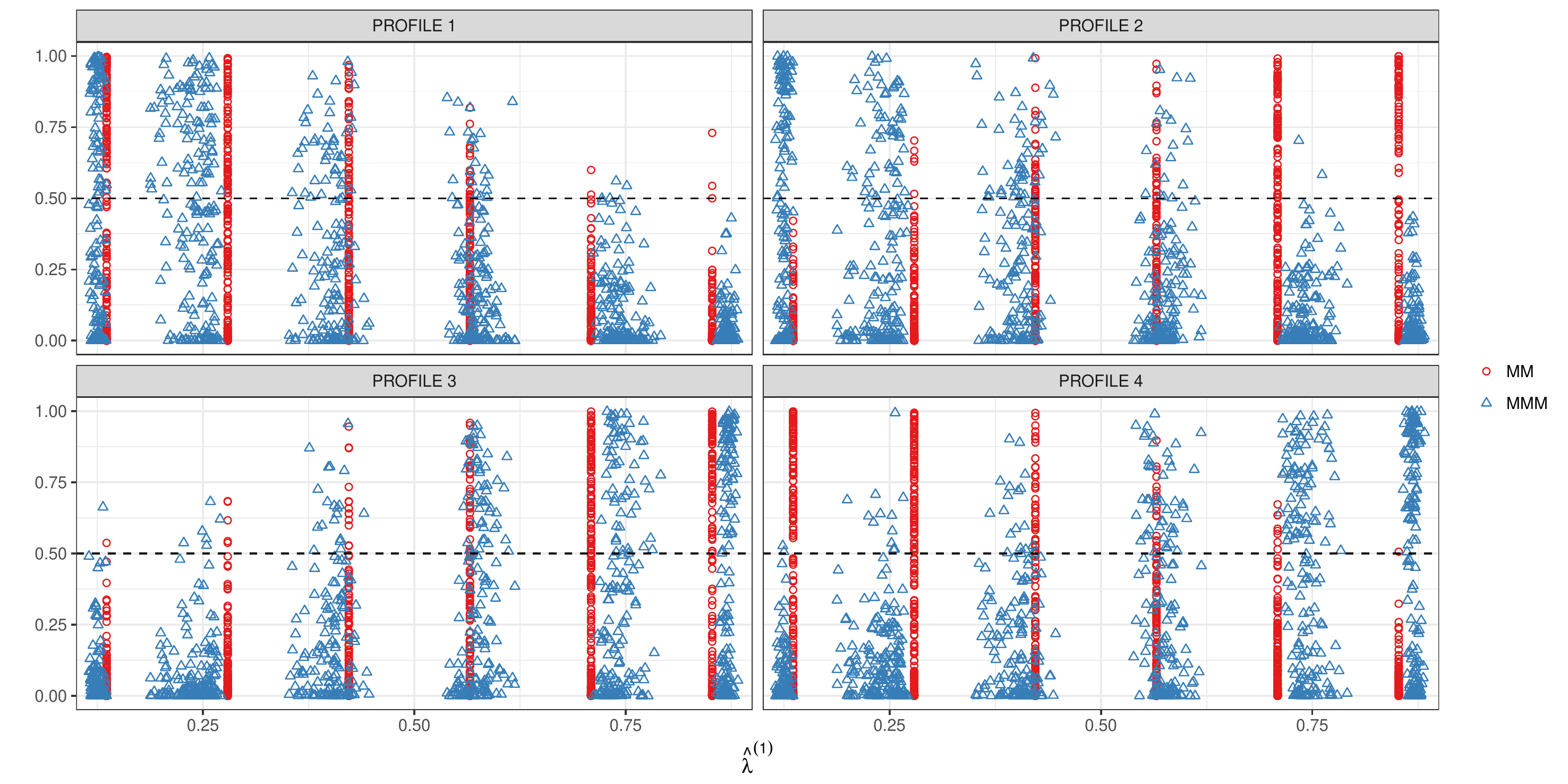}
    \caption{Estimated (x-axis) and `true' (y-axis) values for the score vectors, relying on MMM and \texttt{mixedMem} models for $g=1$. In this case, for each subject $i$, there are four `true' profiles $\lambda^{(1)}_{i1},\ldots,\lambda^{(1)}_{i4},$ represented in the panels y-axis, but just two estimated profiles $(\widehat \lambda^{(1)}_{i1},1 -\widehat \lambda^{(1)}_{i1})$ for the misspecified MM and MMM models.
}
\label{fig:multi_prof}
\end{figure}

Figure~\ref{fig:multi_prof} shows the relations between each component of the  `true' unknown $\lambda^{(1)}_{i1},\ldots,\lambda^{(1)}_{i4}$ and the estimated scores $\widehat \lambda^{(1)}_{i1}$, from our proposed approach and the MM model. As expected, the scores $\widehat \lambda^{(1)}_{i1}$ are strongly correlated with more `true' profiles, for both considered models. Some individual variability is lost in the process as we are projecting a 3-dimensional space to a 1-dimensional one.  This dimensionality reduction leads to `mixed' pure types that can be considered as averages of the `true' ones. For example, MMM model profile 2 is composed of subjects with high values of $\lambda^{(1)}_{i3}$ and $\lambda^{(1)}_{i4}$, and low values of $\lambda^{(1)}_{i1}$ and $\lambda^{(1)}_{i2}$, while in the MM model profile 2 is composed of high values of $\lambda^{(1)}_{i2}$ and $\lambda^{(1)}_{i3}$ and low values of $\lambda^{(1)}_{i1}$ and $\lambda^{(1)}_{i4}$. This can be assessed by looking at the estimated kernels for a representative variable in group 1 (see Table~\ref{tab:sim_kern}).

\begin{table}[h!]
\def\spacingset#1{\renewcommand{\baselinestretch}%
{#1}\small\normalsize} \spacingset{1.0}
    \centering
    \caption{Estimated kernels for MMM and MM model for variable $1$ in group $g_1 = 1$. Numbers in parenthesis are the 0.1 and 0.9 quantiles.}
    \label{tab:sim_kern}
    \scalebox{0.7}{
    \begin{tabular}{lllll}
\toprule
                                                                          & LEVEL 1                  &LEVEL 2                  &LEVEL 3                                &LEVEL 4 \\
\midrule
        MMM $\boldsymbol{\theta}^{(1)}_{1}$                               & 0.487(0.453;0.521) & 0.491(0.458;0.523) & 0.008(0.000;0.024)  & 0.013(0.000;0.037) \\
        $1/2(\boldsymbol{\varphi}^{(1)}_1 +  \boldsymbol{\varphi}^{(1)}_2)$ & 0.450              & 0.450              & 0.050 & 0.050 \\
        mixedMem  $\boldsymbol{\theta}^{(1)}_{1}$                         & 0.504              & 0.000              & 0.000               & 0.496 \\
        $1/2(\boldsymbol{\varphi}^{(1)}_1 +  \boldsymbol{\varphi}^{(1)}_4)$ & 0.450              & 0.050              & 0.050 & 0.450 \\
        \midrule 
        MMM $\boldsymbol{\theta}^{(1)}_{2}$                               & 0.026(0.000;0.059) & 0.011(0.000;0.032) & 0.449(0.412;0.484)  & 0.515(0.478;0.550)\\
        $1/2(\boldsymbol{\varphi}^{(1)}_3 +  \boldsymbol{\varphi}^{(1)}_4)$ & 0.050              & 0.050              & 0.450 & 0.450 \\

        mixedMem  $\boldsymbol{\theta}^{(1)}_{2}$                         & 0.000              & 0.533              & 0.467               & 0.000        \\
        $1/2(\boldsymbol{\boldsymbol{\varphi}}^{(1)}_2 +  \boldsymbol{\varphi}^{(1)}_3)$ & 00450              & 0.450              & 0.450 & 0.050 \\
        \bottomrule
    \end{tabular}
}
\end{table}

To additionally evaluate model performance, we compute the Frobenius norm between the `true' probability tensor $\pi^{(1)}_0 = \{\mbox{pr}(X_1 = x_1, \ldots, X_{p_1} = x_{p_1}); \mbox{ for } x_j=1,\ldots 4, j=1,\ldots,p_1\}$, and $\pi^{(1)}_{\tiny \mbox{MM}}$ and $\pi^{(1)}_{\tiny \mbox{MMM}}$, denoting the estimates from the MMM and MM model, respectively. Leveraging equations~\eqref{eq:sfm} and~\eqref{eq:spem}, for MM we have a closed form expression of the core tensor, while for MMM we use $10^5$ Monte Carlo replicates for the estimation. To estimate uncertainty in the MM case, we rely on 1000 bootstrap replicates.  We obtain a posterior mean of $0.131$ for $\| \pi^{(1)}_0  - \pi^{(1)}_{\tiny \mbox{MMM}}\|_{F}$ with a standard deviation of $0.048$, and a bootstrap mean $0.133$ for $\| \pi^{(1)}_0  - \pi^{(1)}_{\tiny \mbox{MM}}\|_{F}$ with standard deviation $0.029$. As expected, in this scenario having independent scores vectors, separate MM and MMM models share very similar performances in characterizing $\pi^{(1)}_0$.   For comparison, we also estimate the Frobenius norm considering a correctly specified model with $H_1=4$ profiles for group 1, leading to a bootstrap mean of $0.089$ with standard deviation of $0.032$. This slight improvement in estimation accuracy does not justify the greater complexity of interpretation in using the larger $H_g$ value.

\section{Application to malaria risk assessment}
\label{sec:app}
\subsection{Background}
\begin{sloppypar}
Starting from 1981, the World Bank sponsored Polonoroeste Development Project~\citep{worldbank1992}, which included funding for human settlements in previously forested areas~\citep{wade2011}. In these sponsored settlements we find the Machadinho project, in the state of Rond\^onia, where the primary goal was in promoting agricultural development and elevation of living standards by distributing pre-specified plots of land, and favoring migration from outside the area. Land clearance practices at the plots created new areas of partial shade---from cut but not cleared large trees---redefined the boundaries of forest fringe, and led to establishment of new pools of water of relatively high pH. These are precisely the ideal larval development conditions for A. Darlingi mosquitoes, the primary transmitter of malaria in the Brazilian Amazon region~\citep{decastro2006}.
\end{sloppypar}

As part of a field study of the dynamics of the settlement process at Machadinho, a set of household surveys was conducted in $1985$, $1986$, $1987$, and $1995$ at plots with relatively stable occupancy. The surveys were administered to settlers living on $70\%$ of what were regarded as occupied plots in $1985$ and $100\%$ of such plots in $1986$, $1987$, and $1995$.  An occupied plot is one in which settlers cleared some of their land and at least lived part-time in Machadinho. The surveys had as one objective the identification of drivers of malaria risk among the settlers, including some who were ascertained shortly after arrival in Machadinho and others who engaged in early out-migration, largely as a result of difficulties in establishing a productive agricultural site and illness, much of it being due to malaria.

\begin{figure}[h!]
  \centering
  \includegraphics[height=13cm,width=12cm ]{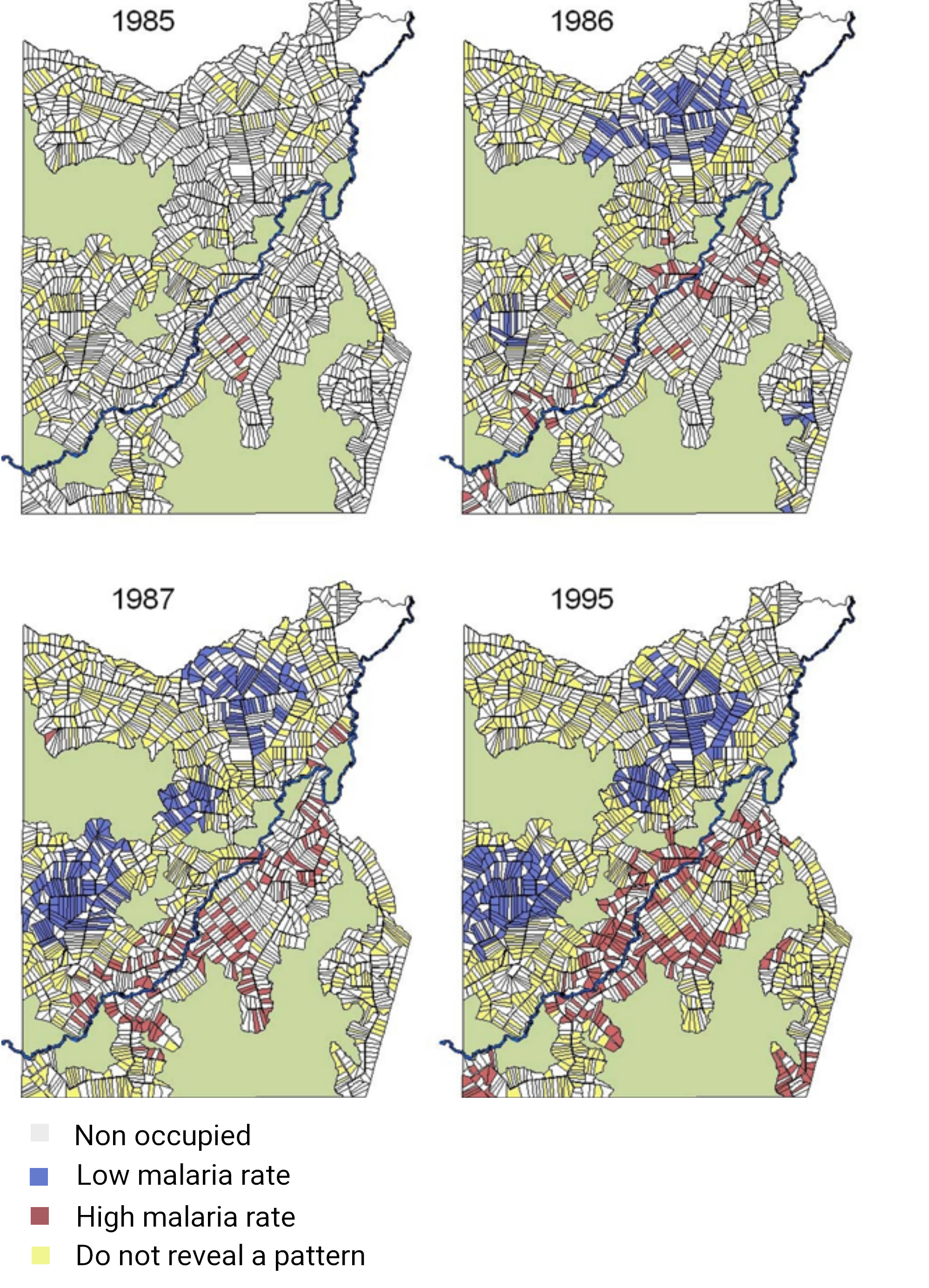}
  \caption{Plots at the Machadinho settlement project, showing occupancy, and clustering of malaria rates using a local indicator of spatial association, $G^*(d)$~\citep{getis1992} (d = 3,500 meters). Plots colored in blue are those significant for a clustering of low malaria rates, while those colored in red are significant for a clustering of high rates. Plots colored in yellow did not reveal a clustering pattern, and those in white were not occupied at the time of the interview. Green areas are protected forest reserved. Detailed ecologically based interpretation of these patterns is given in~\citet{castro2007}.}
  \label{fig:pnas_map}
\end{figure}

Factors that a priori were anticipated to influence exposure of settlers to Anopheles mosquitoes were complex physical environmental conditions and human behavioral conditions. It is natural to focus on extreme risk categories/profiles as ``High'' and ``Low''. Thus each occupied plot would have a numerical degree of similarity score, $\lambda_i$, with value in the unit interval. Values close to $1$ can be associated with proximity to the high risk profile, while values close to $0$ can be associated with low risk conditions. Using these variables in a standard MM analysis for each year, we obtain best fitting models with selected number of profiles $H$ ranging between $5$ and $8$. Since the selected $H$ are greater than $2$, an interpretability problem arises for scoring risk in the unit interval between two extreme profiles. If we force $H = 2$, as in prior analyses~\citep{decastro2006}, we are directly trading off model goodness-of-fit for interpretability in the scoring of malaria risk.  Goodness-of-fit can be improved increasing $H$ and use domain knowledge to map the resulting membership scores into behavioral and environmental risks,  in this case, we have a mixture of environmental and behavioral variables in each profile; hence, it requires some interpretive effort to decide which of the domains is most contributory at particular survey dates~\citep{decastro2006}.
We use an MM model with $H=4$  as a competitor of the proposed MMM, refer to 
Section 2 of~\ref{suppA} for the details on this model.

Interpretability issues  can be alleviated with an MMM analysis where $G = 2$ is the number of subject matter domains and $H_g = 2$ for $g = 1,2$, relying directly on the domain knowledge to partition variables. At Machadinho, the environmental conditions included quality of a house and its proximity to standing water; cut but not cleared trees changing the definition of the forest fringe and producing partial shade; site of initiation of farming near standing water and the forest fringe. Behavioral conditions included wearing of protective clothing, ownership of a chain saw and planter to facilitate land clearance and initiation of crop production, and farming close to the forest fringe. A core of $30$ variables remained common to all the years, while some other questions were gradually added over time. Household spatial locations are also available and will be considered in the analysis. Although the survey was carefully administrated, the composition of the resulting data is highly heterogeneous across time and includes many missing data, which were considered informative for this application; hence we defined a missing category for each variable so that the missingness pattern can inform the analysis results.

The full set of variables in the surveys is displayed in the first column of Tables S5--S8 of~\ref{suppA}, while Table~\ref{tab:samp_dist} shows, at a summary level, the number of subjects and variables included in the analysis by survey year and domain of variables. 

\begin{table}[h!]
\def\spacingset#1{\renewcommand{\baselinestretch}%
{#1}\normalsize} \spacingset{1.0}
    \centering
    \caption{Distribution of the number of subjects and variables included in the analysis.}
    \label{tab:samp_dist}
    \scalebox{0.8}{
    \begin{tabular}{ccccc}
        \toprule
        Year & \# Subjects & \# Behavioral variables & \# Environmental variables& \# Variables \\
        \midrule
        1985 & 269 & 14&28&42 \\
        1986 & 575 & 16&24&40 \\
        1987 & 802 & 14&29&33 \\
        1995 & 1108& 19&36&55 \\
        \midrule
        Total &2754& 63&117 &180\\
        \bottomrule
\end{tabular}
}
\end{table}

A fundamental challenge for identifying the drivers of malaria risk is the fact that there are many environmental and behavioral conditions that contribute to exposure to A. Darlingi, but there is no individual or small combination of such conditions that occurs at high frequency and stands out as a major influence on malaria episodes experienced. This is precisely where MMM can be used to an advantage.

\subsection{Model specification}
Malaria behavioural and environmental risk scores can present distinct time and space evolutions, since we consider data that goes from the beginning of a settlement project to 10 years later. To characterize such evolution we leverage the multivariate Gaussian model in step $5$ of Algorithm~\ref{alg:sim}. Different multivariate spatio-temporal models can be considered~\citep[see for example][]{banerjee2014}, and we rely on a separable model for time and space, assuming no interactions. This assumption leads to a simple and computationally efficient latent model, while accommodating non regular observations in space and time. Indicating with $\lambda^{(\mbox{\tiny B})}_i$ and $\lambda^{(\mbox{\tiny E})}_i$ the behavioral and environmental risk score for subject $i$, space-time dependence can be included in distribution~\eqref{eq:profmod} letting
\begin{eqnarray}
    (\lambda^{(\mbox{\tiny B})}_i,\lambda^{(\mbox{\tiny E})}_i)^T &\sim& \mbox{MLND}(\boldsymbol{\beta}_{t_i} + \boldsymbol{\zeta}_t(\boldsymbol{s}_i),\boldsymbol{\Sigma}_t),
    \label{eq:app_prof_mod}
\end{eqnarray}
where $t_i \in \{1985, 1986,1987,1995\}$, and $\boldsymbol{s}_i = (s_{i1},s_{i2})^T$, are respectively a time indicator and the observed longitude and latitude corresponding to the household of subject $i$.

We account for time dependence through a multivariate Gaussian hierarchical model with common hyperprior. Specifically, $\boldsymbol{\beta}_t = (\beta^{(\mbox{\tiny B})}_t,\beta^{(\mbox{\tiny E})}_t) \sim \mathcal{N}(\boldsymbol{\beta},\boldsymbol{\Sigma})$, $\boldsymbol{\beta} \sim \mathcal{N}(\boldsymbol{\beta_0},\boldsymbol{\Sigma_0})$ and $\boldsymbol{\Sigma} \sim \mathcal{IW}(\nu_\beta,\boldsymbol{\Psi}_\beta)$. This model does not impose a rigid  time evolution, allowing borrowing of information across different years.

For the spatial effect $\boldsymbol{\zeta}_t(\boldsymbol{s}_i)$ we specify a bivariate spatial model. A simple possibility would be to rely on a separable structure for the spatial cross covariance of the process~\citep[e.g.,][]{banerjee2000}. However, this model would imply the same spatial effect for both the environmental and behavioral domain. We expect that behavioral and environmental scores can have a very different spatial evolution, and for this reason we rely on a conditional Gaussian process $p(\boldsymbol{\zeta} | \boldsymbol{\Sigma}_t)$ for the components of ${\boldsymbol{\zeta}_t}(\boldsymbol{s}_i)= (\zeta_t^{(\mbox{\tiny B} )}(\boldsymbol{s}_i),\zeta_t^{(\mbox{\tiny E})}(\boldsymbol{s}_i))^T$. Specifically we consider $\boldsymbol{\Sigma_t} = \boldsymbol{L_t} \boldsymbol{L_t}^T$, where $\boldsymbol{L_t}$ is a lower triangular matrix obtained through Cholesky decomposition, and we let $\tilde{\boldsymbol{\zeta}}_t(\boldsymbol{s}_i) = \boldsymbol{L}^{-1}_t\boldsymbol{\zeta}_t(\boldsymbol{s}_i)$ and $\tilde{\boldsymbol{\zeta}}^{(g)}_t \sim \mbox{GP}(0,\boldsymbol{K}^{(g)}_t)$. This formulation enforces no dependence across time and space for the spatial effects.

Since we are considering standardized data, we parameterize the Gaussian processes in terms of correlation functions $K^{(g)}_t(\boldsymbol{s}_i,\boldsymbol{s}_{i^\prime})$, obtained by normalizing the square exponential form $\exp\{ -1/2\sum_{d=1}^2 \gamma^{(g)}_{td} \mbox{d}^2(s_{id},s_{i^\prime d})\} + \tau I(i=i^\prime) $, where $\gamma^{(g)}_{td}$ are length scale parameters, $\mbox{d}(\cdot,\cdot)$ is the Euclidean distance and $\tau$ is a nugget effect to limit numerical instability. The considered prior induces independent spatial effects for each domain and time, while leading to a computationally efficient model, since small matrices are involved in the Gaussian process computation.

The model can be easily implemented adapting Gibbs sampler  Algorithm~\ref{alg:sim}, with updating of the length scale parameters $\gamma^{(g)}_{td}$ relying on a Metropolis step.

\subsection{Model checking}
\label{model_checking}

We assess goodness-of-fit of the proposed model to the observed data. We are particularly interested in whether the assumption of $H_g = 2$ leads to significant lack of fit.

One  possibility is to compute posterior distributions for some statistics of the considered data and compare them with the corresponding empirical quantities~\citep[e.g.,][]{gelman2013}.  We consider as statistics the marginal and bivariate distributions, that can be obtained as: 
\begin{eqnarray}
	\quad \pi^{(j)}_{x_j} &=& \mbox{pr}(X_j = x_j\mid - ) = \sum_{h=1}^H \bar{a}^{(g_j)}_h  \theta^{(j)}_{hx_j} \nonumber,\\
	\quad \pi^{(j,k)}_{x_j,x_k} &=& \mbox{pr}(X_j = x_j, X_k = x_k \mid - ) = \sum_{h_j=1}^H \sum_{h_k =1}^H \bar{a}^{(g_j,g_k)}_{h_j h_k}  \theta^{(j)}_{h_j x_j} \theta^{(k)}_{h_k x_k},
    \label{eq:marg_biv}
\end{eqnarray}
for $j=1,\ldots,p$ and $k\neq j$, and where $\bar{a}^{(g_j)}_h = \mathbb E[\lambda^{(g_j)}_h]$ and  $\bar{a}^{(g_j,g_k)}_{h_jh_k} = \mathbb E[\lambda^{(g_j)}_{h_j} \lambda^{(g_k)}_{h_k}]$. 
\begin{figure}[h!]
\centering 
    \includegraphics[width=\textwidth]{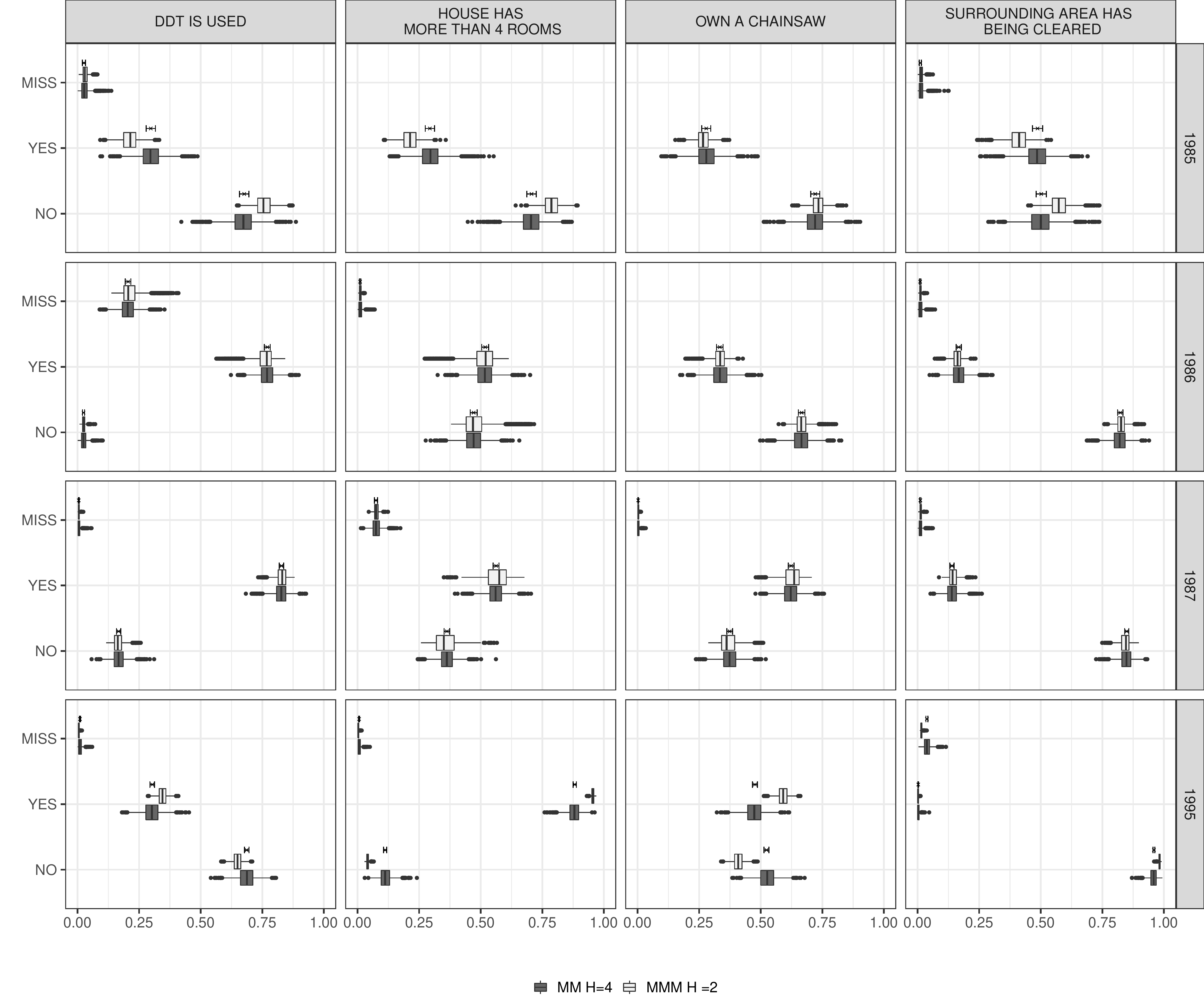}
    \caption{Box-plots of marginal distributions for 4 representative variables over the years. Light gray box-plots indicate the proposed MMM model, while dark gray ones refer to the MM model with $H=4$ profiles described in Section~{2} of~\ref{suppA}. The `x's and error bands above the boxplots are the sample proportions and 0.95  Wald-type confidence intervals.}
    \label{fig:marg_distr}
\end{figure}

Figure~\ref{fig:marg_distr} shows estimated marginal distributions from $2500$ Gibbs samples for 4 representative variables across the years, for the proposed model and the MM described in Section~{2} of~\ref{suppA}, together with the sample proportion with the $0.95$ level Wald type confidence interval. Both models produce estimated marginals that are compatible with the observed ones. We additionally estimated the posterior mean of the L1-norm between the empirical frequencies $\hat{f}_{jc} = n^{-1}\sum_{i = 1}^n I(X_{ij} = c)$ and the estimated ones $\hat{\pi}_{jc}$ obtained by averaging 2500 MCMC samples of the expression in~\eqref{eq:marg_biv}. The L1-norm  has expression $\sum_{c = 1}^{d_j} |\hat{\pi}_{jc} - \hat{f}_{jc}|$. Considering all the variables, we have an average L1-norm of about $0.071$, and  $0.1$ and $0.9$ quantiles of $(0.024, 0.14)$, for the proposed model, and $0.065\mbox{ }(0.039, 0.10)$ for the MM with $H=4$ profiles. These quantities also suggest a strong adherence of the estimated marginals with the empirical ones.

\begin{figure}[h!]
    \includegraphics[ width = \textwidth]{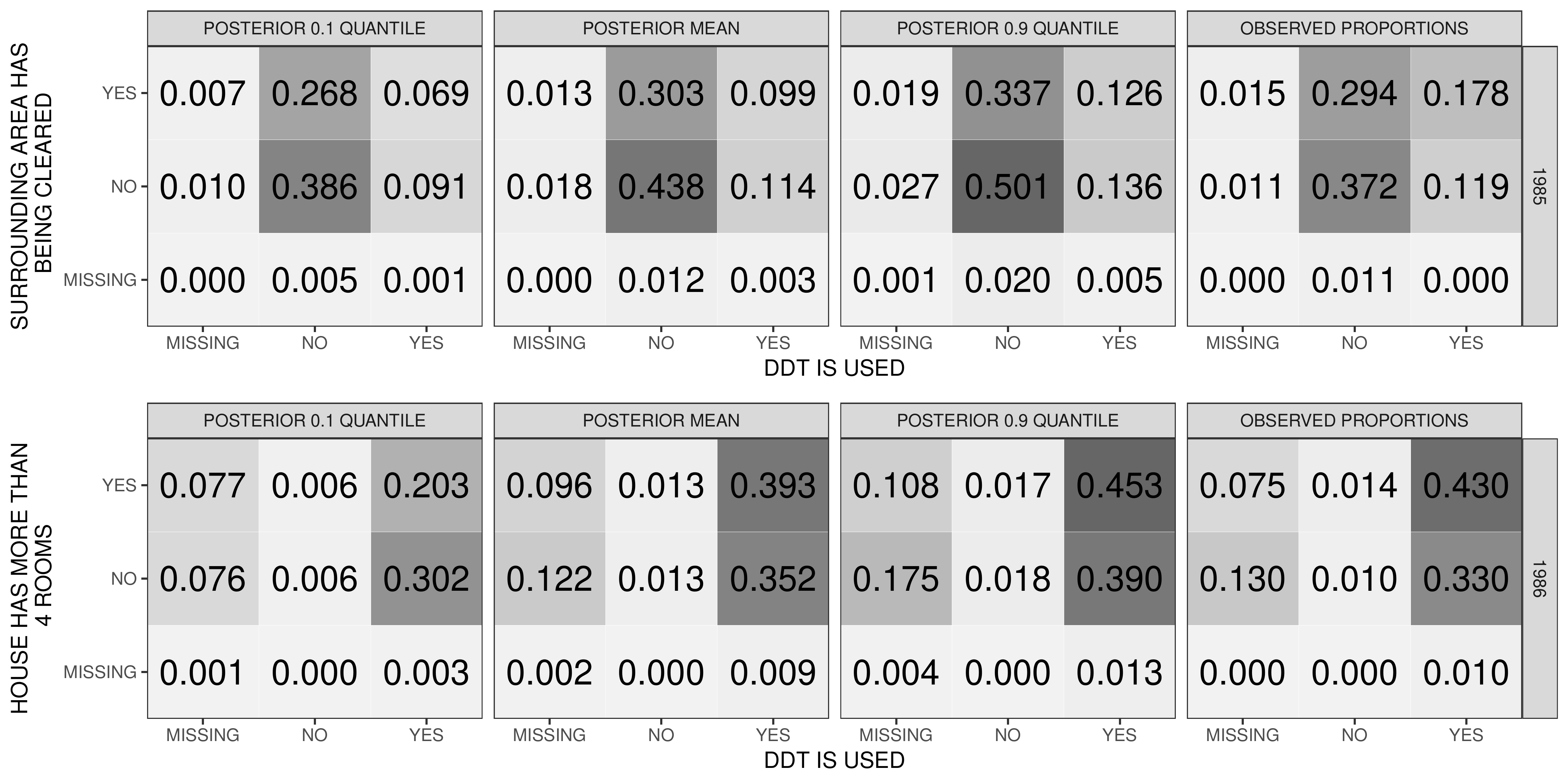}
    \caption{Posterior mean and quantiles for 2 bivariate distributions, compared with the empirical frequencies in the data.} 
    \label{fig:bivariates}
\end{figure}

Figure~\ref{fig:bivariates} shows posterior distributions and quantiles for 2 bivariate distributions (for the MM model refer to Figure~{S7} in \ref{suppA}). Also in this case we observe a satisfactory adherence of the estimated quantities and the empirical ones.

As with the marginals we compute the L1-norm between the empirical bivariate distributions and the estimated ones. We focus on pairs of variables in the same year. We obtain good results also in this case with a mean of $0.13$ and $0.1$ and 0.9 quantiles of $(0.11, 0.18)$, for the proposed MMM model, and $0.012\mbox{ }(0.07, 0.13)$ for the MM model.

In the MMM model as a general tendency, we observed that variables that are sensibly different across the estimated profiles are generally better reconstructed. These are also the most interesting from an interpretation point of view since they characterize the profiles.

The agreement between empirical and estimated quantities suggests that the considered interpretable model with $G=2$ and $H_g = 2$, for $g=1,2$, is sufficiently precise to analyze the profile structure more in depth. The additional latent parameters in the MM model with $H=4$ provide a small benefit in terms of fit to the data, but they make the latent profiles less straightforward to interpret in terms of malaria risk (refer to Section~{2}  of \ref{suppA}). If the MMM model had presented a poor fit to the observed data, we should either have increased $H_g$, or defined a finer partition of variables. Both solutions are viable and, as a consequence of Lemma~\eqref{th1}, might lead to an equally accurate approximation of the underlying probability mass function. Note also that the posterior mean (and standard deviation) of the correlation for the environmental and behavioral score vectors are $0.848\mbox{ }(0.106)$, $0.842\mbox{ }(0.111)$, $0.851\mbox{ }(0.103)$, and $0.866\mbox{ }(0.090)$, for $1985,$ $1986,$ $1987,$ and $1995$ respectively.  These values suggest that two separate MM models would ignore a high correlation structure, while a single MM model with $H=2$ (posterior correlation =1) may be insufficient to characterize the score distribution.

\subsection{The structure and evolution of risk profiles}
All variables that can enter risk profiles take on a discrete set of possible values/levels. Each level of a variable represents what we will refer to as a condition. The conditions that occur in a profile $h$ with substantially greater frequency than in the overall population can be considered as the most relevant to describe the profiles, and will be referred to as admissible~\citep[see][for a detailed discussion]{singer1989}.

To make this precise we say that condition $l$ for variable $X_j$ in vertex $h$ is called admissible if either 

\vspace{-0.45cm}

\begin{subequations}
\begin{tabularx}{\textwidth}{ABC}
\begin{equation}
\theta^{(j)}_{h l} > c_1 \hat{f}_{jl}
\label{eq:admissible_a}
\end{equation}
&
\begin{equation*}
\mbox{or}
\end{equation*}
&\begin{equation}
[\theta^{(j)}_{h l} - \hat{f}_{jl}]/\hat{f}_{jl} > c_2,\label{eq:admissible_b}
\end{equation}
\end{tabularx}
\end{subequations}
\vspace{-0.80cm}

\noindent where  $c_1 = 1.7$, $c_2 = 0.35$ and $\hat{f}_{jl} = n^{-1}\sum_{i = 1}^n I(X_{ij} = l)$ are the marginal empirical frequencies.

Inequality~\eqref{eq:admissible_a} is appropriate for $\hat{f}_{jl}  < 0.5$---i.e. relatively infrequent conditions. Inequality~\eqref{eq:admissible_b} is particularly important in the present survey data, as quite a few conditions occur with high frequency---e.g. $> 0.90$---in the overall population.

The set of conditions $\{l \in \{1,\ldots,d_j\}  : \theta^{(j)}_{hl} \mbox{ satisfies \eqref{eq:admissible_a}  or  \eqref{eq:admissible_b}} \}$ is defined to be an admissible profile. Admissible profiles are described by logical AND statements for the set of admissible conditions. In the proposed Bayesian framework we can compute the posterior probability of \eqref{eq:admissible_a} and \eqref{eq:admissible_b}, and define as admissible the conditions having posterior probability exceeding $0.5$. These conditions are reported in light gray in Tables S5--S8 of~\ref{suppA}. Further categorizing sets of conditions especially relevant for exposure to A. Darlingi mosquitoes in, for example, the environmental profiles, leads to a clear display of the change in such conditions over time as the highly dynamic plot occupancy process evolves (see Figure~\ref{fig:pnas_map}).  %
\begin{table}[!b]
\caption{Admissible profiles for environmental risk}
\label{tab:adm_env_sum}
\setlength\tabcolsep{13pt}
\centering
\resizebox{\textwidth}{!}{
\def\spacingset#1{\renewcommand{\baselinestretch}%
{#1}\small\normalsize} \spacingset{1.0}
    \begin{tabular}{rcccccccc}
   &\multicolumn{8}{c}{Survey years}\\
   Conditions&   \multicolumn{2}{l}{'85}& \multicolumn{2}{l}{'86} &\multicolumn{2}{l}{'87} &\multicolumn{2}{l}{'95}\\
        \midrule\\
                  &\multicolumn{8}{c}{Risk Profiles}\\
                               & Low & High & Low  & High & Low & High & Low & High \\
        \midrule
        House Characteristics  & &&&&&&&\\
        \# rooms$>$ 4          & +&        -&            +&         -&          +&        -&        +&         *\\

        Good quality walls     & +&        -&            +&         -&          +&        -&        *&         * \\
        Good quality roof      & +&        -&            +&         -&          *&        -&        *&         *\\
        Good quality sealing   & +&        *&            *&         *&         +&        *&       +&         *\\
\addlinespace                 \\
Land Clearance \& Water        & &&&&&&&\\
Prior land clearance           & *&        +&            -&         +&         *&        *&        *&         *\\
$>$ 100 m from forest          & +&        -&            +&         -&         *&        *&        *&         *\\
Good water source available    & +&        -&            +&         -&         *&        -&         *&         *\\
Good bathing available         & *&        *&            +&        *&         *&        *&        *&         *\\
Near big pasture area          & *&        *&            +&        *&         *&        *&        +&         *\\
\bottomrule\\
Code:&    \multicolumn{8}{l}{$+=$ stated condition is admissible}\\ 
     &    \multicolumn{8}{l}{$-=$ stated condition is not admissible}\\
     &    \multicolumn{8}{l}{$*=$ no level of the condition is admissible}\\
\midrule\\
\multicolumn{9}{l}{Note: Additional levels for some other variables are admissible, as indicated in Tables S5--S8 of~\ref{suppA}.} 
\end{tabular}

}
\end{table}
Admissible environmental conditions labeled in a high risk profile, summarized in Table~\ref{tab:adm_env_sum}, correspond to situations that facilitate exposure to A. Darlingi mosquitoes (e.g. poor quality of wall and sealing). These high environmental risk conditions, operable during the first three years of the settlement process, disappear by $1995$ when diverse improvements at occupied plots have been incorporated. Such a tendency is clearly highlighted by an increasing trend in the distribution of expected odds ratios of the risk scores $(\lambda^{(\mbox{\tiny B})}_i,\lambda^{(\mbox{\tiny E})}_i)^T$ shown in Figure~{S8} in~\ref{suppA}.

These results are in accordance with current literature on malaria risk, in Amazon areas, reporting that the risk is initially driven by favorable environmental conditions for malaria vectors to proliferate~\citep[e.g.,][]{decastro2006}. Soon after human settlement, there is a phase lasting for about 8 or 10 years, in which environmental risk is high but human behavior is starting to gradually become the predominant risk factor. In the last stage, called the endemic phase, the risk is far more related to behavioral causes. From a spatial perspective, we can study the changes in behavioral and environmental risks by considering the posterior predictive distribution of the ${\boldsymbol{\zeta}_t}$ defined in equation~\eqref{eq:app_prof_mod}. Maps evaluating this distribution over a regular grid of points are available in~\ref{suppA} (Figure~{S9}), showing  that the behavioral risk distribution is constant across time and space; hence the spatial variability is driven by environmental conditions. Environmental risk zones can be mostly explained in term of geographical characteristics of the area; in fact higher risk zones correspond to the forest fringe and the Machadinho river path.

\subsection{Malaria rates and risk profiles}

To assess the relationship between malaria rates and the estimated profiles, for each MCMC sample in each year, we assign households/plots into low, moderate, and high behavioral and environmental risk groups using tertile cut points. We compute the average malaria rate for each of the $9$ groups. Posterior median and quantiles of the malaria rate distribution for the considered clusters are shown in Table~\ref{tab:cross_rate_tertiles}. A similar strategy using tertiles of separate MM models has been used to determine risk profiles for Chagas disease in northern Argentina~\citep{chuit2001}. We expect the obtained distribution to be ordered in such a way that higher malaria rates correspond to high environmental and behavioral risk clusters. Formally considering a table such as Table~\ref{tab:cross_rate_tertiles}, having low risk clusters in the upper left corner, we expect the resulting table to be a double-gradient table, meaning that each row should be non-decreasing from left-to-right and each column should be non-decreasing from top-to-bottom.

\begin{table}[h!]
\def\spacingset#1{\renewcommand{\baselinestretch}%
{#1}\small\normalsize} \spacingset{1.0}
 \caption{Median 0.1 and 0.9 posterior quantiles of the malaria rates for low, medium, and high environmental and behavioral profiles. Groups are specified by using tertiles of $\lambda^{(\mbox{\tiny B})}_i$ and $\lambda^{(\mbox{\tiny E})}_i$ scores. Values at the bottom and right side of the table are obtained using marginal behavioral and environmental scores, respectively. Light gray values indicate violation of the double-gradient hypothesis, while dashes indicate that there are not enough households/plots in the cluster to compute the median and quantiles.} 
\label{tab:cross_rate_tertiles}
\centering 
\resizebox{\textwidth}{!}{
\begin{tabular}{llll|l}
       &               1985 &                    &                    &                    \\
\toprule
       &              Low B & Med B              &             High B &                    \\
Low E  & 0.000(0.000;0.000) & 0.045(0.000;0.197) &              ----- & 0.000(0.000;0.000) \\
Med E  & 0.067(0.000;0.310) & 0.101(0.050;0.143) & 0.115(0.000;0.250) & 0.106(0.067;0.129) \\
High E &              ----- & 0.115(0.000;0.356) & 0.117(0.100;0.125) & 0.117(0.100;0.125) \\
\midrule
       & 0.000(0.000;0.000) & 0.091(0.050;0.125) & 0.117(0.100;0.125) &                    \\
\bottomrule
\end{tabular}
}

\vspace{0.3cm}
\resizebox{\textwidth}{!}{
\begin{tabular}{llll|l}
       &               1986 &                                        &                                        &                    \\
\toprule
       &              Low B & Mod B                                  &                                 High B &                    \\
Low E  & 0.228(0.206;0.250) & 0.231(0.139;0.327)                     &                                  ----- & 0.228(0.206;0.250) \\
Mod E  & 0.238(0.167;0.299) & \cellcolor{gray!25} 0.250(0.200;0.279) & \cellcolor{gray!25} 0.232(0.000;0.658) & 0.250(0.206;0.273) \\
High E &              ----- & 0.250(0.200;0.302)                     &                     0.250(0.250;0.286) & 0.250(0.250;0.279) \\
\midrule
       & 0.229(0.211;0.250) & 0.250(0.217;0.275)                     &                     0.250(0.250;0.286) &                    \\
\bottomrule
\end{tabular}
}

\vspace{0.3cm}

\resizebox{\textwidth}{!}{
\begin{tabular}{llll|l}
       &               1987 &                                        &                                       &                    \\
\toprule
       &              Low B & Mod B                                  &                                High B &                    \\
Low E  & 0.167(0.133;0.180) & 0.216(0.140;0.458)                     &                                 ----- & 0.167(0.133;0.180) \\
Mod E  & 0.177(0.146;0.207) & \cellcolor{gray!25} 0.190(0.167;0.215) & \cellcolor{gray!25}0.180(0.082;0.243) & 0.183(0.167;0.200) \\
High E &              ----- & 0.200(0.171;0.250)                     &                    0.201(0.193;0.227) & 0.201(0.197;0.219) \\
\midrule
       & 0.167(0.150;0.180) & 0.197(0.181;0.209)                     &                    0.201(0.193;0.227) &                    \\
\bottomrule
\end{tabular}
}

\vspace{0.3cm}

\resizebox{\textwidth}{!}{
\begin{tabular}{llll|l}
       &               1995 &                    &                    &                    \\
\toprule
       &              Low B & Mod B              &             High B &                    \\
Low E  & 0.028(0.021;0.042) & -----              &              ----- & 0.028(0.021;0.042) \\
Mod E  & 0.028(0.024;0.028) & -----              &              ----- & 0.028(0.024;0.028) \\
High E & 0.028(0.024;0.033) & 0.033(0.030;0.036) & 0.036(0.029;0.042) & 0.033(0.031;0.033) \\
\midrule
       & 0.028(0.028;0.028) & 0.033(0.030;0.036) & 0.036(0.029;0.042) &                    \\
\bottomrule
\end{tabular}
}  
  
\end{table}
From Table~\ref{tab:cross_rate_tertiles}  we notice that in defining groups with either environmental or behavioral scores we obtain groups sharing almost the same malaria rate. However, a more detailed classification can be obtained leveraging both domains at the same time. In general, we observe higher malaria median rates in high behavioral and environmental risk zones, with only two violations of the expected double gradient assumption if we consider the median; in both cases, however, upper quantiles are still increasing as expected. We provide a similar analysis for the MM model in Section 2 of \ref{suppA}. Although the two models share similar performances in terms of goodness-of-fit, the direct use of domain knowledge in the MMM model lead to results which are simpler to interpret in term of malaria risk. 

\section{Discussion}
We introduced a new family of multivariate mixed membership models (MMM) that facilitate the representation of interpretable shared memberships in classification schemes in settings where good-fitting conventional MM models pose severe interpretation problems. The crux of this issue is that it is virtually impossible to write a coherent sentence describing shared membership among, say, $10$ profiles. However, if a large set of variables can be meaningfully partitioned into separate subject matter domains, for each of which there is a small number (ideally $2$) of domain-specific profiles, then experience to-date indicates that interpretable descriptions of the patterns of shared membership are possible. Further, cross-domain comparisons of shared membership reveal new information that cannot be extracted with MM models that typically incorporate a large number of profiles. This is shown at a most basic level via Table~\ref{tab:cross_rate_tertiles} in our analysis of malaria risk on the Amazon frontier.

In the considered application there is a clear partition of variables into subject matter domains, which simplifies model specification and interpretation of results. Similar sharp partitions of variables can be found in many applications; for example, in testing student language skills we can have questions related to several tasks (i.e., reading, writing, listening, etc.). In other settings, the number and compositions of domains can be just partially known, or unknown. In these cases, it can be useful to let the data inform on the subject matter domains, generalizing model~\eqref{eq:hmmm} with an additional prior on variable groups. Informative priors on the space of partitions can be included adapting recent proposals in \citet{paganin2020} and \citet{smith2019}.  However, additional latent layers make interpretation and computation challenging, inducing a further trade-off between model complexity and interpretability.

The MMM framework that we put forth is quite general and should be applicable in a wide variety of scientific contexts. In the interest of focusing attention on MMM per se, we provided a first top-level illustration of what can be done with this technology that is not feasible with other extant methods. More nuanced spatially explicit analyses that integrate evidence from the surveys used here with satellite imagery and ethnographic appraisal would be a next step for utilization of MMM. An initial pass at this kind of complex data integration in a study of malaria in the Brazilian Amazon region is presented in~\citet{decastro2006}, but the methodology introduced here has the potential to carry this case study much further.

\section*{Acknowledgements}
Data for this study were extracted from the project ``Land Use and Health'' funded by the International Development Research Centre (IDRC), grant \#94-0206-00, awarded to Centro de Desenvolvimento e Planejamento Regional, CEDEPLAR (Belo Horizonte, MG, Brazil), PI: Diana O. Sawyer. Data was cleaned and treated by Marcia C. Castro, and originally used in~\citep{decastro2006}(\url{www.pnas.org/cgi/doi/10.1073/pnas.0510576103}). The work was partially supported by funding from grants R01ES027498 and R01ES028804 from the United States National Institute of Environmental Health Sciences and grant N00014-16-1-2147  from the Office of Naval Research.

\begin{supplement}
\sname{Supplement A}\label{suppA}
\stitle{Malaria risk conditions}  
\slink[url]{url}
\sdescription{The file includes additional simulations and tables showing posterior medians, 0.1 and 0.9 quantiles of the kernels for behavioral and environmental variables for all the survey years. It also includes details of an MM model with $H=4$ profiles applied to the Machadino data, a plot of the expected odds ratios between environmental and behavioral risk scores, and one on spatial risk predictions for the Machadinho area, with some comments.}
\end{supplement}



\newpage
\setcounter{section}{0}
\begin{frontmatter}
\title{Supplementary Material for ``Multivariate mixed membership modeling: Inferring domain-specific risk profiles''}
\runtitle{Multivariate mixed membership modeling}

\begin{aug}
  \author{\fnms{Massimiliano} \snm{Russo}\thanksref{t1}\ead[label=e1]{m\_russp@hms.harvard.edu}},
\author{\fnms{Burton H.} \snm{Singer}\thanksref{t2}\ead[label=e2]{second@somewhere.com}}\\
\and
\author{\fnms{David B.} \snm{Dunson}\thanksref{t3}
\ead[label=e3]{third@somewhere.com}
\ead[label=u1,url]{http://www.foo.com}}

\affiliation{Harvard Medical School, and Dana-Farber Cancer Institute\thanksmark{t1}, University of Florida\thanksmark{t2} and Duke University\thanksmark{t3}}

\end{aug}
\end{frontmatter}

\beginsupplement
\section{Simulation details}
The code to reproduce all the tables and plots in Section 6 is available at 
\url{https://github.com/rMassimiliano/MMM-tutorial}, which includes step-by-step description on how to generate and analyze data from the considered simulation scenarios. To initialize the Variational EM algorithm in the \texttt{mixedMem} package, we used the simulation truth (in Section 6.2 we used true values from the first two profiles). In our MMM model we found no significant difference in the initialization of the parameters. We consider also an MCMC implementation of the MM models as the one proposed in \cite{erosheva2007}, using the software \texttt{NIMBLE}~\citep{nimble}; the code is available in the script \texttt{MM\_nimble.R}. In this implementation we set $(\lambda_{i1},\lambda_{i2}) \sim \mbox{Dirichlet}(\xi \gamma_1,\xi \gamma_2)$, $(\gamma_1,\gamma_2) \sim \mbox{Dirichlet}(1/2,1/2)$, and $\xi \sim \mbox{Gamma}(12,12)$, with mean $1$ and variance $1/12$; with these hyperparameters  the score vector components are centered at $1/2$, with a variance of approximately $0.13$.

    \begin{figure}[h!]
      \centering
      \includegraphics[width=\textwidth]{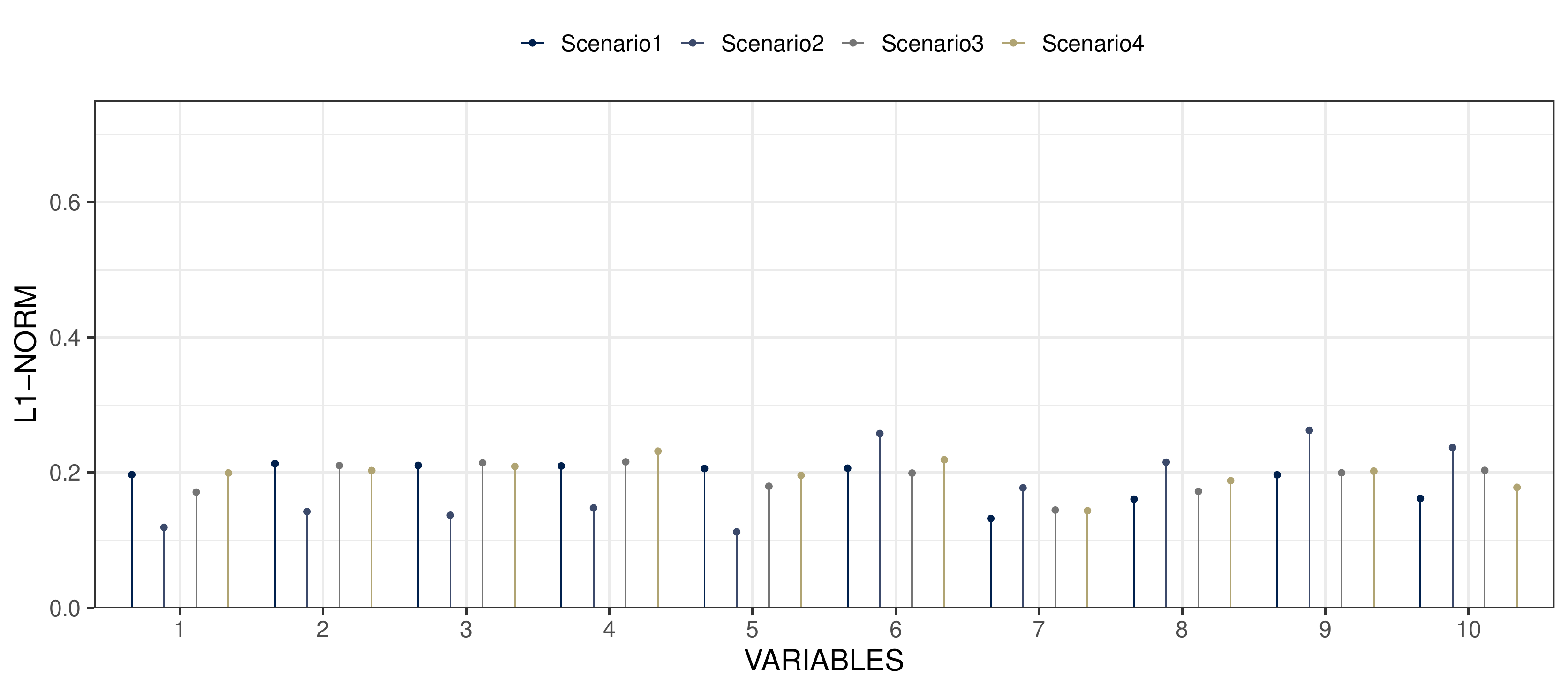}
      \includegraphics[width=\textwidth]{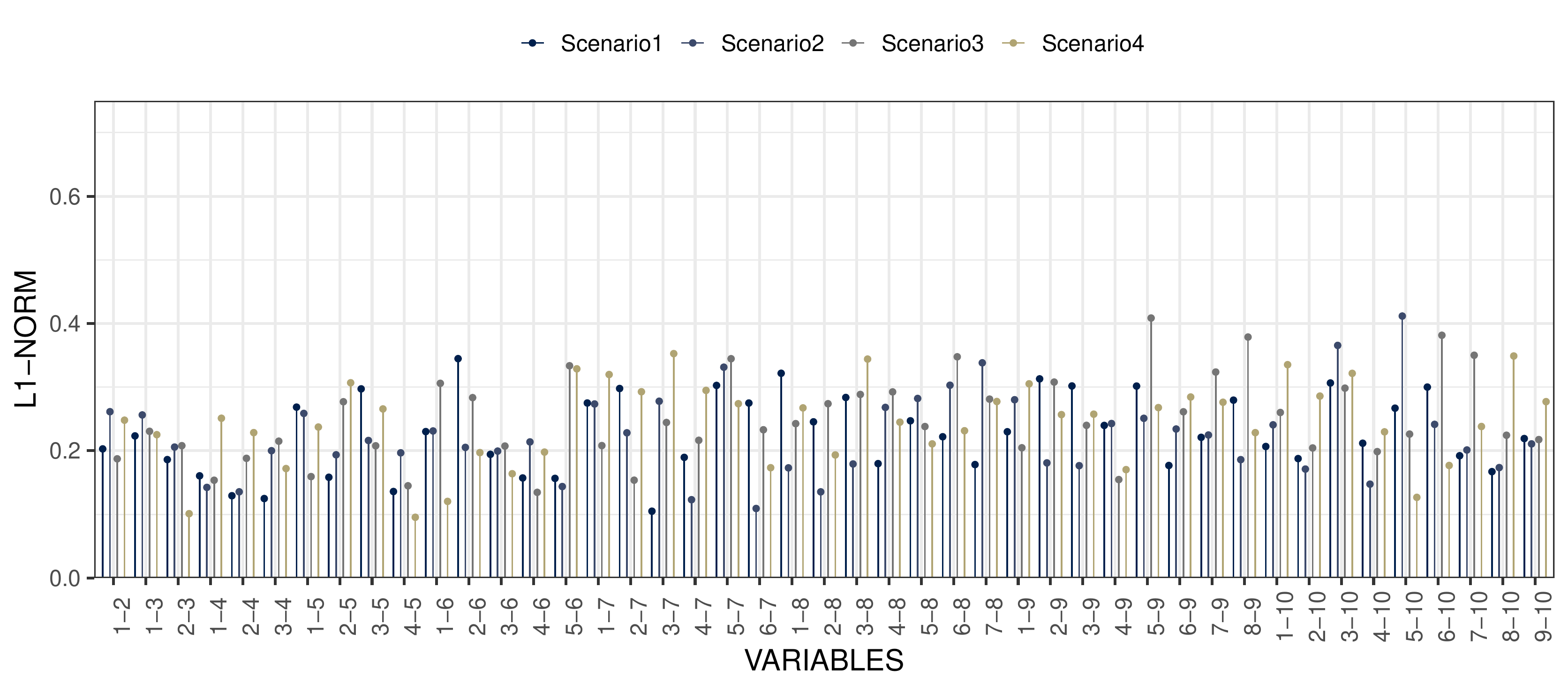}
      \caption{MMM: L1-norm between the empirical frequencies and posterior means for all mariginal (top panel) and bivariate (bottom panel) distribution for all simulated scenarios. Posterior means are obtained averaging $2500$ MCMC samples of the expression in (7.2).  }  
    \end{figure}

    \begin{figure}[h!]
      \centering
      \includegraphics[width = \textwidth]{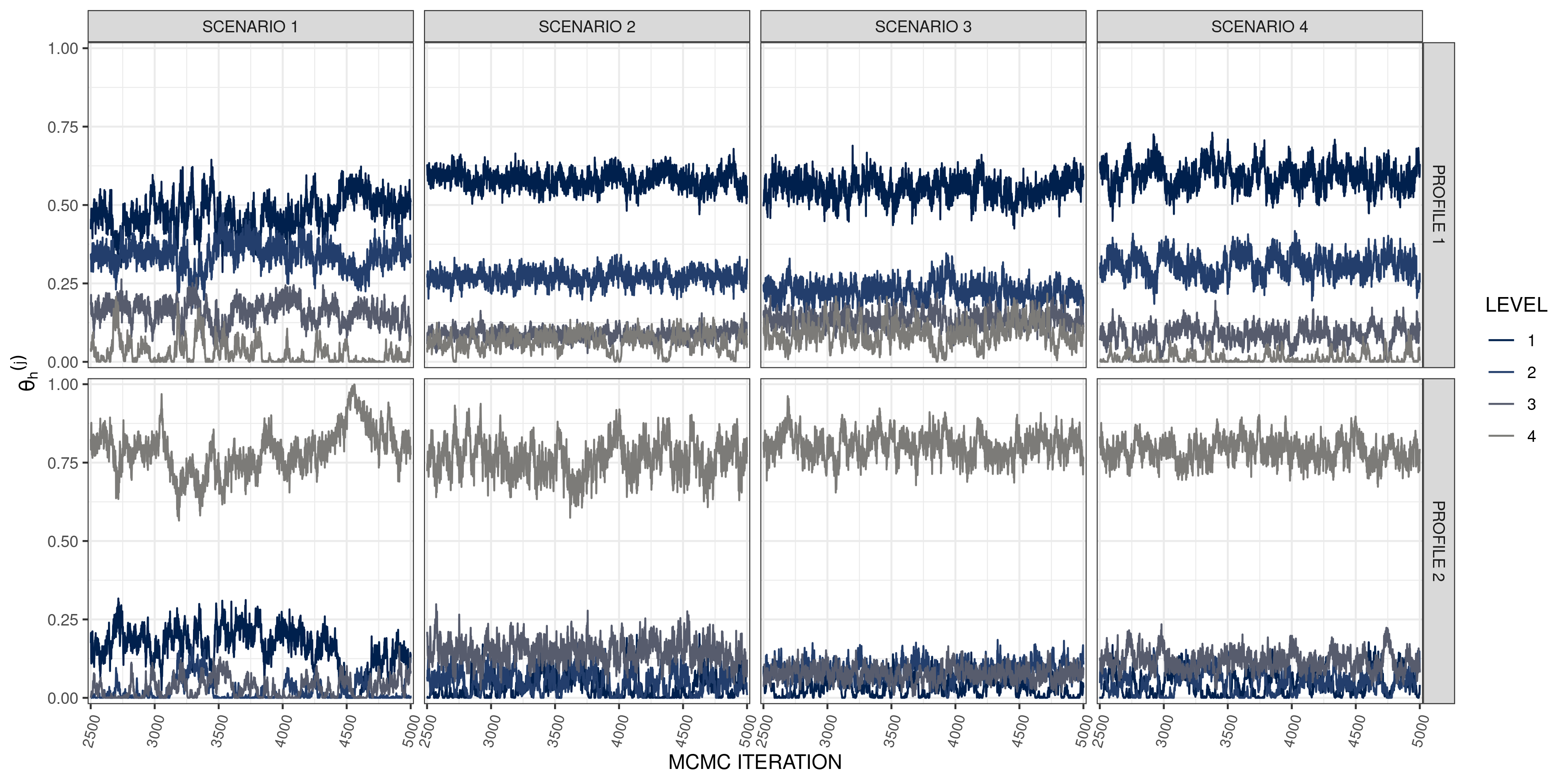}
    \caption{MMM kernel parameters trace plots for a representative variable, showing posterior sample for the  two profiles and all the simulation scenarios described in Section~{6}.
The chains shows no jumps between profiles, which would indicate label switching. Trace plots for the other variables behave similarly, and can be reproduced using the code in \texttt{plot\_and\_tables} at \url{https://github.com/rMassimiliano/MMM-tutorial}.
      }
     \label{fig:traceplot}
    \end{figure}

\begin{figure}[h!]
    \centering
    \includegraphics[width = \textwidth]{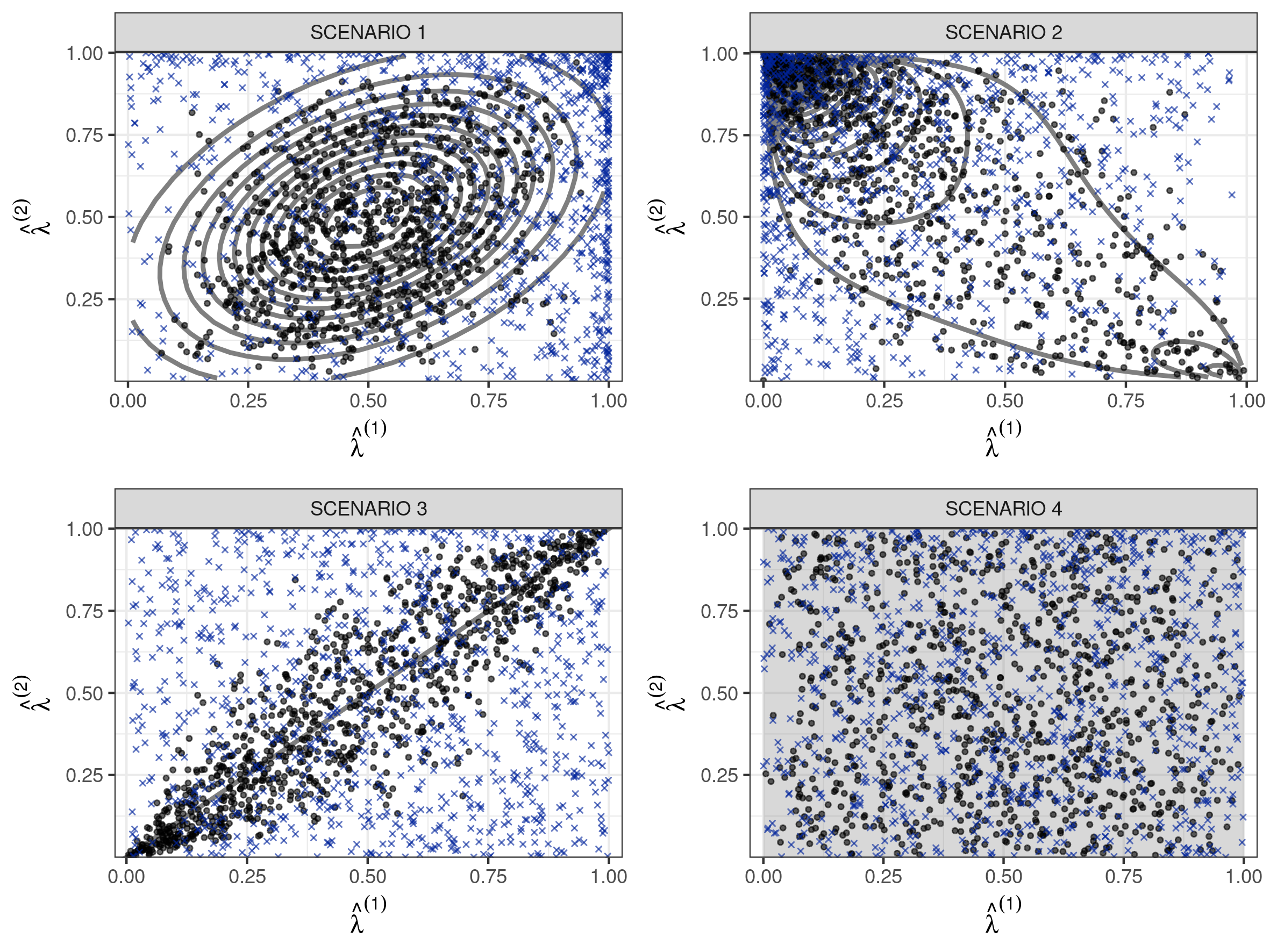}
    \caption{1000 samples from estimated membership scores distribution from model~{3.1} (black dots) and separate MM models estimated with \texttt{NIMBLE} (blue crosses). Grey area represents the contour of the true profiles distribution.}
    \label{fig:mmm1_nimble}
\end{figure}

\begin{table}[h!]
    \centering
\def\spacingset#1{\renewcommand{\baselinestretch}%
{#1}\small\normalsize} \spacingset{1.0}
 \caption{Mean (and standard deviation) of the L1 distance of the individual membership scores $(\lambda^{(1)}_i,\lambda^{(2)}_i)$ and their `true' values in all simulation scenarios.}
    \label{tab:L1_simMCMC}
    \vspace{0.2cm}
\resizebox{\textwidth}{!}{
\begin{tabular}{lllll}
\toprule
                  & SCENARIO 1  & SCENARIO 2  & SCENARIO 3  & SCENARIO 4 \\
\midrule
MMM g = 1         &0.132(0.096) &0.126(0.097) &0.122(0.090) &0.162(0.106) \\
MMM g = 2         &0.130(0.094) &0.134(0.103) &0.117(0.095) &0.138(0.105) \\
MM-MCMC g = 1     &0.233(0.150) &0.131(0.106) &0.156(0.111) &0.157(0.110) \\
MM-MCMC g = 2     &0.153(0.111) &0.147(0.118) &0.147(0.117) &0.139(0.109) \\
\bottomrule
\end{tabular}
}
\end{table}

\begin{figure}[h!]
    \centering

    \includegraphics[width= \textwidth]{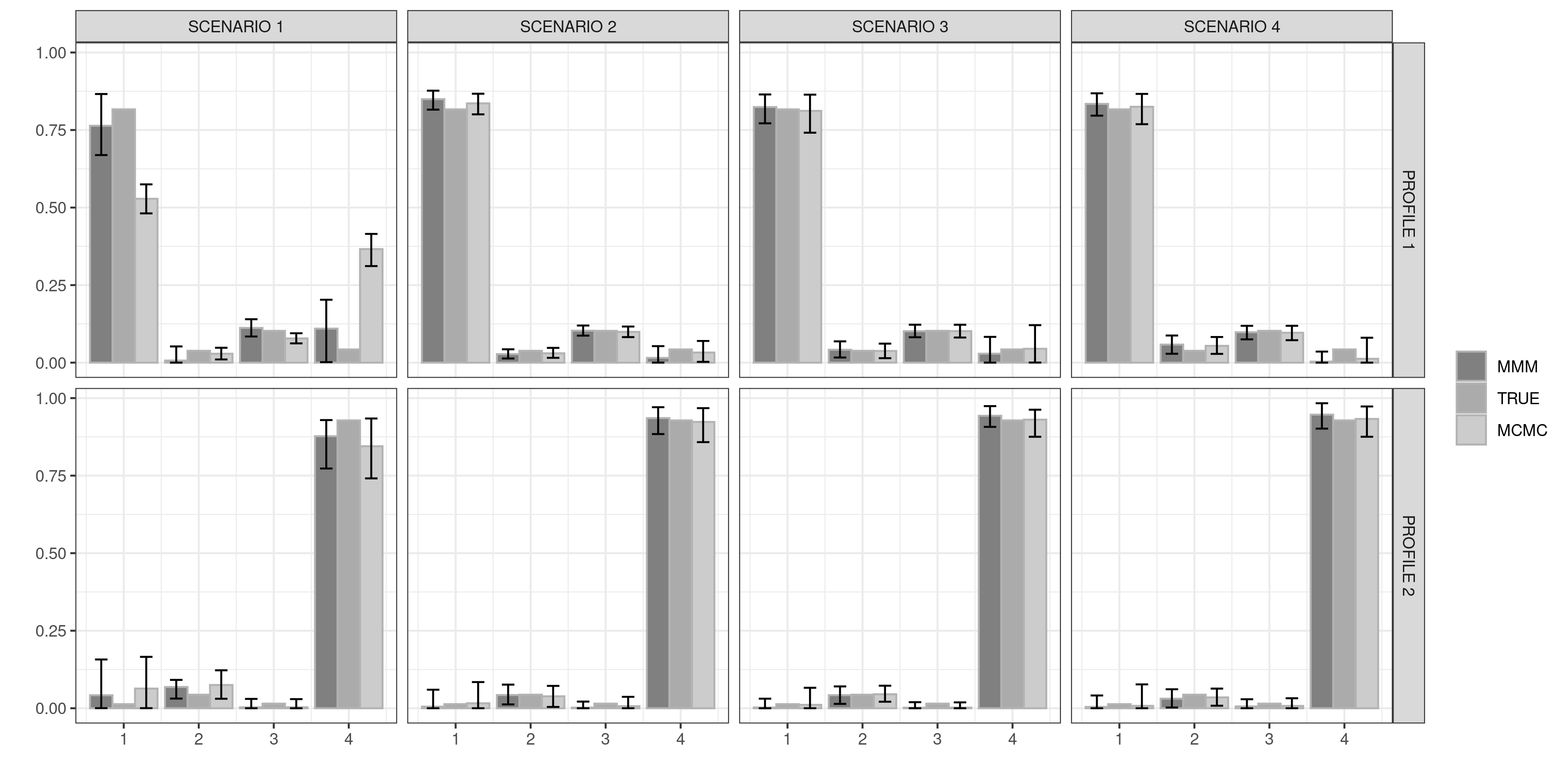}
    \caption{True values of the estimated profiles $\boldsymbol{\theta}^{(j)}_h$ for $h=1,2$ of a representative variable in group $g_j = 1$. Bars represent 0.1 and 0.9  posterior quantiles for our MMM model and for separate MM models estimated using \texttt{NIMBLE}.}
     \label{fig:sim_kern1}
\end{figure}

\begin{figure}[h!]
    \centering
    \includegraphics[width = \textwidth]{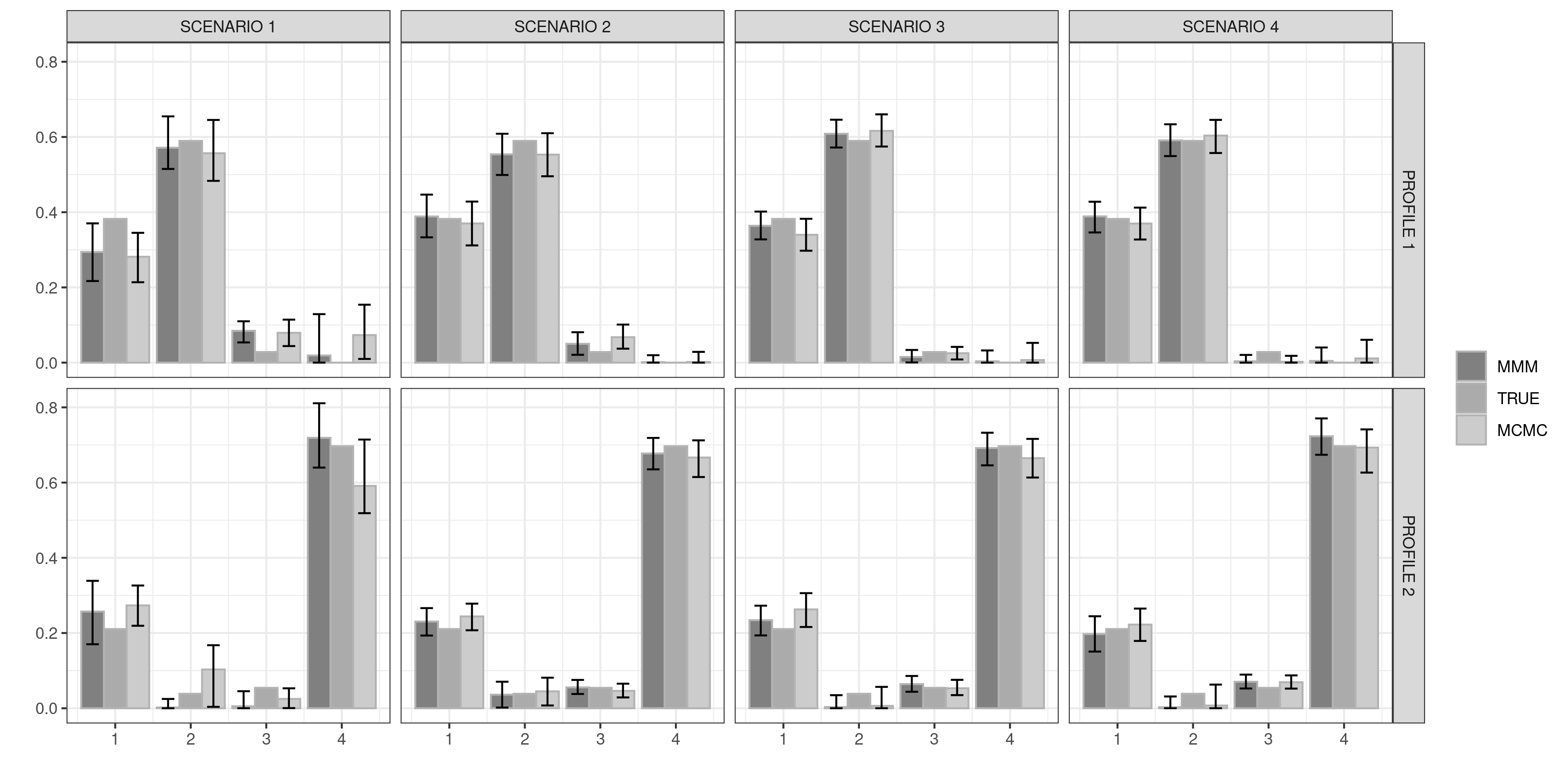}
    \caption{True values of the estimated profiles $\boldsymbol{\theta}^{(j)}_h$ for $h=1,2$ of a representative variable in group $g_j = 2$. Bars represent 0.1 and 0.9  posterior quantiles for our MMM model and for separate MM models estimated using \texttt{NIMBLE}. }
     \label{fig:sim_kern2}
\end{figure}

\begin{figure}[h!]
    \centering
    \includegraphics[width =  \textwidth]{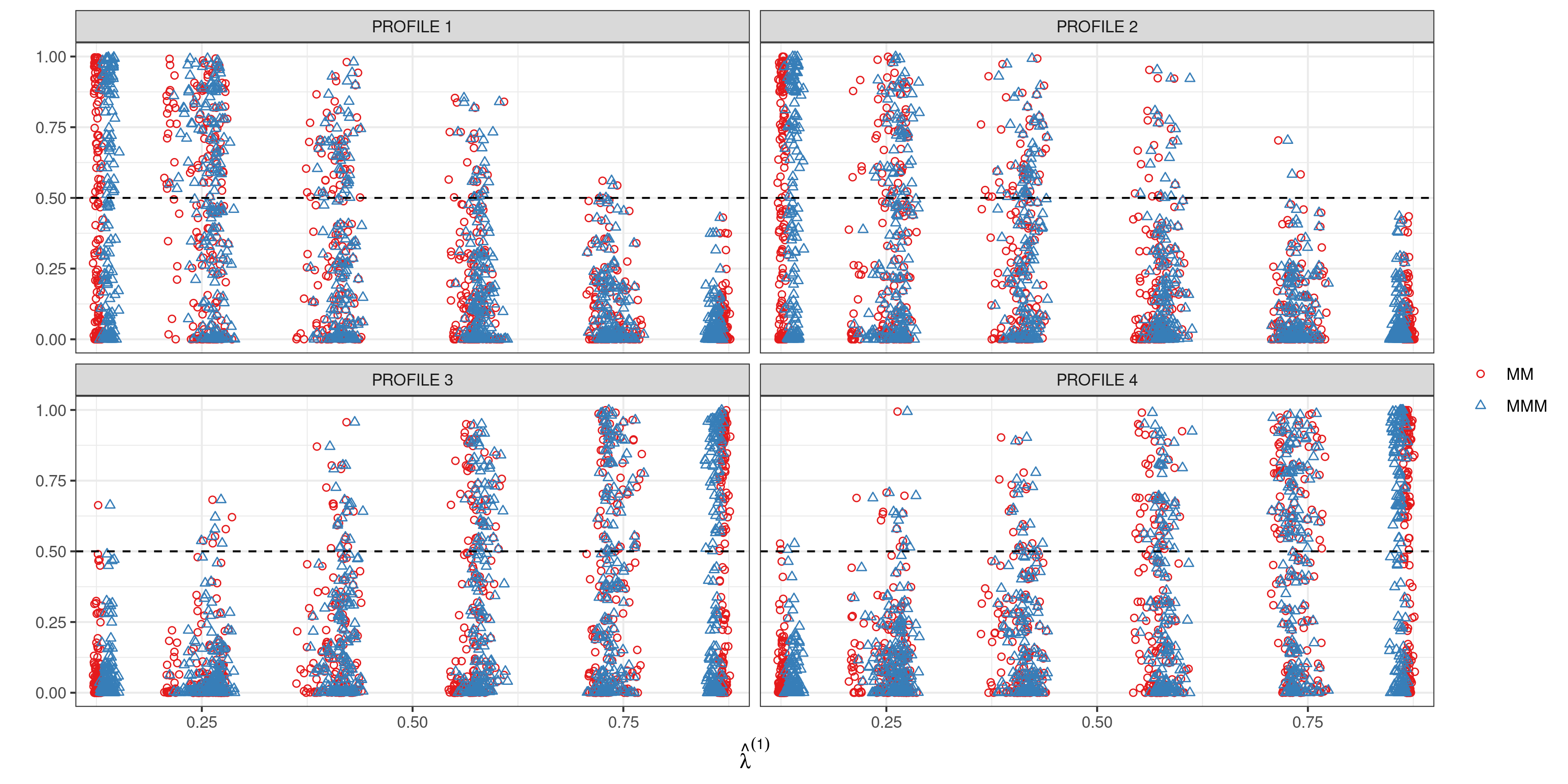}
    \caption{Estimated (x-axis) and `true' (y-axis) values for the score vectors, relying on MMM and MM model (estimated using \texttt{NIMBLE}) for $g=1$. In this case, for each subject $i$, there are four `true' profiles $\lambda^{(1)}_{i1},\ldots,\lambda^{(1)}_{i4},$ represented in the panels y-axis, but just two estimated profiles $(\widehat \lambda^{(1)}_{i1},1 -\widehat \lambda^{(1)}_{i1})$ for the misspecified MM and MMM models.
}
\label{fig:multi_prof_nimble}
\end{figure}

\begin{table}[h!]
\def\spacingset#1{\renewcommand{\baselinestretch}%
{#1}\small\normalsize} \spacingset{1.0}
    \centering
    \caption{Estimated kernels for MMM and MM model (estimated using \texttt{NIMBLE}) for variable $1$ in group $g_1 = 1$. Numbers in parenthesis are the 0.1 and 0.9 quantiles.}
    \label{tab:sim_kern}
    \resizebox{\textwidth}{!}{
    \begin{tabular}{lllll}
\toprule
                                                                          & PROFILE 1                  & PROFILE 2                  & PROFILE 3                                & PROFILE 4 \\
\midrule

        $1/2(\boldsymbol{\varphi}^{(1)}_1 +  \boldsymbol{\varphi}^{(1)}_2)$ & 0.450              & 0.450              & 0.050 & 0.050 \\
        MMM $\boldsymbol{\theta}^{(1)}_{1}$                               & 0.487(0.453;0.521) & 0.491(0.458;0.523) & 0.008(0.000;0.024)  & 0.013(0.000;0.037) \\
	MM-MCMC & 0.486 (0.448;0.522)&  0.493 (0.459;0.527)& 0.008 (0.000;0.024) & 0.013 (0.000;0.036)\\
	\midrule
        $1/2(\boldsymbol{\varphi}^{(1)}_3 +  \boldsymbol{\varphi}^{(1)}_4)$ & 0.050              & 0.050              & 0.450 & 0.450 \\
        MMM $\boldsymbol{\theta}^{(1)}_{2}$                               & 0.026(0.000;0.059) & 0.011(0.000;0.032) & 0.449(0.412;0.484)  & 0.515(0.478;0.550)\\
	MM-MCMC & 0.030 (0.001;0.062) & 0.009 (0.000;0.028) & 0.449 (0.415;0.484) & 0.512 (0.474;0.548)\\
        \bottomrule
    \end{tabular}
}
\end{table}

\clearpage

\section{Mixed Membership model for malaria data}
As a competitor to the MMM model described in Section 7.2, we consider an 
MM model with $H=4$ pure types. This model does not directly incorporate domain knowledge on environmental and behavioral variables, but can potentially represent the same risk structure:
high (low) risk for just one domain and low (high) risk for the other, or high (low) risk for both.  We incorporate space and time information via the following model:
\begin{eqnarray*}
    X_{ij} \mid Z_{ij} = h, \boldsymbol{\theta}_h^{(j)} &\sim& \mbox{Cat}(\theta^{(j)}_{h1},\ldots,\theta^{(j)}_{hd_j}), \nonumber\\
    Z_{ij} \mid \boldsymbol{\lambda}_i &\sim& \mbox{Cat}(\lambda_{i1},\ldots,\lambda_{iH}),
     \\
  \boldsymbol{\lambda}_i &\sim&
	\mbox{LogitNormal}(\boldsymbol{\beta}_{t_i} + \boldsymbol{\zeta}_{t_i}(\mathbf{s}_i), \boldsymbol{\Sigma}_{t_i})
                   , \nonumber
\end{eqnarray*}
where $t_i \in \{1985, 1986,1987,1995\}$ and $\boldsymbol{s}_i = (s_{i1},s_{i2})^T$ are, respectively, a time indicator and the observed longitude and latitude corresponding to the household of subject $i$.

We account for time dependence through a multivariate Gaussian hierarchical model with common hyperprior. Specifically, $\boldsymbol{\beta}_t  \sim \mathcal{N}_{H-1}(\boldsymbol{\beta},\boldsymbol{\Sigma})$, $\boldsymbol{\beta} \sim \mathcal{N}_{H-1}(\boldsymbol{\beta_0},\boldsymbol{\Sigma_0})$ and $\boldsymbol{\Sigma} \sim \mathcal{IW}(\nu_\beta,\boldsymbol{\Psi}_\beta)$. Note that $\boldsymbol{\beta}_t, \boldsymbol{\beta}, \boldsymbol{\beta}_0 \in \mathbb R^{H-1}$, while $\boldsymbol{\Sigma}, \boldsymbol{\Psi}_\beta \in \mathbb R^{(H-1)\times (H-1)}$. For the spatial effect $\boldsymbol{\zeta}_t(\boldsymbol{s}_i) \in \mathbb R^{(H-1)}$ we specify independent Gaussian processes for each profile with the covariance structure described in Section~{7.2}. This model uses more parameters than our proposed MMM specification, which leverages domain-knowledge to parsimoniously characterize space and time variability, using domain-specific parameters  in  place of  profile-specific ones.

We consider here the relation between malaria rates and risk profiles, similarly to what was presented in Section 7.5 for the MMM model.  From table~\ref{tab:marginalRiskMM}, we can notice that all profiles give similar results in terms of malaria rates, and none of them present an increasing (or decreasing) pattern.  
\begin{table}[h!]
	\caption{Median 0.1 and 0.9 posterior quantiles of the malaria rates for
        1st, 2nd and 3rd tertiles} \label{tab:marginalRiskMM}
	\begin{tabular}{cccc}
		\toprule
		{\bf 1985}      & 1st tertile          & 2nd tertile          & 3rd tertile \\ 
		\midrule
		Score 1         & 0.000 (0.000; 0.685) & 0.100 (0.000; 0.628) & 0.046 (0.000; 0.333) \\
                Score 2         & 0.111 (0.000; 0.430) & 0.069 (0.000; 0.571) & 0.271 (0.000; 0.867) \\
                Score 3         & 0.100 (0.000; 0.338) & 0.079 (0.000; 0.642) & 0.111 (0.000; 0.333) \\
		Score 4         & 0.167 (0.000; 0.862) & 0.074 (0.000; 0.500) & 0.085 (0.000; 0.543) \\
		\bottomrule 
		\toprule
		{\bf 1986}      & 1st tertile           & 2nd tertile           & 3rd tertile \\ 
		\midrule
		Score 1         & 0.204 (0.000; 0.667)  & 0.257 (0.000; 0.750)  & 0.198 (0.000; 0.763) \\
                Score 2         & 0.292 (0.000; 0.762)  & 0.250 (0.000; 0.750)  & 0.212 (0.005; 0.768) \\ 
		Score 3         & 0.250 (0.000; 0.691)  & 0.250 (0.000; 0.759)  & 0.167 (0.000; 0.605) \\
		Score 4         & 0.216 (0.026; 0.713)  & 0.250 (0.000; 0.750)  & 0.231 (0.000; 0.703) \\
		\bottomrule
		\toprule
		{\bf 1987}      & 1st tertile           & 2nd tertile           & 3rd tertile\\ 
		\midrule
                Score 1         & 0.167 (0.029; 0.592)  & 0.182 (0.000; 0.596)  & 0.200 (0.000; 0.606) \\
		Score 2         & 0.171 (0.000; 0.598)  & 0.194 (0.000; 0.600)  & 0.143 (0.000; 0.323) \\
		Score 3         & 0.175 (0.012; 0.550)  & 0.183 (0.000; 0.600)  & 0.181 (0.019; 0.624) \\
		Score 4         & 0.167 (0.000; 0.594)  & 0.200 (0.000; 0.596)  & 0.158 (0.004; 0.671) \\
		\bottomrule
		\toprule
		{\bf 1995}      & 1st tertile           & 2nd tertile           & 3rd tertile\\ 
		\midrule
                Score 1         & 0.042 (0.000; 0.206)  & 0.028 (0.000; 0.169)  & 0.031 (0.000; 0.163) \\
		Score 2         & 0.028 (0.000; 0.167)  & 0.030 (0.000; 0.180)  & 0.032 (0.000; 0.180) \\
                Score 3         & 0.028 (0.000; 0.226)  & 0.030 (0.000; 0.167)  & 0.030 (0.000; 0.209) \\
                Score 4         & 0.028 (0.000; 0.180)  & 0.030 (0.000; 0.167)  & 0.033 (0.000; 0.239) \\
		\bottomrule
	\end{tabular}
\end{table}
We also consider a summary of the malaria risk, grouping subjects into `low/high' risk for each of the four profiles using the median of the scores for each profile.  In this way, we divide households into 16 groups. Similarly to Table 5, we would expect 
malaria rates to be non-increasing (or decreasing) reading the tables from top-to-bottom. 
This does not seem to occur in Table~\ref{tab:posetMalariaRisk} at this simple level of analysis.
\begin{table}
\caption{Median 0.1 and 0.9 posterior quantiles of the malaria rates for the profiles. Groups are specified by using medians of $\lambda^{(\mbox{\tiny B})}_{ih}$ scores for $h=1,\dots,4$, where 1 indicates above and 0 below the median value.  Dashes indicate that there are not enough households/plots in the cluster to compute the median and quantiles.} \label{tab:posetMalariaRisk}
\resizebox{\textwidth}{!}{
\begin{tabular}{ccccccc}
& & & {\bf 1985}& & & \\
\toprule
& & & \shortstack{0000 \\ 0.071 (0.014; 0.254) }& & & \\
&\shortstack{1000 \\ 0.046 (0.000; 0.410)} &\shortstack{0100 \\ 0.134 (0.000; 0.535)} & &\shortstack{0010 \\ 0.000 (0.000; 0.479)} & \shortstack{0001 \\ 0.040 (0.000; 0.808)}& \\
\shortstack{1100 \\ 0.133 (0.000; 0.463)} &  \shortstack{1010 \\ 0.141 (0.000; 0.654)} &\shortstack{0110 \\ 0.081 (0.000; 0.724)} & & \shortstack{1001 \\ 0.000 ( 0.000; 0.331)}&
\shortstack{0101 \\ 0.111 (0.000; 0.622)}& \shortstack{0011 \\ 0.159 (000; 0751)}\\
&\shortstack{1110 \\ 0.100 (0.100; 0.100)} &\shortstack{1101 \\ ---}& & \shortstack{1011 \\ 0.250 (0.050; 0.450)} & \shortstack{0111\\ 0.000 (0.000, 0.000) }&\\
&&&\shortstack{1111 \\ ---}& && \\
\bottomrule
\end{tabular}}
\resizebox{\textwidth}{!}{
\begin{tabular}{ccccccc}
& & & {\bf 1986}& & & \\
\toprule
&&&\shortstack{0000 \\ ---} &&&\\
&\shortstack{1000  \\ 0.194 (0.000; 0.714)} 
& \shortstack{0100 \\ 0.294 (0.091; 0.753)}
&
&\shortstack{0010 \\ 0.227 (0.073; 0.950)}
&\shortstack{0001 \\ 0.154 (0.000; 0.688)} 
& \\
\shortstack{1100 \\ 0.182 (0.024; 0.807)} 
&\shortstack{1010 \\ 0.242 (0.000; 0.745)} 
&\shortstack{0110 \\ 0.218 (0.000; 0.903)} 
&
&\shortstack{1001 \\ 0.283 (0.000; 0.770)} 
&\shortstack{0101 \\ 0.315 (0.036; 0.702)} 
&\shortstack{0011 \\ 0.277 (0.034; 0.742)} \\
&\shortstack{1110 \\ 0.451 (0.095; 0.820)} 
&\shortstack{1101 \\ 0.173 (0.000; 0.609)} 
&
&\shortstack{1011 \\ 0.256 (0.047; 0.506)} 
&\shortstack{0111 \\ 0.204 (0.000; 0.401)} 
&\\
&&& 
\shortstack{1111 \\ ---} 
&&&\\

\bottomrule
\end{tabular}
}
\resizebox{\textwidth}{!}{
\begin{tabular}{ccccccc}
& & & {\bf 1987}& & & \\
\toprule
&&& \shortstack{0000 \\ ---}  &&& \\
&\shortstack{1000 \\ 0.194 (0.000; 0.714)} 
&\shortstack{0100 \\ 0.294 (0.091; 0.753)} 
&
&\shortstack{0010 \\ 0.227 (0.073; 0.950)} 
&\shortstack{0001 \\ 0.154 (0.000; 0.688)} 
&\\
\shortstack{1100 \\ 0.182 (0.024; 0.807)} 
&\shortstack{1010 \\ 0.242 (0.000; 0.745)} 
&\shortstack{0110 \\ 0.218 (0.000; 0.903)} 
&
&\shortstack{1001 \\ 0.283 (0.000; 0.770)} 
&\shortstack{0101 \\ 0.315 (0.036; 0.702)} 
&\shortstack{0011 \\ 0.277 (0.034; 0.742)} 
\\
&\shortstack{1110 \\ 0.451 (0.095; 0.820)} 
&\shortstack{1101 \\ 0.173 (0.000; 0.609)} 
&
&\shortstack{1011 \\ 0.256 (0.047; 0.506)} 
&\shortstack{0111 \\ 0.204 (0.000; 0.401)} 
&\\
&&& \shortstack{1111 \\ ---} &&&\\
\bottomrule
\end{tabular}
}
\resizebox{\textwidth}{!}{
\begin{tabular}{ccccccc}
& & & {\bf 1995}& & & \\
\toprule
&&& \shortstack{0000 \\ ---}  &&& \\
&\shortstack{1000 \\ 0.194 (0.000; 0.714)} 
&\shortstack{0100 \\ 0.294 (0.091; 0.753)} 
&
&\shortstack{0010 \\ 0.227 (0.073; 0.950)} 
&\shortstack{0001 \\ 0.154 (0.000; 0.688)} 
& \\
\shortstack{1100 \\ 0.182 (0.024; 0.807)} 
&\shortstack{1010 \\ 0.242 (0.000; 0.745)} 
&\shortstack{0110 \\ 0.218 (0.000; 0.903)} 
&
&\shortstack{1001 \\ 0.283 (0.000; 0.770)} 
&\shortstack{0101 \\ 0.315 (0.036; 0.702)} 
&\shortstack{0011 \\ 0.277 (0.034; 0.742)} 
\\
&\shortstack{1110 \\ 0.451 (0.095; 0.820)} 
&\shortstack{1101 \\ 0.173 (0.000; 0.609)} 
&
&\shortstack{1011 \\ 0.256 (0.047; 0.506)} 
&\shortstack{0111 \\ 0.204 (0.000; 0.401)} 
&\\
&&& \shortstack{1111 \\ ---} &&&\\
\bottomrule

\end{tabular}
}
\end{table}

\begin{figure}[h!]
	\centering
	\includegraphics[width = \textwidth]{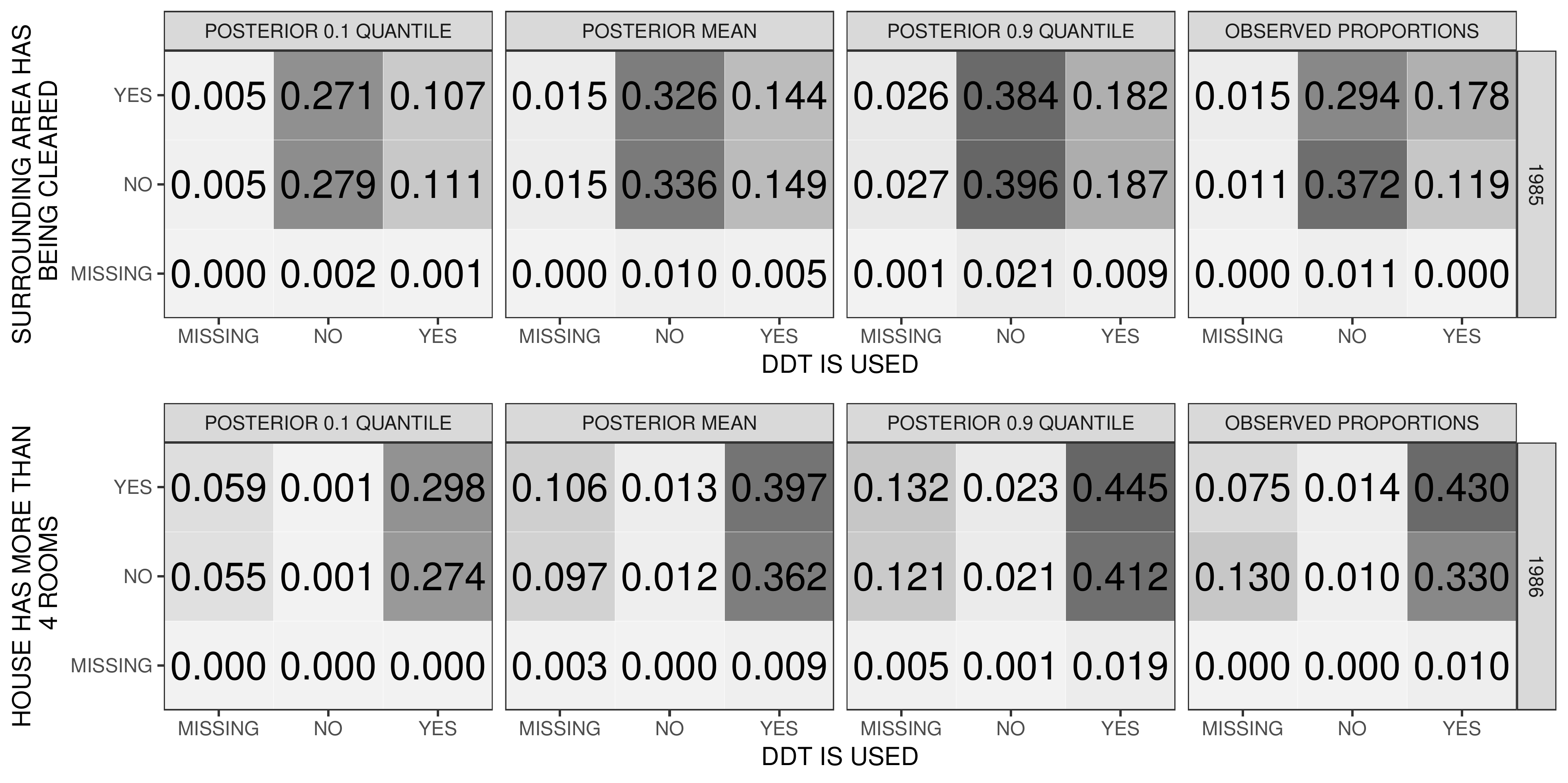}
        \caption{Posterior mean and quantiles for 2 bivariate distributions, compared with the empirical frequencies in the data, for the MM model with $H=4$ profiles described in Section 2.} 
\end{figure}

\clearpage

\section{Time and space domain evolution}
\begin{figure}[h!]
    \centering
    \includegraphics[width = \textwidth]{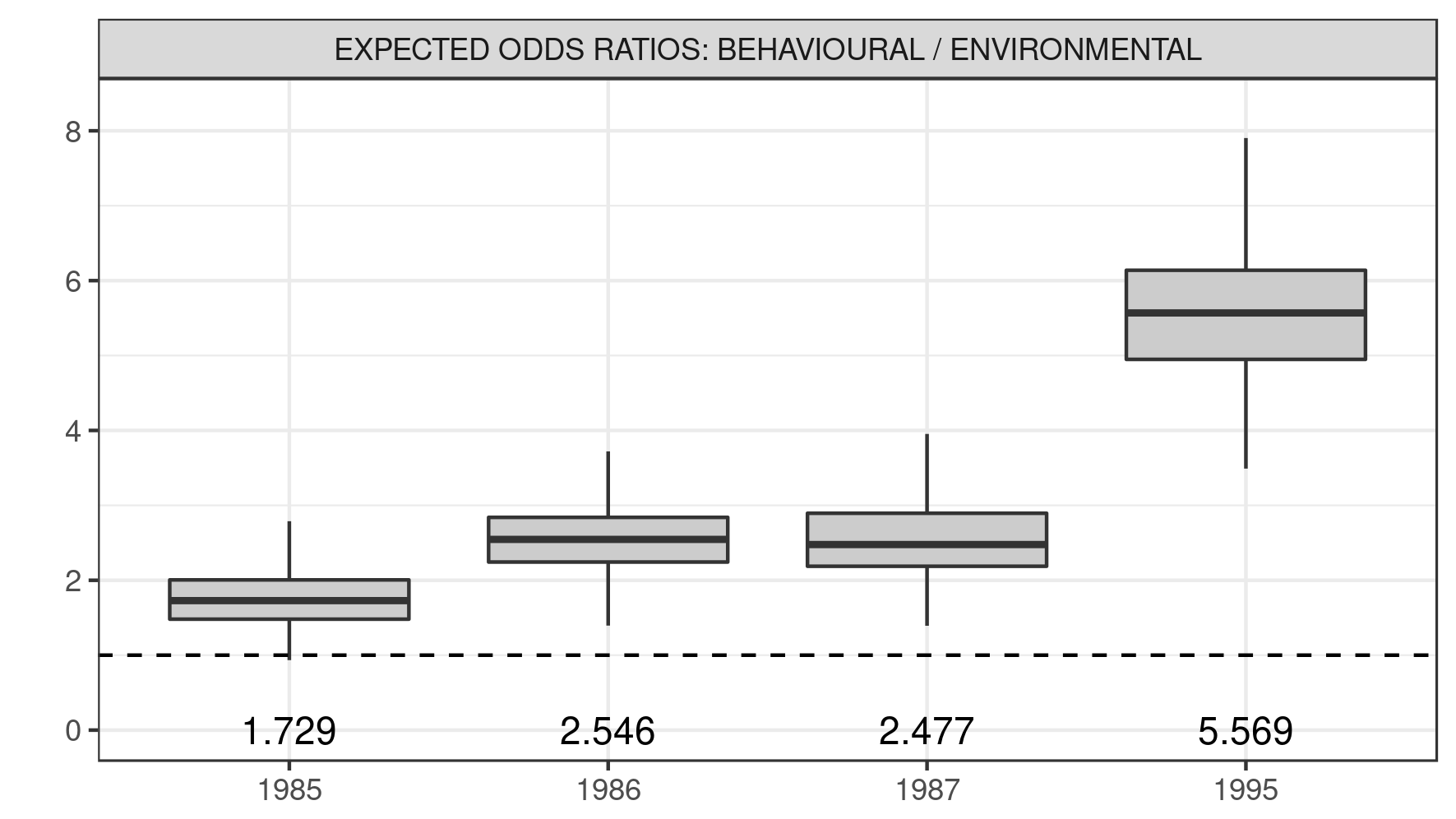}
    \caption{Odds ratios between environmental and behavioral risk scores. Numbers at the bottom of the plots are the value of the median.}
    \label{fig:odds_ratio}
\end{figure}

Figure~\ref{fig:odds_ratio} shows boxplots of the distribution of the expected odds ratios of being high risk in behavioral and environmental domains. Such distributions can be computed relying on equation~(4.2). Specifically, for each iteration of the MCMC algorithm, we can compute the quantity $\exp\{ \beta^{(\mbox{\tiny B})}_t - \beta^{(\mbox{\tiny E} )}_t + 1/2 (\Sigma_{t11}+ \Sigma_{t22} -2\Sigma_{t12})\} $, where $(\Sigma_{t11}, \Sigma_{t22},\Sigma_{t12})$ are the elements of the covariance matrix $\boldsymbol{\Sigma_t}$ in equation~(7.1). 
We notice an increasing trend in the odds ratios across the years; specifically, in $1985$ environmental and behavioral risks coexist, while starting from $1986$ behavioral risk starts to gain more and more importance, as is evident from the fact that the posterior odds ratio is not significantly above one in $1985$ and then it gradually increases. These results are in accordance with current literature on malaria risk, in Amazon areas, reporting that the risk is initially driven by favorable environmental conditions for malaria vectors to proliferate~\citep[e.g.][]{decastro2006}. Soon after human settlement, there is a phase lasting for about 8 or 10 years, in which environmental risk is high but human behavior is starting to gradually become the predominant risk factor. In the last stage, called the endemic phase, the risk is far more related to behavioral causes.

\begin{figure}[h!]
    \centering
    \includegraphics[width = \textwidth]{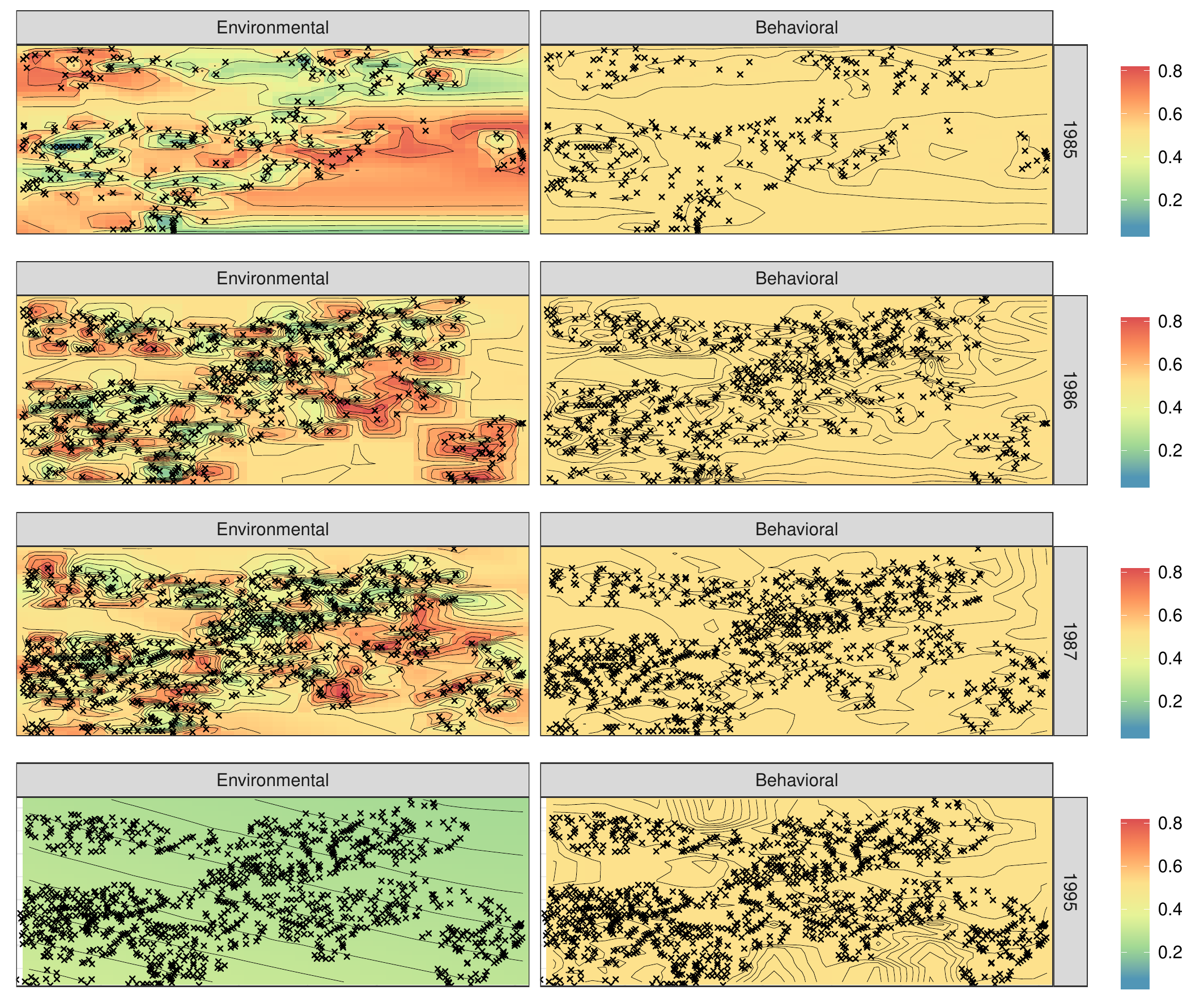}   
    \caption{Spatial risk predictions for the Machadinho area, for both behavioral and environmental domain. Values are expressed on the probability scale.}
    \label{fig:spatial_maps}
\end{figure}

From a spatial perspective, we can consider the posterior predictive distribution of  ${\boldsymbol{\zeta}_t}$ evaluated over a regular grid of values (Figure~\ref{fig:spatial_maps}). We notice that the behavioral risk distribution is constant across time and space; hence, from the considered survey data, it appears that the spatial variability is driven by environmental conditions. Environmental risk zones can be mostly explained in terms of geographical characteristics of the area; in fact higher risk zones correspond to the forest fringe and the Machadinho river path.

\section{Malaria risk conditions}
\Cref{tab:supp_1985,tab:supp_1986,tab:supp_1987,tab:supp_1995} includes posterior medians and 0.1 and 0.9 quantiles of the kernels for all behavioral and environmental variables considered in the analysis, computed using $2500$ posterior samples.  Light gray values in the table highlight admissible conditions as defined in (7.3a) or (7.3b).

\begin{table}[h!]
\def\spacingset#1{\renewcommand{\baselinestretch}%
{#1}\tiny} \spacingset{0.5}
\caption{1985: posterior median and 0.1 and 0.9 quantiles of the kernels for behavioral and environmental variables. Light gray values indicate admissible conditions according to (7.3a) or (7.3b).}
\label{tab:supp_1985}
\begin{minipage}[t]{0.5\textwidth}
\vspace{0cm}
\resizebox{\textwidth}{!}{
\begin{tabular}{lll}
\toprule
 {\bf Behavioral }&  $\boldsymbol{\theta}^{(j)}_1$ &  $\boldsymbol{\theta}^{(j)}_2$ \\
\midrule
Plant Cassava: NO & \cellcolor{gray!25}0.042(0.006;0.107) & 0.990(0.965;0.999)\\
Plant Cassava: YES & 0.933(0.869;0.972) & \cellcolor{gray!25}0.003(0.000;0.025)\\
Plant Cassava: MISSING & 0.021(0.004;0.045) & \cellcolor{gray!25}0.003(0.000;0.017)\\
Lavoura branca: NO & \cellcolor{gray!25}0.105(0.042;0.178) & 0.993(0.977;0.999)\\
Lavoura branca: YES & 0.888(0.814;0.950) & \cellcolor{gray!25}0.002(0.000;0.014)\\
\addlinespace
Lavoura branca: MISSING & 0.005(0.000;0.021) & 0.002(0.000;0.013)\\
DDT is used: NO & \cellcolor{gray!25}0.296(0.211;0.383) & 0.952(0.910;0.976)\\
DDT is used: YES & 0.687(0.598;0.774) & \cellcolor{gray!25}0.007(0.000;0.049)\\
DDT is used: MISSING & 0.013(0.002;0.040) & 0.035(0.016;0.060)\\
Plan to build a new house within a year: NO & 0.562(0.470;0.658) & \cellcolor{gray!25}0.006(0.000;0.040)\\
\addlinespace
Plan to build a new house within a year: YES & 0.343(0.246;0.439) & 0.976(0.936;0.997)\\
Plan to build a new house within a year: MISSING & 0.092(0.052;0.141) & \cellcolor{gray!25}0.011(0.000;0.038)\\
Arrived in Machadino before 1985: NO & 0.655(0.564;0.745) & \cellcolor{gray!25}0.034(0.003;0.093)\\
Arrived in Machadino before 1985: YES & \cellcolor{gray!25}0.345(0.255;0.436) & 0.966(0.907;0.997)\\
Own a planter: NO & 0.138(0.059;0.230) & 0.743(0.674;0.810)\\
\addlinespace
Own a planter: YES & 0.862(0.770;0.941) & \cellcolor{gray!25}0.257(0.190;0.326)\\
Do you own other proprieties: NO & 0.316(0.213;0.411) & 0.761(0.694;0.827)\\
Do you own other proprieties: YES & 0.684(0.589;0.787) & 0.239(0.173;0.306)\\
Lived in current house for more that 1m: NO & 0.580(0.502;0.653) & 0.986(0.963;0.997)\\
Lived in current house for more that 1m: YES & 0.412(0.339;0.488) & \cellcolor{gray!25}0.002(0.000;0.020)\\
\addlinespace
Lived in current house for more that 1m: MISSING & 0.003(0.000;0.024) & 0.009(0.001;0.024)\\
Knowledge of malaria vector: NO & 0.726(0.627;0.819) & 0.330(0.261;0.401)\\
Knowledge of malaria vector: YES & 0.231(0.147;0.330) & 0.495(0.425;0.569)\\
Knowledge of malaria vector: MISSING & 0.029(0.000;0.096) & 0.172(0.125;0.223)\\
Plant cocoa: NO & 0.636(0.559;0.702) & 0.998(0.987;1.000)\\
\addlinespace
Plant cocoa: YES & 0.364(0.298;0.441) & \cellcolor{gray!25}0.002(0.000;0.013)\\
Own more the 4 goods: NO & 0.153(0.072;0.242) & 0.514(0.445;0.589)\\
Own more the 4 goods: YES & 0.847(0.758;0.928) & 0.486(0.411;0.555)\\
Plant coffee: NO & 0.662(0.585;0.728) & 0.993(0.979;0.999)\\
Plant coffee: YES & 0.332(0.267;0.407) & \cellcolor{gray!25}0.001(0.000;0.009)\\
\addlinespace
Plant coffee: MISSING & 0.002(0.000;0.017) & 0.004(0.000;0.014)\\
Do you often go to surrounding cities: NO & 0.003(0.000;0.035) & 0.328(0.274;0.383)\\
Do you often go to surrounding cities: YES & 0.975(0.940;0.992) & 0.669(0.613;0.723)\\
Do you often go to surrounding cities: MISSING & 0.017(0.004;0.038) & \cellcolor{gray!25}0.001(0.000;0.009)\\
HH has high level of education: NO & 0.682(0.581;0.784) & 0.411(0.338;0.487)\\
\addlinespace
HH has high level of education: YES & 0.302(0.199;0.403) & 0.558(0.480;0.630)\\
HH has high level of education: MISSING & 0.010(0.000;0.042) & 0.029(0.010;0.053)\\
Own chickens and/or porks: NO & 0.681(0.590;0.765) & 0.929(0.876;0.981)\\
Own chickens and/or porks: YES & 0.311(0.227;0.401) & \cellcolor{gray!25}0.067(0.015;0.120)\\
Own chickens and/or porks: MISSING & 0.005(0.000;0.021) & 0.002(0.000;0.012)\\
\addlinespace
HH wife has high level of education: $<$ 4 yr & 0.551(0.451;0.655) & 0.452(0.380;0.524)\\
HH wife has high level of education: $>$ 4 yr & 0.279(0.190;0.375) & 0.476(0.407;0.549)\\
HH wife has high level of education: NO-WIFE & 0.013(0.000;0.062) & 0.036(0.006;0.065)\\
HH wife has high level of education: MISSING & 0.141(0.075;0.208) & \cellcolor{gray!25}0.028(0.000;0.078)\\
More than 4 people in the house: NO & 0.652(0.545;0.753) & 0.418(0.344;0.491)\\
\addlinespace
More than 4 people in the house: YES & 0.348(0.247;0.455) & 0.582(0.509;0.656)\\
Spray insecticide: NO & 0.647(0.553;0.738) & 0.841(0.779;0.897)\\
Spray insecticide: YES & 0.353(0.262;0.447) & 0.159(0.103;0.221)\\
Get malaria from dirty water: NO & 0.333(0.240;0.430) & 0.472(0.399;0.546)\\
Get malaria from dirty water: YES & 0.590(0.496;0.689) & 0.427(0.351;0.497)\\
\addlinespace
Get malaria from dirty water: MISSING & 0.070(0.031;0.124) & 0.100(0.064;0.143)\\
Use plant to cure malaria: NO & 0.700(0.604;0.794) & 0.634(0.564;0.705)\\
Use plant to cure malaria: YES & 0.088(0.013;0.171) & 0.235(0.175;0.302)\\
Use plant to cure malaria: MISSING & 0.206(0.138;0.287) & 0.127(0.081;0.180)\\
Own a chainsaw: NO & 0.650(0.558;0.736) & 0.770(0.705;0.828)\\
\addlinespace
Own a chainsaw: YES & 0.350(0.264;0.442) & 0.230(0.172;0.295)\\
Use a bednet: NO & 0.847(0.758;0.934) & 0.762(0.695;0.823)\\
Use a bednet: YES & 0.136(0.048;0.222) & 0.220(0.162;0.286)\\
Use a bednet: MISSING & 0.014(0.000;0.044) & 0.015(0.001;0.034)\\
Use repellent: NO & 0.984(0.940;0.999) & 0.862(0.820;0.900)\\
\addlinespace
Use repellent: YES & 0.016(0.001;0.060) & 0.138(0.100;0.180)\\
Part of family did not come: NO & 0.653(0.561;0.752) & 0.640(0.568;0.711)\\
Part of family did not come: YES & 0.317(0.222;0.409) & 0.351(0.280;0.422)\\
Part of family did not come: MISSING & 0.026(0.002;0.055) & \cellcolor{gray!25}0.004(0.000;0.027)\\
Do you ever go to urban area?: NO & 0.003(0.000;0.027) & 0.090(0.062;0.124)\\
\addlinespace
Do you ever go to urban area?: YES & 0.979(0.949;0.995) & 0.905(0.869;0.934)\\
Do you ever go to urban area?: MISSING & 0.013(0.001;0.034) & \cellcolor{gray!25}0.002(0.000;0.015)\\
Arrived in Rondonia before 1985: NO & 0.943(0.897;0.984) & 0.975(0.940;0.994)\\
Arrived in Rondonia before 1985: YES & 0.002(0.000;0.013) & 0.005(0.000;0.016)\\
Arrived in Rondonia before 1985: MISSING & 0.053(0.013;0.097) & 0.017(0.001;0.051)\\
\addlinespace
Before coming was your occupation rural: NO & \cellcolor{gray!25}0.001(0.000;0.008) & \cellcolor{gray!25}0.001(0.000;0.006)\\
Before coming was your occupation rural: YES & 0.994(0.975;1.000) & 0.982(0.963;0.994)\\
Before coming was your occupation rural: MISSING & 0.003(0.000;0.020) & 0.016(0.005;0.034)\\
Are there rubber tree: NO & 0.982(0.960;0.995) & 0.998(0.988;1.000)\\
Are there rubber tree: YES & 0.018(0.005;0.040) & \cellcolor{gray!25}0.002(0.000;0.012)\\
\bottomrule
\end{tabular}
}
\end{minipage}%
\begin{minipage}[t]{0.5\textwidth}
\vspace{0cm}
\resizebox{\textwidth}{!}{
\begin{tabular}{lll}
\toprule
 {\bf Environmental }&  $\boldsymbol{\theta}^{(j)}_1$ &  $\boldsymbol{\theta}^{(j)}_2$ \\
\midrule
Roof has good quality: NO & \cellcolor{gray!25}0.135(0.031;0.273) & 0.982(0.947;0.996)\\
Roof has good quality: YES & 0.856(0.720;0.960) & \cellcolor{gray!25}0.005(0.000;0.040)\\
Roof has good quality: MISSING & 0.003(0.000;0.022) & 0.009(0.001;0.023)\\
Walls have good quality: NO & \cellcolor{gray!25}0.207(0.080;0.315) & 0.983(0.952;0.996)\\
Walls have good quality: YES & 0.785(0.678;0.911) & \cellcolor{gray!25}0.004(0.000;0.034)\\
\addlinespace
Walls have good quality: MISSING & 0.003(0.000;0.022) & 0.009(0.001;0.023)\\
House has more than 4 rooms: NO & \cellcolor{gray!25}0.235(0.118;0.350) & 0.987(0.949;0.999)\\
House has more than 4 rooms: YES & 0.765(0.650;0.882) & \cellcolor{gray!25}0.013(0.001;0.051)\\
Has the surrounding area being cleared: NO & \cellcolor{gray!25}0.109(0.018;0.213) & 0.736(0.674;0.801)\\
Has the surrounding area being cleared: YES & 0.885(0.782;0.972) & \cellcolor{gray!25}0.246(0.180;0.308)\\
\addlinespace
Has the surrounding area being cleared: MISSING & 0.002(0.000;0.017) & 0.016(0.006;0.033)\\
Good water source available: NO & 0.422(0.316;0.528) & 0.886(0.812;0.961)\\
Good water source available: YES & 0.571(0.463;0.676) & \cellcolor{gray!25}0.103(0.026;0.177)\\
Good water source available: MISSING & 0.003(0.000;0.022) & 0.009(0.001;0.024)\\
Do you have close neighbours (<500mt): NO & 0.620(0.510;0.730) & \cellcolor{gray!25}0.203(0.136;0.273)\\
\addlinespace
Do you have close neighbours (<500mt): YES & 0.380(0.270;0.490) & 0.797(0.727;0.864)\\
More that 100mt from a forest: NO & 0.412(0.317;0.518) & 0.755(0.688;0.817)\\
More that 100mt from a forest: YES & 0.582(0.474;0.676) & 0.239(0.178;0.305)\\
More that 100mt from a forest: MISSING & 0.003(0.000;0.017) & 0.004(0.000;0.014)\\
Distant from stagnant water(no culvert): NO & 0.010(0.000;0.059) & 0.187(0.143;0.236)\\
\addlinespace
Distant from stagnant water(no culvert): YES & 0.982(0.934;0.999) & 0.775(0.724;0.825)\\
Distant from stagnant water(no culvert): MISSING & 0.002(0.000;0.017) & 0.035(0.018;0.057)\\
More than 600mt from a culvert: NO & 0.033(0.001;0.098) & 0.167(0.122;0.218)\\
More than 600mt from a culvert: YES & 0.959(0.892;0.995) & 0.791(0.736;0.841)\\
More than 600mt from a culvert: MISSING & 0.003(0.000;0.022) & 0.040(0.022;0.065)\\
\addlinespace
More than 600mt from a river: NO & 0.498(0.393;0.606) & 0.571(0.496;0.642)\\
More than 600mt from a river: YES & 0.492(0.384;0.595) & 0.420(0.349;0.496)\\
More than 600mt from a river: MISSING & 0.006(0.000;0.030) & 0.007(0.000;0.022)\\
Anybody cleared the area before HH: NO & 0.981(0.946;0.997) & 0.878(0.838;0.912)\\
Anybody cleared the area before HH: YES & 0.006(0.000;0.040) & 0.119(0.085;0.158)\\
\addlinespace
Anybody cleared the area before HH: MISSING & 0.007(0.000;0.025) & \cellcolor{gray!25}0.001(0.000;0.011)\\
More that 10km from an hospital: NO & 0.055(0.005;0.134) & 0.151(0.101;0.201)\\
More that 10km from an hospital: YES & 0.945(0.866;0.995) & 0.849(0.799;0.899)\\
Sealing has good quality: NO & 0.910(0.859;0.944) & 0.987(0.970;0.997)\\
Sealing has good quality: YES & 0.083(0.052;0.130) & \cellcolor{gray!25}0.001(0.000;0.011)\\
\addlinespace
Sealing has good quality: MISSING & 0.003(0.000;0.023) & 0.009(0.001;0.024)\\
Good bathing place is available: NO & 0.979(0.940;0.998) & 0.974(0.950;0.992)\\
Good bathing place is available: YES & 0.012(0.000;0.049) & 0.014(0.001;0.034)\\
Good bathing place is available: MISSING & 0.003(0.000;0.022) & 0.010(0.001;0.024)\\
\bottomrule
\end{tabular}}
\end{minipage}
\end{table}

\begin{table}[h!]
  \caption{1986: posterior median and 0.1 and 0.9 quantiles of the kernels for behavioral and environmental variables. Light gray values indicate admissible conditions according to (7.3a) or (7.3b)}
\label{tab:supp_1986}
\begin{minipage}[t]{0.5\textwidth}
\vspace{0pt}
\resizebox{\textwidth}{!}{
\begin{tabular}{lll}
\toprule
 {\bf Behavioral }&  $\boldsymbol{\theta}^{(j)}_1$ &  $\boldsymbol{\theta}^{(j)}_2$ \\
\midrule
Plant coffee: NO & 0.115(0.068;0.161) & \cellcolor{gray!25}0.957(0.915;0.977)\\
Plant coffee: YES & \cellcolor{gray!25}0.882(0.837;0.930) & 0.005(0.000;0.052)\\
Plant coffee: MISSING & 0.001(0.000;0.006) & \cellcolor{gray!25}0.031(0.017;0.050)\\
Cultivate rice: NO & 0.007(0.000;0.041) & \cellcolor{gray!25}0.743(0.661;0.835)\\
Cultivate rice: YES & \cellcolor{gray!25}0.988(0.956;0.999) & 0.213(0.115;0.294)\\
\addlinespace
Cultivate rice: MISSING & 0.001(0.000;0.009) & \cellcolor{gray!25}0.044(0.026;0.068)\\
Own a planter: NO & 0.060(0.008;0.111) & \cellcolor{gray!25}0.710(0.624;0.805)\\
Own a planter: YES & 0.940(0.889;0.992) & 0.290(0.195;0.376)\\
Own chickens and/or porks: NO & 0.001(0.000;0.012) & \cellcolor{gray!25}0.566(0.498;0.635)\\
Own chickens and/or porks: YES & 0.997(0.985;1.000) & 0.408(0.338;0.479)\\
\addlinespace
Own chickens and/or porks: MISSING & 0.001(0.000;0.006) & \cellcolor{gray!25}0.023(0.011;0.041)\\
DDT is used: NO & 0.024(0.005;0.042) & 0.022(0.001;0.061)\\
DDT is used: YES & 0.969(0.948;0.991) & 0.399(0.321;0.467)\\
DDT is used: MISSING & 0.003(0.000;0.020) & \cellcolor{gray!25}0.574(0.508;0.650)\\
More than 4 people in the house: NO & \cellcolor{gray!25}0.723(0.671;0.772) & 0.148(0.062;0.239)\\
\addlinespace
More than 4 people in the house: YES & 0.277(0.228;0.329) & \cellcolor{gray!25}0.852(0.761;0.938)\\
Own more the 4 goods: NO & 0.032(0.002;0.078) & \cellcolor{gray!25}0.601(0.522;0.693)\\
Own more the 4 goods: YES & 0.968(0.922;0.998) & 0.399(0.307;0.478)\\
Plant cocoa: NO & 0.424(0.379;0.468) & \cellcolor{gray!25}0.971(0.930;0.989)\\
Plant cocoa: YES & 0.573(0.529;0.619) & 0.006(0.000;0.052)\\
\addlinespace
Plant cocoa: MISSING & 0.001(0.000;0.007) & \cellcolor{gray!25}0.017(0.006;0.033)\\
Are there rubber tree: NO & 0.528(0.485;0.570) & \cellcolor{gray!25}0.980(0.960;0.993)\\
Are there rubber tree: YES & 0.469(0.428;0.512) & 0.001(0.000;0.015)\\
Are there rubber tree: MISSING & 0.001(0.000;0.008) & \cellcolor{gray!25}0.016(0.005;0.031)\\
Before coming was your occupation rural: NO & 0.820(0.770;0.871) & 0.364(0.283;0.449)\\
\addlinespace
Before coming was your occupation rural: YES & 0.176(0.127;0.227) & \cellcolor{gray!25}0.633(0.548;0.714)\\
Before coming was your occupation rural: MISSING & 0.002(0.000;0.007) & 0.001(0.000;0.008)\\
Do you own other proprieties: NO & 0.401(0.350;0.455) & \cellcolor{gray!25}0.851(0.773;0.926)\\
Do you own other proprieties: YES & 0.599(0.545;0.650) & 0.149(0.074;0.227)\\
Lived in current house for more that 1m: NO & 0.524(0.478;0.571) & \cellcolor{gray!25}0.931(0.863;0.964)\\
\addlinespace
Lived in current house for more that 1m: YES & 0.473(0.427;0.519) & 0.024(0.000;0.095)\\
Lived in current house for more that 1m: MISSING & 0.001(0.000;0.007) & \cellcolor{gray!25}0.041(0.025;0.063)\\
Plan to build a new house within a year: NO & 0.689(0.635;0.740) & 0.267(0.183;0.354)\\
Plan to build a new house within a year: YES & 0.268(0.219;0.320) & 0.627(0.540;0.716)\\
Plan to build a new house within a year: MISSING & 0.041(0.010;0.074) & 0.103(0.053;0.162)\\
\addlinespace
HH wife has high level of education: $<$ 4 yr & 0.525(0.474;0.575) & 0.235(0.150;0.316)\\
HH wife has high level of education: $<$ 4 yr & 0.454(0.405;0.506) & 0.348(0.264;0.429)\\
HH wife has high level of education: NO-WIFE & 0.010(0.000;0.022) & 0.008(0.000;0.034)\\
HH wife has high level of education: MISSING & 0.003(0.000;0.029) & \cellcolor{gray!25}0.406(0.343;0.471)\\
Working in the plot from more than 1 month: NO & 0.001(0.000;0.014) & \cellcolor{gray!25}0.392(0.335;0.453)\\
\addlinespace
Working in the plot from more than 1 month: YES & 0.966(0.945;0.983) & 0.568(0.504;0.630)\\
Working in the plot from more than 1 month: MISSING & 0.030(0.014;0.048) & 0.037(0.012;0.072)\\
Part of family did not come: NO & 0.750(0.698;0.799) & 0.348(0.265;0.432)\\
Part of family did not come: YES & 0.247(0.199;0.299) & 0.570(0.487;0.654)\\
Part of family did not come: MISSING & 0.001(0.000;0.007) & \cellcolor{gray!25}0.081(0.057;0.109)\\
\addlinespace
Own a chainsaw: NO & 0.540(0.488;0.594) & 0.888(0.813;0.963)\\
Own a chainsaw: YES & 0.460(0.406;0.512) & 0.112(0.037;0.187)\\
Spray insecticide: NO & 0.549(0.496;0.601) & 0.795(0.709;0.874)\\
Spray insecticide: YES & 0.444(0.391;0.496) & 0.192(0.109;0.279)\\
Spray insecticide: MISSING & 0.007(0.000;0.017) & 0.011(0.001;0.029)\\
\addlinespace
Knowledge of malaria vector: NO & 0.472(0.421;0.523) & 0.386(0.305;0.469)\\
Knowledge of malaria vector: YES & 0.298(0.251;0.346) & 0.406(0.329;0.488)\\
Knowledge of malaria vector: MISSING & 0.229(0.186;0.273) & 0.206(0.140;0.275)\\
Arrived in Machadino before 1985: NO & 0.228(0.191;0.270) & 0.085(0.032;0.147)\\
Arrived in Machadino before 1985: YES & 0.772(0.730;0.809) & 0.915(0.853;0.968)\\
\addlinespace
Get malaria from dirty water: NO & 0.417(0.369;0.467) & 0.332(0.252;0.409)\\
Get malaria from dirty water: YES & 0.535(0.486;0.583) & 0.639(0.562;0.721)\\
Get malaria from dirty water: MISSING & 0.047(0.027;0.068) & 0.025(0.001;0.058)\\
Arrived in Rondonia before 1985: NO & 0.860(0.833;0.889) & 0.884(0.838;0.918)\\
Arrived in Rondonia before 1985: YES & 0.001(0.000;0.006) & \cellcolor{gray!25}0.096(0.070;0.128)\\
\addlinespace
Arrived in Rondonia before 1985: MISSING & 0.137(0.110;0.165) & 0.014(0.000;0.057)\\
Use plant to cure malaria: NO & 0.814(0.770;0.858) & 0.918(0.847;0.989)\\
Use plant to cure malaria: YES & 0.177(0.134;0.219) & 0.076(0.006;0.147)\\
Use plant to cure malaria: MISSING & 0.009(0.002;0.017) & 0.002(0.000;0.015)\\
HH has high level of education: NO & 0.522(0.472;0.570) & 0.428(0.350;0.505)\\
\addlinespace
HH has high level of education: YES & 0.473(0.424;0.523) & 0.559(0.481;0.636)\\
HH has high level of education: MISSING & 0.003(0.000;0.014) & 0.011(0.000;0.028)\\
\bottomrule
\end{tabular}
}
\end{minipage}%
\begin{minipage}[t]{0.5\textwidth}
\vspace{0pt}
\resizebox{\textwidth}{!}{
\begin{tabular}{lll}
\toprule
 {\bf Environmental }&  $\boldsymbol{\theta}^{(j)}_1$ &  $\boldsymbol{\theta}^{(j)}_2$ \\
\midrule
House has more than 4 rooms: NO & 0.043(0.001;0.097) & \cellcolor{gray!25}0.991(0.970;0.999)\\
House has more than 4 rooms: YES & \cellcolor{gray!25}0.940(0.888;0.983) & 0.002(0.000;0.022)\\
House has more than 4 rooms: MISSING & 0.015(0.006;0.027) & 0.003(0.000;0.016)\\
Walls have good quality: NO & 0.003(0.000;0.015) & \cellcolor{gray!25}0.891(0.818;0.964)\\
Walls have good quality: YES & \cellcolor{gray!25}0.997(0.985;1.000) & 0.109(0.036;0.182)\\
\addlinespace
Roof has good quality: NO & 0.020(0.001;0.071) & \cellcolor{gray!25}0.763(0.696;0.829)\\
Roof has good quality: YES & \cellcolor{gray!25}0.980(0.929;0.999) & 0.237(0.171;0.304)\\
Good water source available: NO & 0.183(0.126;0.242) & \cellcolor{gray!25}0.864(0.778;0.946)\\
Good water source available: YES & \cellcolor{gray!25}0.815(0.755;0.872) & 0.132(0.050;0.218)\\
Good water source available: MISSING & 0.001(0.000;0.006) & 0.002(0.000;0.009)\\
\addlinespace
Anybody cleared the area before HH: NO & 0.982(0.945;0.997) & 0.498(0.441;0.558)\\
Anybody cleared the area before HH: YES & 0.009(0.000;0.048) & \cellcolor{gray!25}0.484(0.426;0.541)\\
Anybody cleared the area before HH: MISSING & 0.005(0.000;0.017) & 0.016(0.003;0.033)\\
Do you have close neighbours (<500mt): NO & 0.630(0.567;0.692) & 0.301(0.220;0.384)\\
Do you have close neighbours (<500mt): YES & 0.370(0.308;0.433) & \cellcolor{gray!25}0.699(0.616;0.780)\\
\addlinespace
Far from permanent water: NO & 0.439(0.376;0.496) & 0.735(0.663;0.812)\\
Far from permanent water: YES & 0.544(0.488;0.607) & 0.261(0.184;0.333)\\
Far from permanent water: MISSING & 0.016(0.007;0.028) & 0.001(0.000;0.013)\\
More that 100mt from a forest: NO & 0.691(0.650;0.732) & 0.977(0.932;0.997)\\
More that 100mt from a forest: YES & 0.307(0.266;0.347) & 0.020(0.000;0.064)\\
\addlinespace
More that 100mt from a forest: MISSING & 0.001(0.000;0.007) & 0.002(0.000;0.009)\\
Is topography bottom: NO & 0.659(0.596;0.717) & 0.406(0.323;0.482)\\
Is topography bottom: YES & 0.335(0.277;0.399) & 0.566(0.490;0.647)\\
Is topography bottom: MISSING & 0.003(0.000;0.014) & \cellcolor{gray!25}0.027(0.013;0.044)\\
Has the surrounding area being cleared: NO & 0.722(0.679;0.767) & 0.956(0.905;0.981)\\
\addlinespace
Has the surrounding area being cleared: YES & 0.276(0.231;0.319) & 0.019(0.000;0.071)\\
Has the surrounding area being cleared: MISSING & 0.001(0.000;0.008) & \cellcolor{gray!25}0.022(0.011;0.037)\\
Near big planted area: NO & 0.832(0.799;0.864) & 0.957(0.927;0.978)\\
Near big planted area: YES & 0.161(0.132;0.193) & 0.006(0.000;0.033)\\
Near big planted area: MISSING & 0.003(0.000;0.019) & \cellcolor{gray!25}0.033(0.015;0.053)\\
\addlinespace
Sealing has good quality: NO & 0.879(0.854;0.903) & 0.995(0.980;1.000)\\
Sealing has good quality: YES & \cellcolor{gray!25}0.114(0.092;0.139) & 0.002(0.000;0.015)\\
Sealing has good quality: MISSING & 0.005(0.000;0.012) & 0.001(0.000;0.009)\\
More that 10km from an hospital: NO & 0.149(0.110;0.189) & 0.045(0.008;0.093)\\
More that 10km from an hospital: YES & 0.851(0.811;0.890) & 0.955(0.907;0.992)\\
\addlinespace
Good bathing place is available: NO & 0.871(0.838;0.903) & 0.967(0.929;0.992)\\
Good bathing place is available: YES & 0.127(0.095;0.160) & 0.024(0.001;0.061)\\
Good bathing place is available: MISSING & 0.001(0.000;0.005) & \cellcolor{gray!25}0.007(0.002;0.017)\\
Far from temporary water: NO & 0.903(0.868;0.933) & 0.943(0.903;0.977)\\
Far from temporary water: YES & 0.087(0.059;0.118) & 0.031(0.004;0.069)\\
\addlinespace
Far from temporary water: MISSING & 0.009(0.000;0.027) & 0.023(0.003;0.044)\\
Near big pasture area: NO & 0.994(0.979;0.999) & 0.963(0.944;0.980)\\
Near big pasture area: YES & 0.002(0.000;0.009) & 0.005(0.000;0.014)\\
Near big pasture area: MISSING & 0.002(0.000;0.017) & \cellcolor{gray!25}0.031(0.014;0.048)\\
\bottomrule
\end{tabular}}
\end{minipage}%
\end{table}

\begin{table}[h!]
\caption{1987: posterior median and 0.1 and 0.9 quantiles of the kernels for behavioral and environmental variables. Light gray values indicate admissible conditions according to (7.3a) or (7.3b).}
\label{tab:supp_1987}
\begin{minipage}[t]{0.5\textwidth}
\vspace{0pt}
\resizebox{\textwidth}{!}{
\begin{tabular}{lll}
\toprule
 {\bf Behavioral }&  $\boldsymbol{\theta}^{(j)}_1$ &  $\boldsymbol{\theta}^{(j)}_2$ \\
\midrule
Plant coffee: NO & 0.002(0.000;0.014) & \cellcolor{gray!25}0.883(0.799;0.936)\\
Plant coffee: YES & 0.994(0.981;0.999) & 0.043(0.001;0.132)\\
Plant coffee: MISSING & 0.002(0.000;0.009) & \cellcolor{gray!25}0.068(0.042;0.099)\\
Plant banana: NO & 0.001(0.000;0.008) & \cellcolor{gray!25}0.664(0.588;0.756)\\
Plant banana: YES & 0.998(0.990;1.000) & 0.248(0.145;0.328)\\
\addlinespace
Plant banana: MISSING & 0.000(0.000;0.004) & \cellcolor{gray!25}0.088(0.062;0.121)\\
Own more the 4 goods: NO & 0.002(0.000;0.012) & \cellcolor{gray!25}0.750(0.669;0.833)\\
Own more the 4 goods: YES & 0.998(0.988;1.000) & 0.250(0.167;0.331)\\
Own a chainsaw: NO & 0.206(0.174;0.242) & \cellcolor{gray!25}0.944(0.833;0.995)\\
Own a chainsaw: YES & 0.792(0.756;0.824) & 0.051(0.001;0.161)\\
\addlinespace
Own a chainsaw: MISSING & 0.002(0.000;0.006) & 0.003(0.000;0.015)\\
More than 4 people in the house: NO & 0.621(0.590;0.652) & 0.004(0.000;0.041)\\
More than 4 people in the house: YES & 0.298(0.267;0.331) & \cellcolor{gray!25}0.990(0.953;0.999)\\
More than 4 people in the house: MISSING & 0.080(0.066;0.095) & 0.001(0.000;0.014)\\
Plan to build a new house within a year: NO & 0.822(0.787;0.858) & 0.219(0.115;0.324)\\
\addlinespace
Plan to build a new house within a year: YES & 0.172(0.136;0.208) & \cellcolor{gray!25}0.740(0.634;0.843)\\
Plan to build a new house within a year: MISSING & 0.004(0.000;0.014) & \cellcolor{gray!25}0.042(0.014;0.071)\\
Own chickens and/or porks: NO & 0.001(0.000;0.005) & \cellcolor{gray!25}0.413(0.351;0.478)\\
Own chickens and/or porks: YES & 0.999(0.994;1.000) & 0.399(0.323;0.471)\\
Own chickens and/or porks: MISSING & 0.000(0.000;0.002) & \cellcolor{gray!25}0.187(0.147;0.231)\\
\addlinespace
HH wife has high level of education: $<$ 4 yr & 0.570(0.533;0.607) & 0.088(0.002;0.190)\\
HH wife has high level of education: $>$ 4 yr & 0.398(0.365;0.434) & 0.309(0.217;0.398)\\
HH wife has high level of education: NO-WIFE & 0.001(0.000;0.008) & \cellcolor{gray!25}0.051(0.027;0.077)\\
HH wife has high level of education: MISSING & 0.027(0.003;0.054) & \cellcolor{gray!25}0.547(0.462;0.629)\\
Plant cocoa: NO & 0.452(0.421;0.483) & \cellcolor{gray!25}0.942(0.914;0.962)\\
\addlinespace
Plant cocoa: YES & 0.547(0.516;0.577) & 0.002(0.000;0.015)\\
Plant cocoa: MISSING & 0.000(0.000;0.004) & \cellcolor{gray!25}0.053(0.034;0.078)\\
Do you go often to urban area: NO & 0.429(0.393;0.468) & \cellcolor{gray!25}0.802(0.695;0.892)\\
Do you go often to urban area: YES & 0.535(0.496;0.573) & 0.108(0.009;0.215)\\
Do you go often to urban area: MISSING & 0.035(0.021;0.050) & \cellcolor{gray!25}0.090(0.048;0.138)\\
\addlinespace
Do you own other proprieties: NO & 0.487(0.453;0.523) & \cellcolor{gray!25}0.887(0.798;0.971)\\
Do you own other proprieties: YES & 0.513(0.477;0.547) & 0.113(0.029;0.202)\\
Arrived in Rondonia before 1985: NO & 0.956(0.934;0.976) & 0.574(0.496;0.656)\\
Arrived in Rondonia before 1985: YES & 0.023(0.006;0.044) & \cellcolor{gray!25}0.369(0.293;0.451)\\
Arrived in Rondonia before 1985: MISSING & 0.020(0.009;0.032) & \cellcolor{gray!25}0.053(0.019;0.094)\\
\addlinespace
DDT is used: NO & 0.091(0.066;0.119) & \cellcolor{gray!25}0.419(0.335;0.508)\\
DDT is used: YES & 0.907(0.880;0.933) & 0.560(0.471;0.646)\\
DDT is used: MISSING & 0.000(0.000;0.004) & \cellcolor{gray!25}0.019(0.006;0.036)\\
Before coming was your occupation rural: NO & 0.800(0.765;0.836) & 0.452(0.354;0.555)\\
Before coming was your occupation rural: YES & 0.200(0.164;0.235) & \cellcolor{gray!25}0.548(0.445;0.646)\\
\addlinespace
Use plant to cure malaria: NO & 0.448(0.410;0.485) & \cellcolor{gray!25}0.782(0.676;0.884)\\
Use plant to cure malaria: YES & 0.538(0.499;0.576) & 0.198(0.099;0.307)\\
Use plant to cure malaria: MISSING & 0.015(0.005;0.024) & 0.011(0.000;0.049)\\
Use protective clothes: NO & 0.855(0.823;0.884) & 0.514(0.412;0.603)\\
Use protective clothes: YES & 0.137(0.108;0.167) & 0.294(0.211;0.392)\\
\addlinespace
Use protective clothes: MISSING & 0.005(0.000;0.021) & \cellcolor{gray!25}0.190(0.137;0.245)\\
Spray insecticide: NO & 0.603(0.569;0.640) & \cellcolor{gray!25}0.917(0.815;0.991)\\
Spray insecticide: YES & 0.397(0.360;0.431) & 0.083(0.009;0.185)\\
Are there rubber tree: NO & 0.693(0.667;0.719) & 0.942(0.913;0.963)\\
Are there rubber tree: YES & 0.306(0.280;0.331) & 0.001(0.000;0.013)\\
\addlinespace
Are there rubber tree: MISSING & 0.000(0.000;0.004) & \cellcolor{gray!25}0.054(0.035;0.079)\\
Part of family did not come: NO & 0.639(0.603;0.674) & 0.359(0.260;0.458)\\
Part of family did not come: YES & 0.358(0.324;0.394) & 0.612(0.515;0.711)\\
Part of family did not come: MISSING & 0.001(0.000;0.006) & \cellcolor{gray!25}0.027(0.013;0.048)\\
Use a bednet: NO & 0.789(0.755;0.822) & 0.559(0.462;0.657)\\
\addlinespace
Use a bednet: YES & 0.204(0.172;0.238) & 0.250(0.151;0.342)\\
Use a bednet: MISSING & 0.004(0.000;0.018) & \cellcolor{gray!25}0.192(0.140;0.243)\\
Plant guarana: NO & 0.769(0.745;0.792) & 0.916(0.886;0.941)\\
Plant guarana: YES & 0.229(0.206;0.253) & 0.001(0.000;0.011)\\
Plant guarana: MISSING & 0.000(0.000;0.004) & \cellcolor{gray!25}0.081(0.056;0.110)\\
\addlinespace
HH has high level of education: NO & 0.631(0.596;0.667) & 0.415(0.313;0.513)\\
HH has high level of education: YES & 0.365(0.329;0.400) & 0.524(0.428;0.626)\\
HH has high level of education: MISSING & 0.002(0.000;0.012) & \cellcolor{gray!25}0.059(0.029;0.091)\\
Get malaria from dirty water: NO & 0.467(0.429;0.503) & 0.289(0.189;0.393)\\
Get malaria from dirty water: YES & 0.497(0.460;0.536) & 0.655(0.548;0.755)\\
\addlinespace
Get malaria from dirty water: MISSING & 0.035(0.021;0.051) & 0.054(0.014;0.102)\\
Do you ever go to city through BR364: NO & 0.576(0.538;0.612) & 0.602(0.503;0.698)\\
Do you ever go to city through BR364: YES & 0.407(0.370;0.444) & 0.287(0.187;0.380)\\
Do you ever go to city through BR364: MISSING & 0.017(0.003;0.032) & \cellcolor{gray!25}0.110(0.065;0.168)\\
Arrived in Machadino before 1985: NO & 0.168(0.144;0.191) & 0.025(0.002;0.084)\\
\addlinespace
Arrived in Machadino before 1985: YES & 0.832(0.809;0.856) & 0.975(0.916;0.998)\\
Knowledge of malaria vector: NO & 0.504(0.466;0.540) & 0.452(0.348;0.550)\\
Knowledge of malaria vector: YES & 0.327(0.292;0.360) & 0.332(0.243;0.430)\\
Knowledge of malaria vector: MISSING & 0.169(0.141;0.197) & 0.215(0.139;0.294)\\
Worked in rural area for more tha 1 year: NO & 0.032(0.006;0.049) & 0.059(0.011;0.141)\\
\addlinespace
Worked in rural area for more tha 1 year: YES & 0.967(0.950;0.993) & 0.892(0.807;0.942)\\
Worked in rural area for more tha 1 year: MISSING & 0.000(0.000;0.003) & \cellcolor{gray!25}0.047(0.029;0.071)\\
Lived in rural area for more than 1 year: NO & 0.028(0.003;0.045) & 0.046(0.001;0.129)\\
Lived in rural area for more than 1 year: YES & 0.972(0.954;0.997) & 0.902(0.819;0.953)\\
Lived in rural area for more than 1 year: MISSING & 0.000(0.000;0.002) & \cellcolor{gray!25}0.049(0.030;0.074)\\
\addlinespace
Own a planter: NO & 0.656(0.619;0.691) & 0.698(0.596;0.799)\\
Own a planter: YES & 0.343(0.308;0.380) & 0.297(0.198;0.397)\\
Own a planter: MISSING & 0.000(0.000;0.003) & \cellcolor{gray!25}0.004(0.000;0.014)\\
\bottomrule
\end{tabular}
}
\end{minipage}%
\begin{minipage}[t]{0.5\textwidth}
\vspace{0pt}
\resizebox{\textwidth}{!}{
\begin{tabular}{lll}
\toprule
 {\bf Environmental }&  $\boldsymbol{\theta}^{(j)}_1$ &  $\boldsymbol{\theta}^{(j)}_2$ \\
\midrule
House has more than 4 rooms: NO & 0.089(0.049;0.132) & \cellcolor{gray!25}0.905(0.851;0.954)\\
House has more than 4 rooms: YES & \cellcolor{gray!25}0.841(0.799;0.884) & 0.003(0.000;0.024)\\
House has more than 4 rooms: MISSING & 0.069(0.043;0.095) & 0.086(0.042;0.137)\\
Walls have good quality: NO & 0.166(0.123;0.213) & \cellcolor{gray!25}0.916(0.836;0.987)\\
Walls have good quality: YES & \cellcolor{gray!25}0.832(0.785;0.875) & 0.082(0.011;0.163)\\
\addlinespace
Walls have good quality: MISSING & 0.001(0.000;0.005) & 0.001(0.000;0.006)\\
Roof has good quality: NO & 0.003(0.000;0.026) & \cellcolor{gray!25}0.671(0.603;0.743)\\
Roof has good quality: YES & 0.995(0.971;0.999) & 0.325(0.253;0.394)\\
Roof has good quality: MISSING & 0.001(0.000;0.004) & 0.002(0.000;0.009)\\
Near big planted area: NO & 0.542(0.507;0.577) & 0.698(0.641;0.749)\\
\addlinespace
Near big planted area: YES & 0.444(0.411;0.478) & 0.001(0.000;0.018)\\
Near big planted area: NO-PLOT & 0.012(0.000;0.030) & \cellcolor{gray!25}0.091(0.055;0.128)\\
Near big planted area: MISSING & 0.000(0.000;0.002) & \cellcolor{gray!25}0.204(0.168;0.246)\\
Good water source available: NO & 0.241(0.200;0.282) & \cellcolor{gray!25}0.647(0.558;0.724)\\
Good water source available: YES & 0.758(0.717;0.799) & 0.349(0.271;0.438)\\
\addlinespace
Good water source available: MISSING & 0.000(0.000;0.003) & \cellcolor{gray!25}0.003(0.000;0.010)\\
Do you have close neighbours (<500mt): NO & 0.666(0.619;0.718) & 0.305(0.208;0.396)\\
Do you have close neighbours (<500mt): YES & 0.334(0.282;0.381) & 0.695(0.604;0.792)\\
Has the surrounding area being cleared: NO & 0.950(0.927;0.978) & 0.618(0.544;0.685)\\
Has the surrounding area being cleared: YES & 0.048(0.022;0.072) & \cellcolor{gray!25}0.345(0.282;0.416)\\
\addlinespace
Has the surrounding area being cleared: MISSING & 0.000(0.000;0.004) & \cellcolor{gray!25}0.035(0.021;0.053)\\
Is topography bottom: NO & 0.435(0.395;0.478) & 0.152(0.080;0.222)\\
Is topography bottom: YES & 0.528(0.485;0.570) & 0.801(0.722;0.875)\\
Is topography bottom: MISSING & 0.036(0.021;0.054) & 0.047(0.018;0.081)\\
Near big pasture area: NO & 0.950(0.934;0.965) & 0.721(0.667;0.769)\\
\addlinespace
Near big pasture area: YES & 0.048(0.034;0.064) & 0.010(0.000;0.044)\\
Near big pasture area: NO-PLOT & 0.000(0.000;0.005) & \cellcolor{gray!25}0.051(0.034;0.072)\\
Near big pasture area: MISSING & 0.000(0.000;0.002) & \cellcolor{gray!25}0.210(0.172;0.253)\\
More that 100mt from a forest: NO & 0.803(0.778;0.829) & 0.960(0.924;0.980)\\
More that 100mt from a forest: YES & 0.194(0.168;0.220) & 0.009(0.000;0.048)\\
\addlinespace
More that 100mt from a forest: MISSING & 0.002(0.000;0.008) & \cellcolor{gray!25}0.026(0.013;0.043)\\
Sealing has good quality: NO & 0.836(0.813;0.857) & 0.993(0.977;0.999)\\
Sealing has good quality: YES & 0.164(0.142;0.185) & 0.002(0.000;0.017)\\
Sealing has good quality: MISSING & 0.000(0.000;0.003) & \cellcolor{gray!25}0.003(0.000;0.010)\\
Distance from coop >200mt: NO & 0.990(0.976;0.999) & 0.895(0.860;0.929)\\
\addlinespace
Distance from coop >200mt: YES & 0.005(0.000;0.019) & \cellcolor{gray!25}0.065(0.035;0.094)\\
Distance from coop >200mt: MISSING & 0.002(0.000;0.011) & \cellcolor{gray!25}0.040(0.023;0.061)\\
More that 10km from an hospital: NO & 0.128(0.103;0.152) & 0.030(0.002;0.072)\\
More that 10km from an hospital: YES & 0.872(0.848;0.897) & 0.970(0.928;0.998)\\
Good bathing place is available: NO & 0.925(0.909;0.940) & 0.994(0.978;0.999)\\
\addlinespace
Good bathing place is available: YES & 0.073(0.059;0.089) & 0.003(0.000;0.018)\\
Good bathing place is available: MISSING & 0.001(0.000;0.004) & 0.001(0.000;0.008)\\
\bottomrule
\end{tabular}}
\end{minipage}
\end{table}

\begin{table}[h!]
  \caption{1995: posterior median and 0.1 and 0.9 quantiles of the kernels for behavioral and environmental variables. Light gray values indicate admissible conditions according to (7.3a) or (7.3b).}
\label{tab:supp_1995}

\begin{minipage}[t]{0.5\textwidth}
\vspace{0pt}
\resizebox{\textwidth}{!}{
\begin{tabular}{lll}
\toprule
 {\bf Behavioral }&  $\boldsymbol{\theta}^{(j)}_1$ &  $\boldsymbol{\theta}^{(j)}_2$ \\
\midrule
Planted Corn: NO & \cellcolor{gray!25}0.969(0.949;0.980) & 0.041(0.003;0.087)\\
Planted Corn: YES & 0.002(0.000;0.018) & \cellcolor{gray!25}0.958(0.912;0.996)\\
Planted Corn: MISSING & \cellcolor{gray!25}0.027(0.018;0.038) & 0.000(0.000;0.002)\\
Plant Cassava: NO & \cellcolor{gray!25}0.951(0.932;0.967) & 0.090(0.046;0.137)\\
Plant Cassava: YES & 0.001(0.000;0.012) & \cellcolor{gray!25}0.906(0.859;0.951)\\
\addlinespace
Plant Cassava: MISSING & \cellcolor{gray!25}0.045(0.030;0.062) & 0.002(0.000;0.010)\\
Plant banana: NO & \cellcolor{gray!25}0.942(0.876;0.966) & 0.187(0.147;0.232)\\
Plant banana: YES & 0.015(0.000;0.085) & \cellcolor{gray!25}0.812(0.767;0.852)\\
Plant banana: MISSING & \cellcolor{gray!25}0.038(0.027;0.052) & 0.000(0.000;0.003)\\
Cultivate rice: NO & \cellcolor{gray!25}0.740(0.670;0.823) & 0.001(0.000;0.005)\\
\addlinespace
Cultivate rice: YES & 0.231(0.144;0.304) & 0.999(0.994;1.000)\\
Cultivate rice: MISSING & \cellcolor{gray!25}0.027(0.018;0.040) & 0.000(0.000;0.002)\\
Planted Bean: NO & \cellcolor{gray!25}0.967(0.951;0.978) & 0.330(0.287;0.370)\\
Planted Bean: YES & 0.001(0.000;0.009) & 0.669(0.629;0.711)\\
Planted Bean: MISSING & \cellcolor{gray!25}0.030(0.020;0.044) & 0.001(0.000;0.004)\\
\addlinespace
Own a chainsaw: NO & \cellcolor{gray!25}0.947(0.869;0.998) & 0.298(0.250;0.346)\\
Own a chainsaw: YES & 0.053(0.002;0.131) & 0.702(0.654;0.750)\\
Active in community organization: NO & \cellcolor{gray!25}0.990(0.965;0.998) & 0.401(0.359;0.438)\\
Active in community organization: YES & 0.005(0.000;0.030) & 0.597(0.561;0.640)\\
Active in community organization: MISSING & \cellcolor{gray!25}0.003(0.000;0.009) & 0.001(0.000;0.004)\\
\addlinespace
More than 4 people in the house: NO & 0.167(0.091;0.249) & \cellcolor{gray!25}0.681(0.634;0.730)\\
More than 4 people in the house: YES & 0.833(0.751;0.909) & 0.319(0.270;0.366)\\
Own a planter: NO & \cellcolor{gray!25}0.461(0.410;0.517) & 0.001(0.000;0.007)\\
Own a planter: YES & 0.539(0.483;0.590) & 0.999(0.993;1.000)\\
Plant coffee: NO & \cellcolor{gray!25}0.403(0.355;0.457) & 0.001(0.000;0.006)\\
\addlinespace
Plant coffee: YES & 0.566(0.511;0.617) & 0.999(0.993;1.000)\\
Plant coffee: MISSING & \cellcolor{gray!25}0.029(0.019;0.042) & 0.000(0.000;0.002)\\
HH wife has high level of education: $<$ 4 yr & 0.148(0.077;0.210) & 0.468(0.428;0.505)\\
HH wife has high level of education: $>$ 4 yr & 0.503(0.438;0.572) & 0.502(0.462;0.540)\\
HH wife has high level of education: NO-WIFE & \cellcolor{gray!25}0.008(0.001;0.016) & 0.001(0.000;0.005)\\
\addlinespace
HH wife has high level of education: MISSING & \cellcolor{gray!25}0.343(0.279;0.407) & 0.027(0.000;0.061)\\
Own more the 4 goods: NO & \cellcolor{gray!25}0.340(0.300;0.385) & 0.001(0.000;0.008)\\
Own more the 4 goods: YES & 0.660(0.615;0.700) & 0.999(0.992;1.000)\\
Plant cocoa: NO & 0.965(0.946;0.977) & 0.660(0.630;0.688)\\
Plant cocoa: YES & 0.002(0.000;0.017) & 0.339(0.312;0.370)\\
\addlinespace
Plant cocoa: MISSING & \cellcolor{gray!25}0.030(0.021;0.043) & 0.000(0.000;0.002)\\
Are there rubber tree: NO & 0.975(0.959;0.985) & 0.680(0.652;0.707)\\
Are there rubber tree: YES & 0.002(0.000;0.016) & 0.319(0.293;0.348)\\
Are there rubber tree: MISSING & \cellcolor{gray!25}0.020(0.012;0.031) & 0.000(0.000;0.002)\\
Arrived in Rondonia before 1985: NO & 0.543(0.476;0.605) & 0.839(0.803;0.876)\\
\addlinespace
Arrived in Rondonia before 1985: YES & 0.396(0.334;0.461) & 0.150(0.117;0.184)\\
Arrived in Rondonia before 1985: MISSING & \cellcolor{gray!25}0.063(0.034;0.089) & 0.009(0.000;0.026)\\
Use plant to cure malaria: NO & 0.853(0.795;0.918) & 0.577(0.536;0.617)\\
Use plant to cure malaria: YES & 0.145(0.080;0.203) & 0.422(0.382;0.463)\\
Use plant to cure malaria: MISSING & 0.001(0.000;0.006) & 0.001(0.000;0.003)\\
\addlinespace
Plant guarana: NO & 0.979(0.942;0.998) & 0.737(0.709;0.766)\\
Plant guarana: YES & 0.021(0.002;0.058) & 0.263(0.234;0.291)\\
DDT is used: NO & 0.822(0.758;0.881) & 0.613(0.575;0.652)\\
DDT is used: YES & 0.149(0.092;0.215) & 0.386(0.347;0.424)\\
DDT is used: MISSING & \cellcolor{gray!25}0.027(0.017;0.040) & 0.000(0.000;0.003)\\
\addlinespace
Do you own other proprieties: NO & 0.784(0.712;0.851) & 0.565(0.522;0.611)\\
Do you own other proprieties: YES & 0.216(0.149;0.288) & 0.435(0.389;0.478)\\
Got a loan for pasture: NO & 0.963(0.947;0.974) & 0.788(0.765;0.809)\\
Got a loan for pasture: YES & 0.001(0.000;0.009) & 0.211(0.189;0.234)\\
Got a loan for pasture: MISSING & \cellcolor{gray!25}0.034(0.024;0.049) & 0.000(0.000;0.003)\\
\addlinespace
Planted Nut: NO & 0.947(0.920;0.964) & 0.801(0.776;0.824)\\
Planted Nut: YES & 0.003(0.000;0.026) & 0.197(0.174;0.221)\\
Planted Nut: MISSING & \cellcolor{gray!25}0.046(0.032;0.063) & 0.001(0.000;0.006)\\
Own chickens and/or porks: NO & \cellcolor{gray!25}0.188(0.158;0.221) & 0.001(0.000;0.003)\\
Own chickens and/or porks: YES & 0.812(0.779;0.842) & 0.999(0.997;1.000)\\
\addlinespace
Knowledge of malaria vector: NO & 0.310(0.240;0.379) & 0.461(0.419;0.504)\\
Knowledge of malaria vector: YES & 0.573(0.504;0.645) & 0.474(0.430;0.516)\\
Knowledge of malaria vector: MISSING & 0.117(0.072;0.163) & 0.064(0.041;0.091)\\
Planted Pepper: NO & 0.975(0.962;0.984) & 0.842(0.822;0.861)\\
Planted Pepper: YES & 0.001(0.000;0.008) & 0.157(0.138;0.178)\\
\addlinespace
Planted Pepper: MISSING & \cellcolor{gray!25}0.022(0.014;0.033) & 0.000(0.000;0.002)\\
Get malaria from dirty water: NO & 0.507(0.433;0.582) & 0.379(0.335;0.421)\\
Get malaria from dirty water: YES & 0.462(0.388;0.535) & 0.602(0.560;0.645)\\
Get malaria from dirty water: MISSING & 0.028(0.007;0.055) & 0.020(0.006;0.033)\\
Got a loan for agriculture: NO & 0.963(0.948;0.975) & 0.855(0.835;0.872)\\
\addlinespace
Got a loan for agriculture: YES & 0.001(0.000;0.006) & 0.144(0.127;0.164)\\
Got a loan for agriculture: MISSING & \cellcolor{gray!25}0.034(0.023;0.049) & 0.000(0.000;0.003)\\
Spray insecticide: NO & 0.850(0.789;0.909) & 0.712(0.674;0.749)\\
Spray insecticide: YES & 0.148(0.089;0.210) & 0.286(0.250;0.324)\\
Spray insecticide: MISSING & 0.001(0.000;0.005) & 0.001(0.000;0.003)\\
\addlinespace
Do you go often to urban area: NO & 0.492(0.405;0.580) & 0.618(0.567;0.666)\\
Do you go often to urban area: YES & 0.497(0.409;0.582) & 0.380(0.332;0.431)\\
Do you go often to urban area: MISSING & \cellcolor{gray!25}0.010(0.003;0.019) & 0.001(0.000;0.006)\\
Lived in rural area for more than 1 year: NO & \cellcolor{gray!25}0.140(0.101;0.191) & 0.038(0.013;0.059)\\
Lived in rural area for more than 1 year: YES & 0.847(0.797;0.888) & 0.960(0.939;0.986)\\
\addlinespace
Lived in rural area for more than 1 year: MISSING & \cellcolor{gray!25}0.011(0.004;0.021) & 0.000(0.000;0.004)\\
Arrived in Machadino before 1985: NO & 0.001(0.000;0.013) & 0.096(0.081;0.112)\\
Arrived in Machadino before 1985: YES & 0.968(0.943;0.991) & 0.891(0.870;0.910)\\
Arrived in Machadino before 1985: MISSING & 0.028(0.007;0.051) & 0.012(0.001;0.026)\\
Use a bednet: NO & 0.880(0.835;0.920) & 0.963(0.941;0.984)\\
\addlinespace
Use a bednet: YES & \cellcolor{gray!25}0.104(0.068;0.146) & 0.029(0.008;0.048)\\
Use a bednet: MISSING & 0.014(0.001;0.033) & 0.008(0.000;0.018)\\
Have another rural plot: NO & 0.839(0.779;0.898) & 0.763(0.727;0.800)\\
Have another rural plot: YES & 0.159(0.100;0.218) & 0.236(0.200;0.272)\\
Have another rural plot: MISSING & 0.001(0.000;0.006) & 0.001(0.000;0.003)\\
\addlinespace
HH has high level of education: NO & 0.392(0.313;0.462) & 0.441(0.401;0.486)\\
HH has high level of education: YES & 0.597(0.527;0.675) & 0.557(0.513;0.597)\\
HH has high level of education: MISSING & \cellcolor{gray!25}0.010(0.003;0.020) & 0.001(0.000;0.005)\\
Go to main urban area from treatment: NO & 0.199(0.131;0.277) & 0.165(0.122;0.206)\\
Go to main urban area from treatment: YES & 0.790(0.711;0.859) & 0.834(0.794;0.877)\\
\addlinespace
Go to main urban area from treatment: MISSING & \cellcolor{gray!25}0.010(0.004;0.017) & 0.000(0.000;0.002)\\
Go to secondary urban area from treatment: NO & 0.829(0.753;0.898) & 0.829(0.789;0.871)\\
Go to secondary urban area from treatment: YES & 0.158(0.089;0.233) & 0.170(0.129;0.211)\\
Go to secondary urban area from treatment: MISSING & \cellcolor{gray!25}0.013(0.006;0.022) & 0.000(0.000;0.002)\\
Got a loan for equipment: NO & 0.963(0.949;0.975) & 0.981(0.973;0.987)\\
\addlinespace
Got a loan for equipment: YES & 0.000(0.000;0.005) & 0.018(0.012;0.025)\\
Got a loan for equipment: MISSING & \cellcolor{gray!25}0.036(0.024;0.049) & 0.000(0.000;0.003)\\
\bottomrule
\end{tabular}
}
\end{minipage}%
\begin{minipage}[t]{0.5\textwidth}
\vspace{0pt}
\resizebox{\textwidth}{!}{
\begin{tabular}{lll}
\toprule
 {\bf Environmental }&  $\boldsymbol{\theta}^{(j)}_1$ &  $\boldsymbol{\theta}^{(j)}_2$ \\
\midrule
House has more than 4 rooms: NO & \cellcolor{gray!25}0.923(0.781;0.969) & 0.001(0.000;0.010)\\
House has more than 4 rooms: YES & 0.024(0.000;0.184) & 0.996(0.988;1.000)\\
House has more than 4 rooms: MISSING & \cellcolor{gray!25}0.040(0.013;0.072) & 0.001(0.000;0.006)\\
More that 10km from an hospital: NO & \cellcolor{gray!25}0.568(0.432;0.710) & 0.007(0.000;0.024)\\
More that 10km from an hospital: YES & 0.432(0.290;0.568) & 0.993(0.976;1.000)\\
\addlinespace
Anybody cleared the area before HH: NO & 0.006(0.000;0.050) & 0.332(0.311;0.354)\\
Anybody cleared the area before HH: YES & 0.697(0.609;0.779) & 0.667(0.645;0.687)\\
Anybody cleared the area before HH: MISSING & \cellcolor{gray!25}0.285(0.212;0.370) & 0.001(0.000;0.004)\\
Has the surrounding area being cleared: NO & 0.673(0.576;0.760) & 0.996(0.990;0.999)\\
Has the surrounding area being cleared: YES & \cellcolor{gray!25}0.006(0.000;0.031) & 0.002(0.000;0.005)\\
\addlinespace
Has the surrounding area being cleared: MISSING & \cellcolor{gray!25}0.316(0.232;0.407) & 0.001(0.000;0.007)\\
Do you have close neighbours (<500mt): NO & 0.427(0.234;0.615) & 0.661(0.630;0.690)\\
Do you have close neighbours (<500mt): YES & 0.367(0.188;0.551) & 0.309(0.280;0.339)\\
Do you have close neighbours (<500mt): MISSING & \cellcolor{gray!25}0.205(0.095;0.324) & 0.030(0.016;0.046)\\
Is topography bottom: NO & 0.937(0.799;0.997) & 0.688(0.664;0.711)\\
\addlinespace
Is topography bottom: YES & 0.057(0.001;0.193) & 0.305(0.283;0.329)\\
Is topography bottom: MISSING & 0.002(0.000;0.018) & 0.007(0.004;0.011)\\
Is road quality good: NO & 0.005(0.000;0.047) & 0.094(0.082;0.107)\\
Is road quality good: YES & 0.534(0.330;0.730) & 0.644(0.612;0.677)\\
Is road quality good: MISSING & 0.452(0.257;0.646) & 0.261(0.231;0.291)\\
\addlinespace
Near big pasture area: NO & 0.801(0.735;0.858) & 0.998(0.994;1.000)\\
Near big pasture area: YES & \cellcolor{gray!25}0.010(0.000;0.029) & 0.001(0.000;0.003)\\
Near big pasture area: MISSING & \cellcolor{gray!25}0.187(0.131;0.251) & 0.000(0.000;0.005)\\
Distant from stagnant water: NO & 0.796(0.708;0.873) & 0.975(0.965;0.984)\\
Distant from stagnant water: YES & 0.002(0.000;0.024) & 0.018(0.013;0.024)\\
\addlinespace
Distant from stagnant water: MISSING & \cellcolor{gray!25}0.196(0.122;0.278) & 0.006(0.000;0.016)\\
More than 600mt from a river: NO & 0.558(0.345;0.772) & 0.635(0.602;0.667)\\
More than 600mt from a river: YES & 0.363(0.149;0.580) & 0.365(0.332;0.397)\\
More than 600mt from a river: MISSING & \cellcolor{gray!25}0.076(0.048;0.116) & 0.000(0.000;0.002)\\
Roof has good quality: NO & \cellcolor{gray!25}0.200(0.120;0.282) & 0.008(0.001;0.019)\\
\addlinespace
Roof has good quality: YES & 0.800(0.718;0.880) & 0.992(0.981;0.999)\\
Sealing has good quality: NO & 0.499(0.291;0.686) & 0.641(0.609;0.671)\\
Sealing has good quality: YES & 0.501(0.314;0.709) & 0.359(0.329;0.391)\\
Distance from coop >200mt: NO & 0.802(0.719;0.860) & 0.908(0.895;0.922)\\
Distance from coop >200mt: YES & 0.013(0.000;0.083) & 0.091(0.078;0.104)\\
\addlinespace
Distance from coop >200mt: MISSING & \cellcolor{gray!25}0.174(0.126;0.233) & 0.000(0.000;0.003)\\
Walls have good quality: NO & 0.387(0.211;0.598) & 0.262(0.232;0.290)\\
Walls have good quality: YES & 0.613(0.402;0.789) & 0.738(0.710;0.768)\\
Distant from to well: NO & 0.852(0.689;0.929) & 0.878(0.860;0.900)\\
Distant from to well: YES & 0.052(0.001;0.218) & 0.120(0.099;0.138)\\
\addlinespace
Distant from to well: MISSING & \cellcolor{gray!25}0.085(0.049;0.130) & 0.001(0.000;0.005)\\
Good water source available: NO & 0.217(0.049;0.400) & 0.210(0.184;0.238)\\
Good water source available: YES & 0.777(0.593;0.945) & 0.789(0.761;0.815)\\
Good water source available: MISSING & \cellcolor{gray!25}0.003(0.000;0.016) & 0.000(0.000;0.002)\\
More that 100mt from a forest: NO & 0.917(0.809;0.962) & 0.854(0.836;0.871)\\
\addlinespace
More that 100mt from a forest: YES & 0.031(0.000;0.145) & 0.146(0.128;0.164)\\
More that 100mt from a forest: MISSING & \cellcolor{gray!25}0.045(0.024;0.075) & 0.000(0.000;0.002)\\
Good bathing place is available: NO & 0.985(0.924;0.999) & 0.873(0.857;0.888)\\
Good bathing place is available: YES & 0.015(0.001;0.076) & 0.127(0.112;0.143)\\
More than 500mt from health unit: NO & \cellcolor{gray!25}0.086(0.016;0.181) & 0.037(0.024;0.049)\\
More than 500mt from health unit: YES & 0.914(0.819;0.984) & 0.963(0.951;0.976)\\
\bottomrule
\end{tabular}}
\end{minipage}
\end{table}


\clearpage
\bibliographystyle{imsart-nameyear}
{
\def\spacingset#1{\renewcommand{\baselinestretch}{#1}\small } \spacingset{0.95}
\bibliography{bibliography}
}
\end{document}